\documentclass{article}

\PassOptionsToPackage{square,sort&compress}{natbib}
\PassOptionsToPackage{table}{xcolor}

\usepackage[final]{neurips_2024}

\newif\ifshowextra
\showextrafalse

\usepackage[utf8]{inputenc} 
\usepackage[T1]{fontenc}    
\usepackage{hyperref}       
\usepackage{url}            
\usepackage{booktabs}       
\usepackage{amsfonts}       
\usepackage{nicefrac}       
\usepackage{microtype}      

\usepackage{amsmath}
\usepackage{array}
\usepackage[english]{babel}
\usepackage{csquotes}
\usepackage{enumitem} 
\usepackage{etoolbox}
\usepackage{float}
\usepackage{fontawesome}  
\usepackage{graphicx} 
\usepackage{import}
\usepackage{lipsum}     
\usepackage{longtable}
\usepackage{pdflscape} 
\usepackage{tabularx} 
\usepackage[most]{tcolorbox}
\usepackage{tikz}
\usetikzlibrary{arrows.meta,positioning,fit,shapes.misc}
\usepackage[colorinlistoftodos]{todonotes}
\usepackage{wrapfig} 
\usepackage[table]{xcolor}

\usepackage[acronym,toc]{glossaries}
\setacronymstyle{long-short}
\makeglossaries

\newglossaryentry{nowcasting}{
    name={nowcasting},
    description={The use of mathematical models for the quantitative estimation of a current situation that include adjustments to imperfect data}
}

\newacronym[
    description={A metric for comparing statistical models that balances how well a model fits the data against its complexity}
]{aic}{AIC}{Akaike Information Criterion}

\newacronym[
    description={A forecasting method that predicts future values based on patterns in past values, including trends and momentum}
]{arima}{ARIMA}{Autoregressive Integrated Moving Average}

\newglossaryentry{backcasting}{
    name={backcasting},
    description={A planning method that starts with defining a desirable future and then works backwards to identify policies and programs that will connect that specified future to the present}
}

\newacronym[
    description={A metric similar to AIC for comparing statistical models, but with a stronger penalty for model complexity}
]{bic}{BIC}{Bayesian Information Criterion}

\newglossaryentry{changepoint_detection}{
    name={changepoint detection},
    description={A statistical method for identifying points in time when the underlying pattern in data fundamentally shifts}   
}

\newglossaryentry{endemic}{
    name={endemic},
    description={A condition or harm that persists at a stable, ongoing level within a population or system rather than appearing in isolated outbreaks}
}

\newglossaryentry{exposure_normalization}{
    name={exposure normalization},
    description={The process of adjusting raw counts by a measure of how widely something is deployed or used, to enable fair comparison of risk rates}
}

\newglossaryentry{exposure_offset}{
    name={exposure offset},
    description={A technical adjustment in statistical models that incorporates deployment scale into the analysis, enabling calculation of per-unit-of-exposure rates}
}

\newacronym[
    description={A statistical adjustment applied when testing multiple hypotheses simultaneously to control the proportion of false positive findings}
]{fdr}{FDR}{False Discovery Rate correction}

\newglossaryentry{gaussian_emissions}{
    name={Gaussian emissions},
    description={An assumption in Hidden Markov Models that the observable data generated by each hidden state follows a normal (bell-shaped) distribution}
}

\newacronym[
    description={A flexible statistical method for making predictions that also quantifies uncertainty in those predictions}
]{gpr}{GPR}{Gaussian Process Regression}

\newacronym[
    description={A statistical model that infers unobservable underlying states from observable data, assuming the system transitions between states over time}   
]{hmm}{HMM}{Hidden Markov Model}

\newglossaryentry{kmeans}{
    name={K-means cluster analysis},
    description={A method for grouping observations into a specified number of clusters based on their similarity to each other}
}

\newglossaryentry{latent_state}{
    name={latent-state},
    description={An underlying condition or status that cannot be directly observed but must be inferred from available evidence}
}

\newglossaryentry{lognormal}{
    name={log-normal},
    description={A distribution for positive quantities where most values cluster at lower levels but rare cases are much larger, a pattern that arises when the logarithm of the quantity follows a normal distribution}   
}

\newacronym[
    description={A statistical method for analyzing count data that accommodates high variability in the counts}
]{negative_binomial}{NB}{negative binomial regression}

\newglossaryentry{overdispersion}{
    name={overdispersion},
    description={A situation where the variability in data is greater than what a standard statistical model would predict}
}

\newacronym[
    description={A specific algorithm for detecting multiple changepoints in data efficiently}
]{pelt}{PELT}{Pruned Exact Linear Time}

\newglossaryentry{proxy}{
    name={proxy},
    plural={proxies},
    description={A variable that is not in itself directly relevant, but that serves in place of an unobservable or immeasurable variable, ideally having a close correlation with the variable of interest}
}

\newglossaryentry{reporting_lag}{
    name={reporting lag},
    description={The delay between when an incident occurs and when it is recorded or published in a database}
}

\newglossaryentry{silhouette}{
    name={Silhouette score},
    description={A measure of how well-separated and internally cohesive clusters are, ranging from -1 to 1, with higher values indicating better-defined groupings}
}

\newglossaryentry{ssmodel}{
    name={state-space model},
    description={A mathematical model of a physical system that uses state variables to track how inputs shape system behavior over time}
}

\newglossaryentry{standard_deviation}{
    name={standard deviation},
    description={A measure of how spread out values are from the average; a higher standard deviation means greater variability}
}

\newglossaryentry{standardized_risk}{
    name={standardized risk},
    description={Risk expressed in terms of standard deviations from the mean, allowing comparison across different scales or time periods}
}

\newglossaryentry{tseries}{
    name={time series},
    description={A series of data points indexed in time order, commonly taken at successive equally spaced points in time}
}

\newglossaryentry{zscore}{
    name={$z$-score},
    description={A measure expressing how many standard deviations a value is above or below the average}
}

\newcommand{\keywords}[1]{\textbf{Keywords:} #1}

\newcolumntype{C}[1]{>{\centering\arraybackslash}p{#1}}

\title{\texorpdfstring{AI Incident Monitoring through a Public Health Lens}}

\author{%
    Sophia Abraham\thanks{denotes the equal contribution} \\Arcadia Impact AI\\Governance Taskforce
    \And
    Taiye Chen$^*$ \\Arcadia Impact AI\\Governance Taskforce
    \And
    Cyril Chhun \\Arcadia Impact AI\\Governance Taskforce
    \And
    Giovanna Jaramillo-Gutierrez \\Arcadia Impact AI\\Governance Taskforce
    \And
    Simon Mylius \\Arcadia Impact AI\\Governance Taskforce, Arcola AI
    \And
    Sayash Raaj\\Arcadia Impact AI\\Governance Taskforce
    \And
    Peter Slattery\\MIT FutureTech
    \And
    Sean McGregor\\Responsible AI Collaborative
}

\date{November 2025}

\begin{document}

\maketitle

\begin{abstract}
Artificial intelligence systems are now deployed at scale across sectors, accompanied by a growing number of real-world incidents ranging from misinformation and cybercrime to autonomous-system failures. Databases of AI incidents index these events, but they cannot measure ``risk'' (i.e., a joint measure of likelihood and severity) without additional data regarding the prevalence of risk-associated systems and their incident reporting rates. As a result, policymakers, companies, and the general public lack a means to weigh the benefits of AI against their in-context risks. Inspired by public-health processes, which presume noisy and incomplete disease surveillance, we identify six phases of incident emergence. We demonstrate the framework through a detailed case study of autonomous vehicles, whose mandatory reporting requirements produces reliable incident-rate ground truth expressed in distance traveled. The case study shows that an informed panel of domain experts (e.g., self-driving experts) can combine their domain expertise, incident data, and a collection of statistical and visualization tools to arrive at incident phase determinations serving public needs. We further demonstrate the approach with a deepfake incident case study and chart a path for future research in incident phase determination.
\end{abstract}

\vspace*{1em}


\clearpage

\newpage

\section{Introduction}
\label{sec:1_introduction}
Artificial intelligence (AI) systems are increasingly embedded in social and economic infrastructure: they moderate information flows, assist decision-making, and operate physical systems. As deployments scale, a growing number of real-world incidents ranging from misinformation and fraud enabled by generative models to failures in autonomous systems enter public databases and media accounts. Repositories such as the AI Incident Database (AIID) \citep{mcgregor2021preventing} and the OECD AI Incidents and Hazards Monitor (AIM) \citep{perset2025framework} aggregate these accounts, but their data sources provide an incomplete view of underlying risk. Uncorrected incident counts from incident databases that do not benefit from mandatory reporting requirements are confounded by at least three distinct factors: (1) the unknown true underlying frequency of harmful events per unit of exposure, (2) the scale of system deployment (exposure), and (3) the propensity for incidents to be observed and reported (\gls{reporting_lag} and media amplification). Without disentangling these processes, decision-makers \textit{cannot reliably discern} whether an increase in reports reflects a genuine rise in per-exposure risk, a growth in use, or a media-driven reporting spike \citep{paeth2025lessons}.

To solve the risk identification gap, we adapt concepts and tools from public-health surveillance. Public-health agencies have long handled imperfect surveillance data-streams distorted by reporting delays, variable testing, and media attention by combining delay-adjustment, exposure denominators, and phase-based reasoning to create actionable situational awareness \citep{declich1994public}. We translate this logic into a phase-based framework for AI incidents: instead of treating incident counts as a direct measure of harm, we treat them as a noisy signal from which latent lifecycle \emph{phases} can be inferred and tracked over time. These phases (defined in Section~\ref{sub:2_phase_definitions}) are intentionally governance-oriented: each corresponds to a different set of plausible policy responses, from early-warning monitoring to mandatory reporting, audits, and sustained enforcement.

Methodologically, our contribution is twofold. First, we specify a data-processing and modeling pipeline that combines reporting-delay correction, exposure normalization, media-adjustment, and \gls{latent_state} or \glslink{changepoint_detection}{changepoint} inference. We find this and all similar model-centric pipelines to be inadequate for risk determinations. We therefore next demonstrate the utility of applying the data-processing pipelines to inform public health-styled declarations made by one or more domain experts. As case studies, we make preliminary phase declarations for autonomous vehicle (AV) incidents and deepfake-enabled (DF) fraud.

While the case studies demonstrate it is not feasible under current incident reporting regimes to ascertain exact risk likelihood, when informed by expert judgment it is possible to make \emph{phase determinations}: classifying the latent state of a harm (e.g., rare, expanding, endemic) by humans knowledgeable in the state of technology and its deployment contexts rather than algorithmic count-based triggers.

The importance of model-augmented expert determination of incident phases was recently illustrated by the impact of a recent blog post (\cite{atherton2026incidentroundup}) published by the AI Incident Database. The blog's prominent statement of ``Deepfake-enabled fraud is now a default business model'' led to an article published in the Guardian four days later with the headline, ``Deepfake fraud taking place on an industrial scale, study finds'' (\cite{guardian2026deepfake}). For an informal blog post to be given the imprimatur of a ``study'' is a sign of the demand for such declarations by the public -- and the responsibility in correctly making such declarations. The purpose of this paper is to lay out the challenge in improving such declarations and suggest a best path forward: \textit{public health-inspired phase declarations.}

The paper unfolds as follows. In Section~\ref{sec:2_conceptual_framework}, we lay out our conceptual framework, defining our model phases as well as our statistical methods. In Section~\ref{sec:3_autonomous_vehicles}, we present the AV case study and its results. In Section~\ref{sec:4_deepfakes}, we apply the framework to deepfake-enabled incidents, revealing a different trajectory of irreversible phase transition. Finally, in Section~\ref{sec:5_discussion}, we derive implications from our research and suggest directions for future work.

\section{Conceptual Framework}
\label{sec:2_conceptual_framework}
We adapt epidemiological surveillance techniques to AI incident data by combining incident-rate models with exposure offsets, \gls{nowcasting} procedures to correct reporting delays, and \glspl{ssmodel} to characterize and forecast phases of AI incident emergence across domains and jurisdictions. In public health, these approaches formally separate a stable background (``\gls{endemic}'') level from episodic surges (``epidemics''), or use regime-switching models to detect structural shifts in transmission and reporting dynamics \citep{declich1994public}. We apply the same logic to AI incidents to distinguish routine baseline harms from spikes driven by new model capabilities, deployment contexts, or governance shocks.

Methodologically, our contribution is to apply these \gls{tseries} and state-space tools to noisy AI incident data. This allows us to move beyond raw incident counts and instead infer latent risk dynamics probabilistically, incorporating uncertainty from incomplete observability. By integrating exposure \glslink{proxy}{proxies}, delay-adjustment methods, and phase-switching models, our framework provides a principled basis for identifying emergence, acceleration, stabilization, and mitigation phases even under imperfect surveillance conditions.


The framework presented below is intended as an exploratory application of epidemiological surveillance concepts to AI incident data. Several important caveats apply throughout. First, the AI Incident Database (AIID) captures media-reported incidents, which systematically underrepresents minor events and overrepresents high-profile cases; our analysis characterizes \textit{reporting patterns} rather than ground-truth incident rates. Second, the mapping between statistical model outputs (e.g., \gls{hmm} states, \gls{pelt} segments) and governance-relevant phases is interpretive rather than deterministic---different analysts might draw phase boundaries differently. Third, exposure proxies are imperfect. Where AI deployments may be global, information about those deployments may be geographically or contextually heterogeneous.
We present quantitative results as indicative of plausible dynamics rather than precise measurements, and encourage readers to focus on qualitative patterns (e.g., oscillatory vs. monotonic trajectories).

\subsection{Phase Definitions for AI Incident Emergence}
\label{sub:2_phase_definitions}

AI incident data shares many of the structural challenges long recognized in public-health surveillance: reporting delays, uneven observability, heterogeneous data quality, media-driven visibility spikes, and rapidly shifting underlying conditions. During the COVID-19 pandemic, such imperfect and unstable data could not support precise real-time estimates, yet governments still required actionable decision frameworks. Public-health agencies therefore relied on phase classifications emergence, acceleration, peak, decline, and endemicity to structure uncertainty and to match interventions to the evolving risk landscape, even when case counts were incomplete or biased \citep{haldane2021health}.

AI incident surveillance faces a similar dilemma. The available datasets (AIID \citep{mcgregor2021preventing}, AIM \citep{perset2025framework}) provide noisy but informative signals rather than comprehensive surveillance of harm events. Perfect measurement is not feasible, yet governance decisions whether to intervene, escalate oversight, or maintain monitoring cannot wait for perfect data. For this reason, we adopt a phase-based framework analogous to epidemic intelligence: a structured, interpretable, and governance-aligned method for inferring where a risk sits within its lifecycle.

\subsubsection*{Linking Phases to Governance: Phase-Specific Policy Levers}

The six phases below reflect the theoretical lifecycle of an AI incident type, analogous to the lifecycle of an infectious hazard in epidemiology. These phases are not tied to any specific dataset; they form the conceptual foundation on which statistical modeling and governance interpretation build. 

\begin{figure}[ht]
    \centering
    \includegraphics[width=\linewidth]{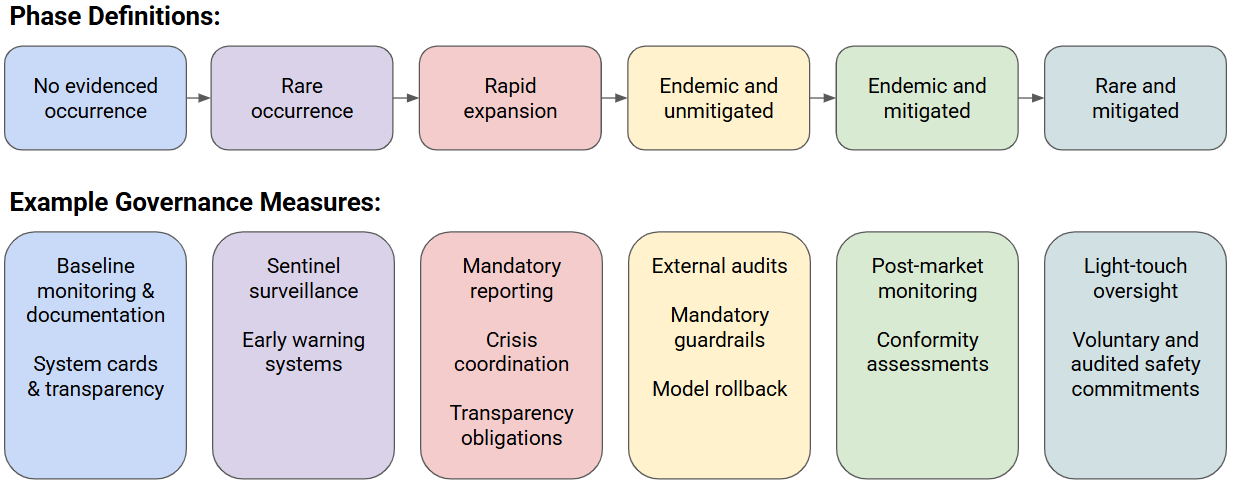}
    \caption{AI Incident Lifecycle Phases and Corresponding Example Governance Measures.}
    \label{fig:ai_incident_lifecycle}
\end{figure}

Figure~\ref{fig:ai_incident_lifecycle} presents our proposed six conceptual phases of AI incident emergence. Besides the obvious progression forwards through the phases, other transitions may be possible, For example, if a mitigation ceases to be effective, an incident type could move ``back'' from endemic mitigated to endemic unmitigated.

Because each phase reflects a distinct pattern of underlying risk, it can be paired with phase-specific governance measures,  ranging from early-warning monitoring and sandboxing to post-market surveillance, mandatory incident reporting, audits, and recertification. This mirrors the logic used in pandemic response, where different phases activate different layers of intervention \citep{migus2020covid}.

This introduces the interpretive structure used throughout the paper for classifying AI-incident-type trajectories.

\begin{enumerate}
    \item \textbf{No Evidenced Occurrence}: no incidents have yet been observed. This reflects limitations in detection and reporting rather than guaranteed absence of harm; it should be interpreted as ``absence of evidence'', not ``evidence of absence''.
    \item \textbf{Rare Occurrence}: incidents appear sporadically and remain at very low exposure-adjusted rates. This is an early-warning stage indicating that a harm type has emerged but has not yet demonstrated sustained growth.
    \item \textbf{Rapid Expansion and Unmitigated}: incident rates increase faster than underlying system deployment. Severity may rise, clusters may appear, and media amplification may further accelerate visibility. Effective controls are not yet in place.
    \item \textbf{Endemic and Unmitigated}: incidents stabilize at a persistently high baseline after adjusting for exposure. The harm becomes a routine feature of the system, and mitigation efforts have not meaningfully reduced frequency or severity.
    \item \textbf{Endemic and Mitigated}: incidents continue to occur but at lower exposure-adjusted rates and with reduced severity. Technical, organizational, or regulatory mitigation measures are visibly containing the harm.
    \item \textbf{Rare and Mitigated}: incidents occur infrequently and with low severity relative to exposure. Controls are functioning effectively, though underlying risk is not eliminated. This represents a plausible long-term target for governance.
\end{enumerate}

\subsection{Model Inputs}
\label{sub:2_model_inputs}

\begin{figure}[ht]
    \centering
    \includegraphics[width=\linewidth]{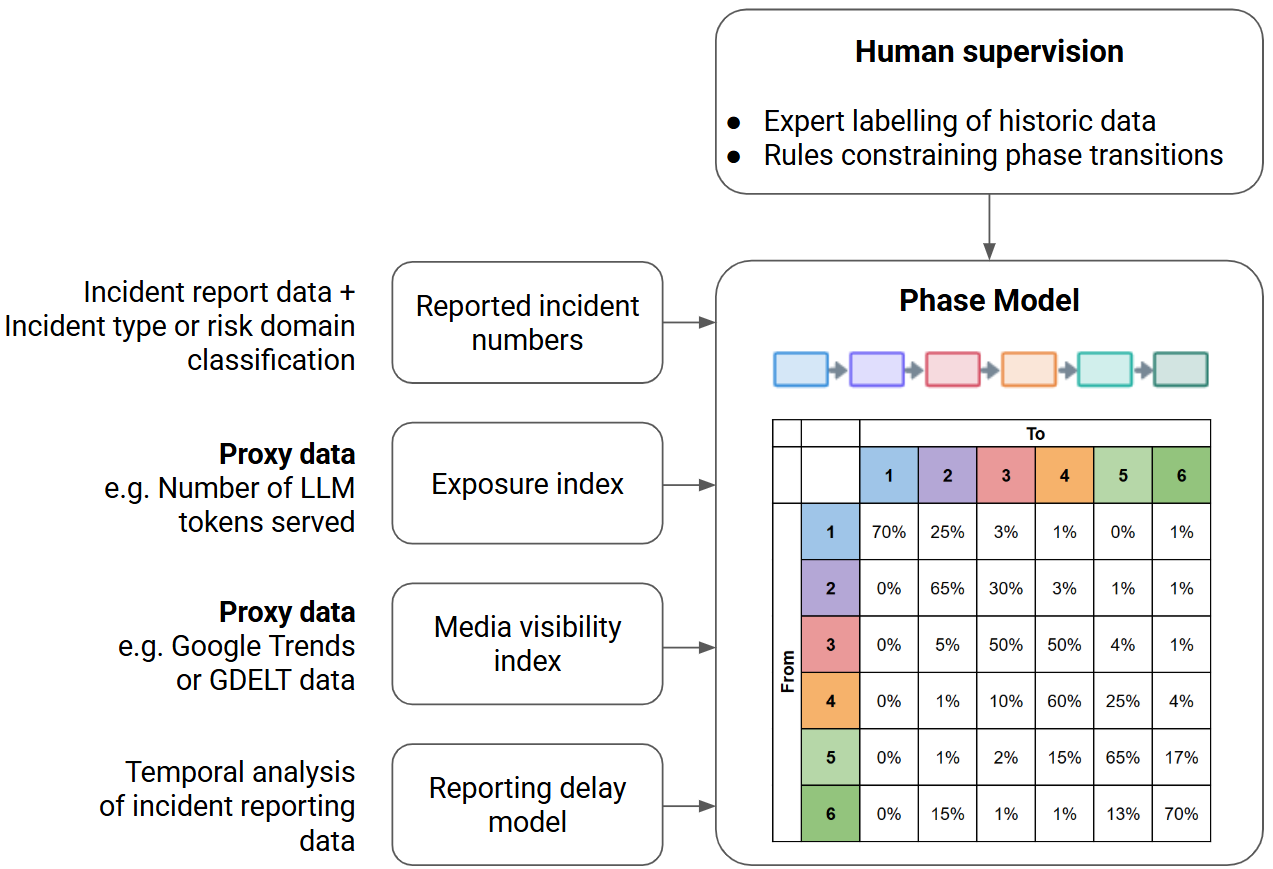}
    \caption{Inputs to the Phase Model}
    \label{fig:model_inputs}
\end{figure}

Our demonstration model combines four types of signals:

\begin{enumerate}
    \item \textbf{AI incident data}: documented incidents involving a specific system type (e.g., emotional manipulation, suicide-related harms, or unsafe advice from mental-health chatbots). We have collected the incident reporting data available from the AI Incident Database (AIID), a crowdsourced incident collection \citep{mcgregor2021preventing}. As of October 13th 2025, it contained 1,200 incidents with 5,300 corresponding reports. All incidents have been automatically classified with an LLM-based tool \citep{mylius2024scalable} according to the MIT Risk Repository’s causal and domain taxonomies \citep{slattery2024ai}. 
    \item \textbf{Exposure \glspl{proxy}}: domain-specific denominators that convert raw counts into rates. For AVs, we use California DMV records of autonomous miles driven; for deepfakes, we construct a depreciated installed base from GitHub Stars of major open-source repositories (see Table~\ref{tab:exposure_media_cases}).
    \item \textbf{Reporting-delay model}: AI incidents are typically recorded by their report date, which can lag days to months behind the event date. To correct for this distortion, we estimate a reporting-delay distribution using available metadata and apply a \gls{nowcasting} procedure to reconstruct the underlying event-time series. The delay-corrected \gls{tseries}, rather than the raw report-time counts, is used in phase inference.
    \item \textbf{Media-derived indicators}: Google Trends data serve as a covariate that captures fluctuations in public attention and the visibility of incidents, allowing the model to distinguish true changes in underlying risk from attention-driven reporting spikes.
\end{enumerate}

\subsection{Model Outputs}
\label{sub:2_model_outputs}

\begin{figure}[ht]
    \centering
    \includegraphics[width=\linewidth]{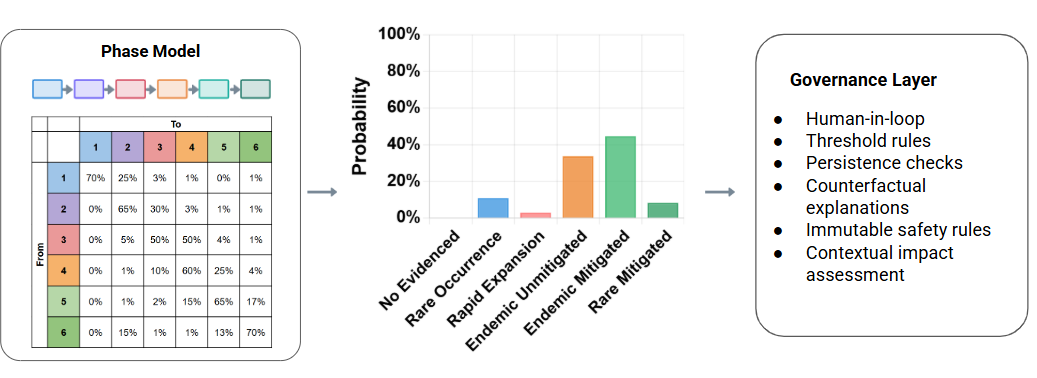}
    \caption{Outputs from the Phase Model}
    \label{fig:model_outputs}
\end{figure}

The phase model is designed to produce two primary quantitative outputs. First, a transition matrix captures the estimated probabilities of moving between phases over the forecast horizon, where rows represent current states and columns represent future states. The diagonal elements indicate persistence probability (\textit{the likelihood that an incident category remains in its current phase}) while off-diagonal elements capture the probability of phase transitions. Figure \ref{fig:model_outputs} illustrates this structure with a hypothetical example. A category currently in ``Rare Occurrence'' (row 2) might show a 65\% probability of remaining in that phase, a 30\% probability of transitioning to ``Rapid Expansion,'' and smaller probabilities of moving to later stages. The matrix structure enforces plausible dynamics—early-phase categories can only progress forward or persist, while late-phase categories (e.g., ``Rare Mitigated'') may show non-zero probability of regression, reflecting potential mitigation failure or incident recurrence.

Second, the model outputs a marginal probability distribution across all six phases, representing the forecast-period distribution of incident categories. These quantitative outputs feed into a governance layer that translates probabilistic assessments into actionable recommendations. Categories with high probability of Rapid Expansion could trigger human-in-the-loop review requirements and threshold-based escalation rules. Persistence checks flag categories that remain in concerning phases across multiple assessment periods. For categories approaching Endemic status, counterfactual explanations identify which input features most influenced the phase assignment. Finally, immutable safety rules provide hard constraints for categories involving critical infrastructure or vulnerable populations.

\subsection{Modeling Framework} 
\label{sub:2_modeling_framework}

Our proposed modeling framework combines real AI incident data with exposure proxies and media indicators to adjust for three main sources of bias: exposure growth, media visibility, and reporting delays. The aim is to separate genuine shifts in underlying risk from patterns created by attention, scale, or timing artifacts.

To estimate the phase of emergence for each class of AI incidents, we model incident trajectories using a latent-state time-series framework that treats phases as unobserved regimes inferred from corrected incident data. The pipeline consists of three sequential adjustments:

\begin{enumerate}
    \item \textbf{Delay correction}: observed incident counts are adjusted for reporting delays using a fitted delay distribution and nowcasting procedures to approximate the true event timeline.
    \item \textbf{Exposure normalization}: counts are divided by domain-specific exposure proxies to obtain an exposure-adjusted incident rate.
    \item \textbf{Media adjustment}: incident counts are regressed on media-intensity covariates to remove amplification effects and isolate a latent signal of underlying harm.
\end{enumerate}

The adjusted rate series is then passed to regime-detection methods which attempt to classify each time point into one of the six lifecycle phases. Each phase corresponds to a distinct generative pattern: sporadic low-intensity counts for rare phases, superlinear growth for rapid expansion, high stable plateaus for unmitigated endemicity, and declining or stabilized rates for mitigated phases.

Exposure data and media dynamic variation necessitate adapting the framework to each case study's data availability and risk structure (Table~\ref{tab:exposure_media_cases}).

\begin{table}
    \caption{Exposure and Media Adjustment Strategies by Case Study.}
    \centering
    \small
    \begin{tabular}{p{2.5cm}p{4.2cm}p{4.3cm}}
    \toprule
    Case & Exposure Proxy & Media Adjustment \\
    \midrule
    \textbf{Autonomous Vehicles} &
    California DMV autonomous miles driven (empirical, 2014--2025). Used as generalized linear model offset to compute per-mile incident rates. &
    Media spikes (e.g., post-fatality) are short-lived and do not systematically bias reporting. \\
    \addlinespace
    \textbf{Deepfakes} &
    GitHub Stars for top 8 repositories with 24-month half-life depreciation to create ``Depreciated Installed Base.'' Used as log-offset in Negative Binomial GLM. &
    Google Trends index for ``deepfake'' (worldwide, monthly) used as covariate in Negative Binomial GLM. \\
    \bottomrule
    \end{tabular}
    \label{tab:exposure_media_cases}
\end{table}

The cleaned rate series is then analyzed using complementary regime-detection methods to infer phase transitions. Because the two case studies exhibit different structural properties---oscillatory crisis-recovery cycles (AV) versus a unidirectional phase transition (deepfakes)---we employed different primary methods while maintaining a common validation framework. 

\paragraph{Autonomous Vehicles: Probabilistic State Inference via HMM.}
For the AV case, we fitted a Gaussian \acrfull{hmm} with six latent states to capture the oscillatory risk dynamics. The \gls*{hmm} estimates posterior state probabilities $P(z_t = k \mid \text{data}_{1:T})$ at each time point using Bayesian filtering and smoothing, where $z_t \in \{1,\dots,6\}$ represents unobserved risk regimes. This probabilistic approach allows the model to quantify uncertainty in state assignments and to detect transitions between baseline, elevated, and extreme-spike regimes even under noisy reporting conditions. The six \gls*{hmm} states were then approximately mapped to the epidemiological phase framework, though the correspondence is interpretive rather than exact (see Appendix~\ref{sec:b_autonomous_vehicles}).

\paragraph{Deepfakes: Structural Break Detection via PELT.}
For the deepfake case, we applied the \gls{pelt} algorithm to identify structural breaks in the media-adjusted risk signal. \gls*{pelt} detects multiple changepoints by minimizing a penalized cost function, partitioning the time series into segments with distinct mean levels and trends. Unlike HMM, \gls*{pelt} is deterministic and produces sharp temporal boundaries rather than probabilistic state assignments. We selected \gls*{pelt} over HMM for deepfakes because the domain exhibited an irreversible regime shift rather than oscillatory dynamics, making temporal boundary detection more interpretable than state-space modeling.

\paragraph{Common Validation}
Both cases employed \glslink{kmeans}{K-means clustering} as an independent validation method, clustering months based on \gls{standardized_risk} level and local trend. Additionally, both cases applied phase classification rules (Table~\ref{tab:phase_criteria}) to translate continuous risk/trend metrics into the six epidemiological phases.


Each phase corresponds to a distinct generative pattern for the latent incident rate: sporadic low-intensity counts for rare phases, superlinear growth for rapid expansion, high stable plateaus for unmitigated endemicity, and declining or stabilized rates for mitigated phases. The model updates phase probabilities as new reports arrive, producing a time-varying distribution that captures uncertainty and supports continuous reclassification. This approach allows the system to infer phase transitions even under noisy, incomplete, or media-biased conditions, providing an interpretable statistical basis. 

Detailed statistical methods including method selection rationale, model assumptions, parameter selection procedures, data processing pipeline, and sensitivity analyses are provided in Appendix~\ref{sec:a_methods}.

\textbf{Next we present the two case studies. Both case studies demonstrate the inadequacy of model-centered phase determination, while showing the capacity for the model to inform human-centered phase determination.}

\section{Case Study: Autonomous Vehicles}
\label{sec:3_autonomous_vehicles}
We now consider what phase determination we would make for an incident type where there are alternate sources of information providing strong ground truths. By comparing the incident determination with and without the strong ground truth, we can gain confidence in the capacity for human expert panels to arrive at similar determinations for incident types lacking ground truth.

We analyzed 75 Autonomous Vehicle (AV) incidents (565 reports) from July 2014 to February 2025 following the pipeline in Appendix~\ref{sub:a_data_processing_pipeline}. Table~\ref{tab:av_data_overview} summarizes data features for our AV case study, and domain-specific parameters are reported in Table~\ref{tab:av_model_parameters}.

\begin{table}[htbp]
\centering
\caption{Autonomous Vehicles: Data Overview.}
\label{tab:av_data_overview}
\begin{tabular}{lc}
\toprule
Feature & Autonomous Vehicles \\
\midrule
Unique incidents & 75 \\
Media reports & 565 \\
Observation period & Jul 2014 -- Feb 2025 \\
Months observed & 128 \\
Mean reporting delay (days) & 170 \\
Zero-incident months & 57.0\% \\
Exposure data source & CA DMV miles \\
\bottomrule
\end{tabular}
\end{table}

\begin{table}[htbp]
\centering
\caption{Autonomous Vehicles: Model Parameters.}
\label{tab:av_model_parameters}
\small
\begin{tabular}{lll}
\toprule
Model & Parameter & Value \\
\midrule
Delay model & $\mu$, $\sigma$ & 2.84, 2.59 \\
HMM & States & 6 \\
\gls*{pelt} & Penalty & 2.43 \\
\bottomrule
\end{tabular}
\end{table}

\noindent The six-state HMM specification is chosen for alignment with the six-phase classification framework; the sparse AV count distribution is not suited to formal Gaussian HMM model selection (see Appendix~\ref{sub:b_hmm_selection}).

\subsection{The Insufficiency of Model-Centric Phase Determination}
\label{sub:3_insufficiency}

We compared the AIID media-reported incident series against California DMV mandatory collision reports to assess how well the AIID data alone recovers the ground-truth signal. The California DMV autonomous vehicle program provides a rare ground-truth benchmark for this analysis. Under state law, every company testing autonomous vehicles on public roads must submit collision reports within ten days of any incident and annual reports of total autonomous miles driven, with penalties for non-compliance. This yields two data streams unavailable elsewhere in the AI incident landscape: a mandatory-reporting collision record with standardized criteria, and a deployment-scale denominator. Neither is distorted by media attention cycles, neither depends on voluntary disclosure, and the miles-driven denominator makes per-exposure rates calculable. The comparison that follows treats DMV collision counts as the closest available approximation of ground truth, against which AIID media-reported incidents can be benchmarked.

During the overlapping period (January 2019 -- February 2025, 85 months), AIID records 0--4 unique incidents per month while the DMV registers 0--22 collisions per month --- reflecting the fundamental difference between media salience and regulatory reporting.

\begin{figure}[htbp]
    \centering
    \includegraphics[width=\textwidth]{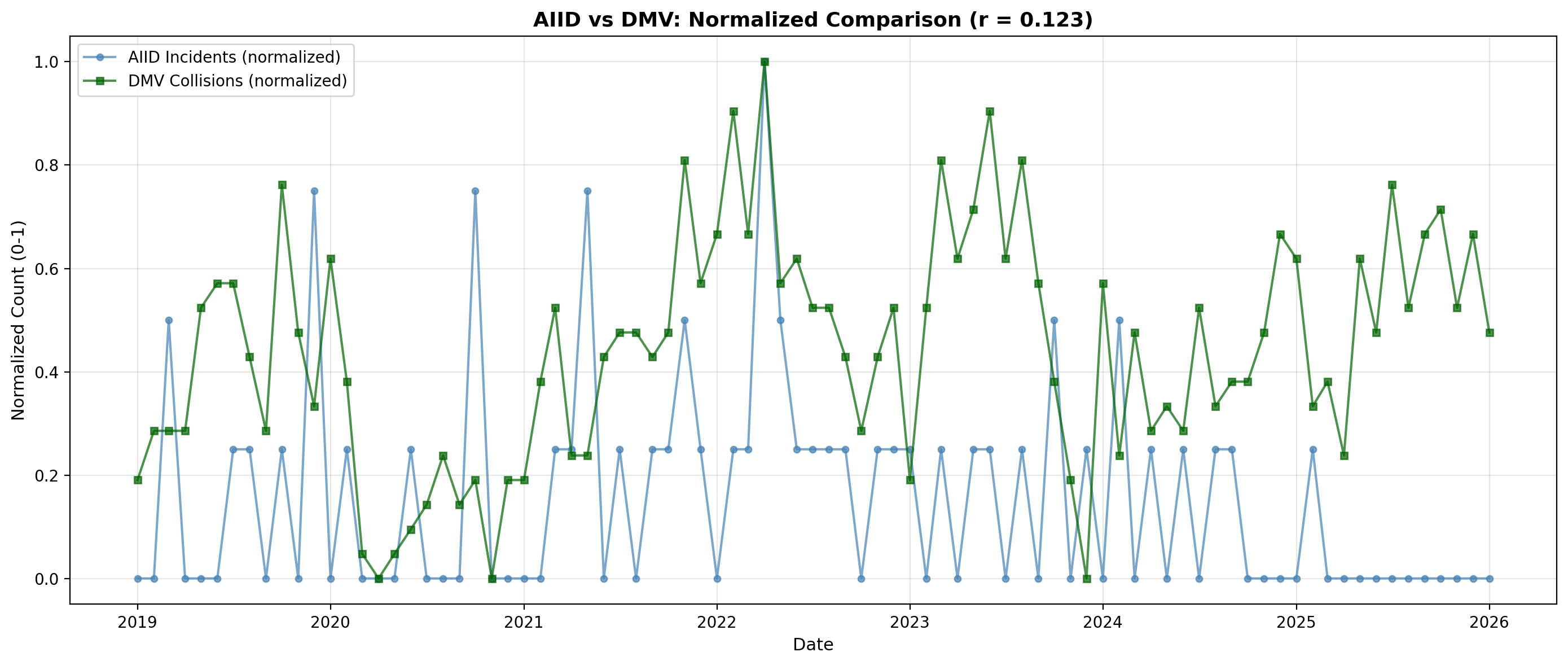}
    \caption{Normalized comparison of AIID incidents and DMV collisions ($r = 0.123$). Both series are min-max scaled to [0, 1] for visual comparison.}
    \label{fig:aiid_v_dmv_norm}
\end{figure}

The contemporaneous Pearson correlation between the two series is weak ($r = 0.123$; Figure~\ref{fig:aiid_v_dmv_norm}). A lagged cross-correlation analysis (Appendix~\ref{sub:b_ccf}) shows a maximum of $r = 0.199$ at lag $+2$ months, consistent with media reporting latency, but no lag within $\pm 6$ months achieves statistical significance, ruling out reporting delay as the primary source of divergence.

\begin{figure}[htbp]
    \centering
    \includegraphics[width=\textwidth]{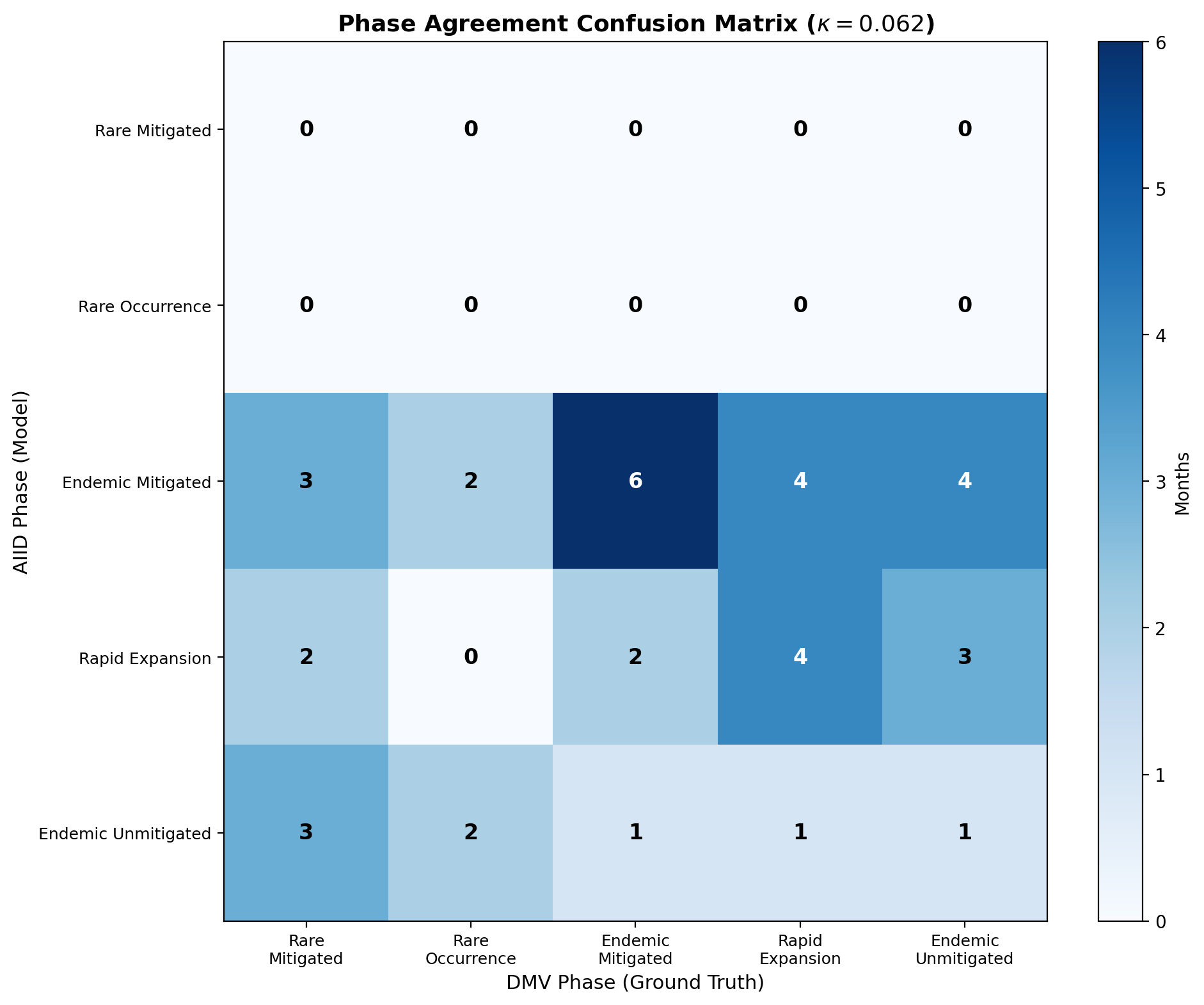}
    \caption{Phase agreement confusion matrix between AIID-derived phases (rows) and DMV ground-truth phases (columns). Cell values indicate the number of months assigned to each phase pair. The near-uniform distribution across columns within each AIID row confirms that model-centric phase classification applied to AIID alone does not recover the ground-truth phase structure ($\kappa = 0.062$). The 46 months classified as ``No Evidenced Occurrence'' by AIID are omitted as they have no DMV counterpart.}
    \label{fig:confusion}
\end{figure}

Applying the six-phase classification framework independently to each series produces near-chance agreement. Observed phase agreement is 15.3\% across the 85 overlapping months; given the skewed marginal distributions (AIID assigns 54\% of months in the 85-month overlap window to ``No Evidenced Occurrence'' --- compared with 57\% over the full 128-month series --- while DMV assigns none, since California autonomous testing produces at least one reported collision every month from January 2019 onward), the expected chance agreement is 9.7\%, yielding a Cohen's $\kappa = 0.062$ --- effectively no agreement beyond chance. The confusion matrix (Figure~\ref{fig:confusion}) reveals that disagreement is not concentrated in particular phase pairs: AIID's ``Endemic Mitigated'' months (the largest non-zero category) are distributed nearly uniformly across all five DMV phases, indicating that the AIID signal carries essentially no information about the DMV ground-truth phase.

\begin{figure}[htbp]
    \centering
    \includegraphics[width=\textwidth]{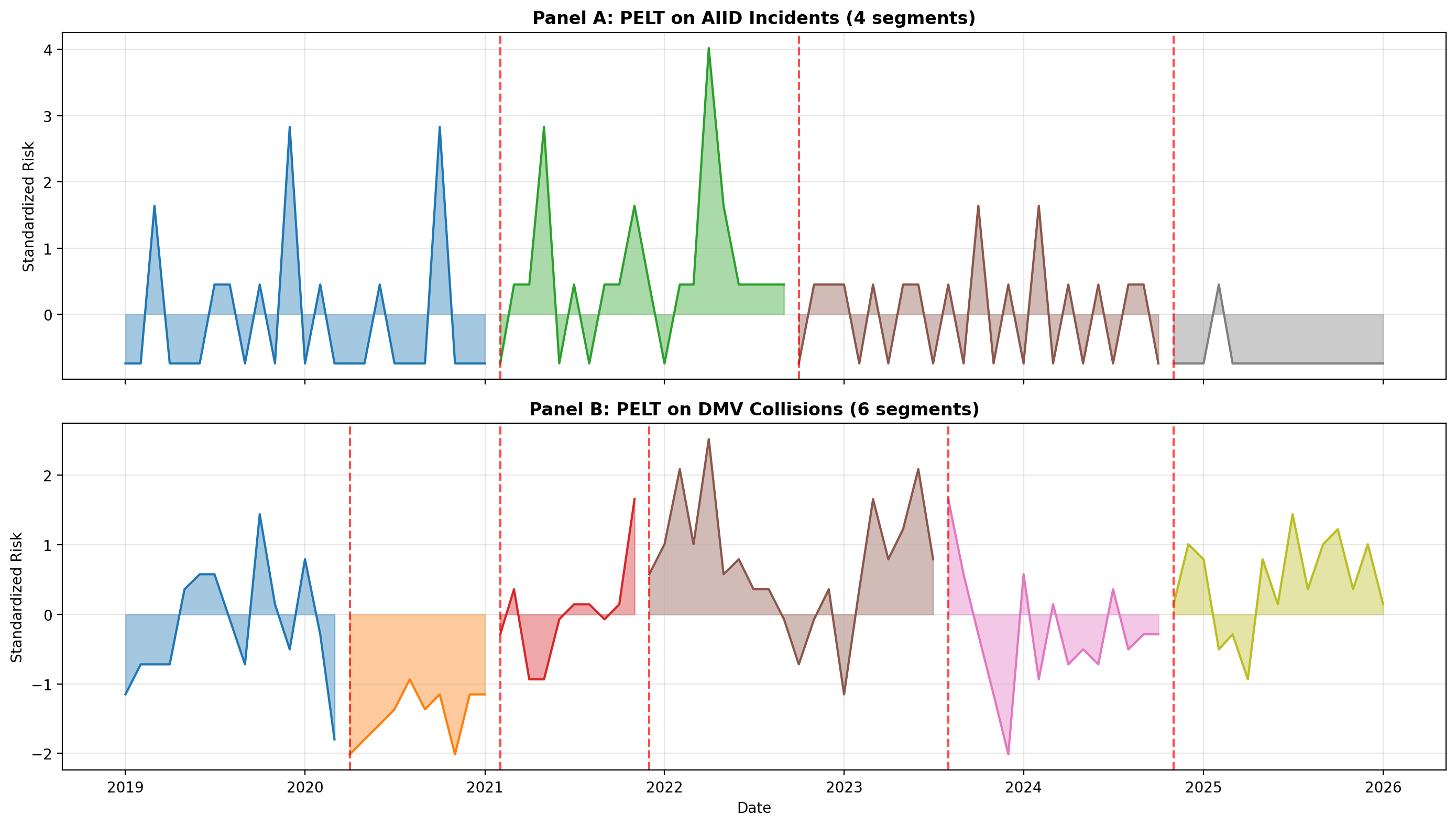}
    \caption{\gls*{pelt} changepoint detection applied independently to AIID incidents (Panel A, 4 segments) and DMV collisions (Panel B, 6 segments) over the shared period 2019--2025. Red dashed lines indicate detected changepoints. Only two of the DMV's five changepoints appear in the AIID segmentation, confirming that model-centric analysis of media-reported data misses governance-relevant regime shifts visible in mandatory reporting.}
    \label{fig:pelt_comparison}
\end{figure}

The structural disagreement extends to regime detection. Applying \gls*{pelt} changepoint detection with the same penalty ($\rho = 2.43$) to both series over their shared period (Figure~\ref{fig:pelt_comparison}) yields four segments in AIID (changepoints at February 2021, October 2022, and November 2024) versus six segments in DMV (changepoints at April 2020, February 2021, December 2021, August 2023, and November 2024). Only two changepoints coincide (February 2021 and November 2024); the DMV series detects regime shifts in early 2020, late 2021, and mid-2023 --- likely corresponding to pandemic-era testing changes and the Cruise operational suspension --- that leave no trace in the AIID record.

\begin{figure}[htbp]
    \centering
    \includegraphics[width=\textwidth]{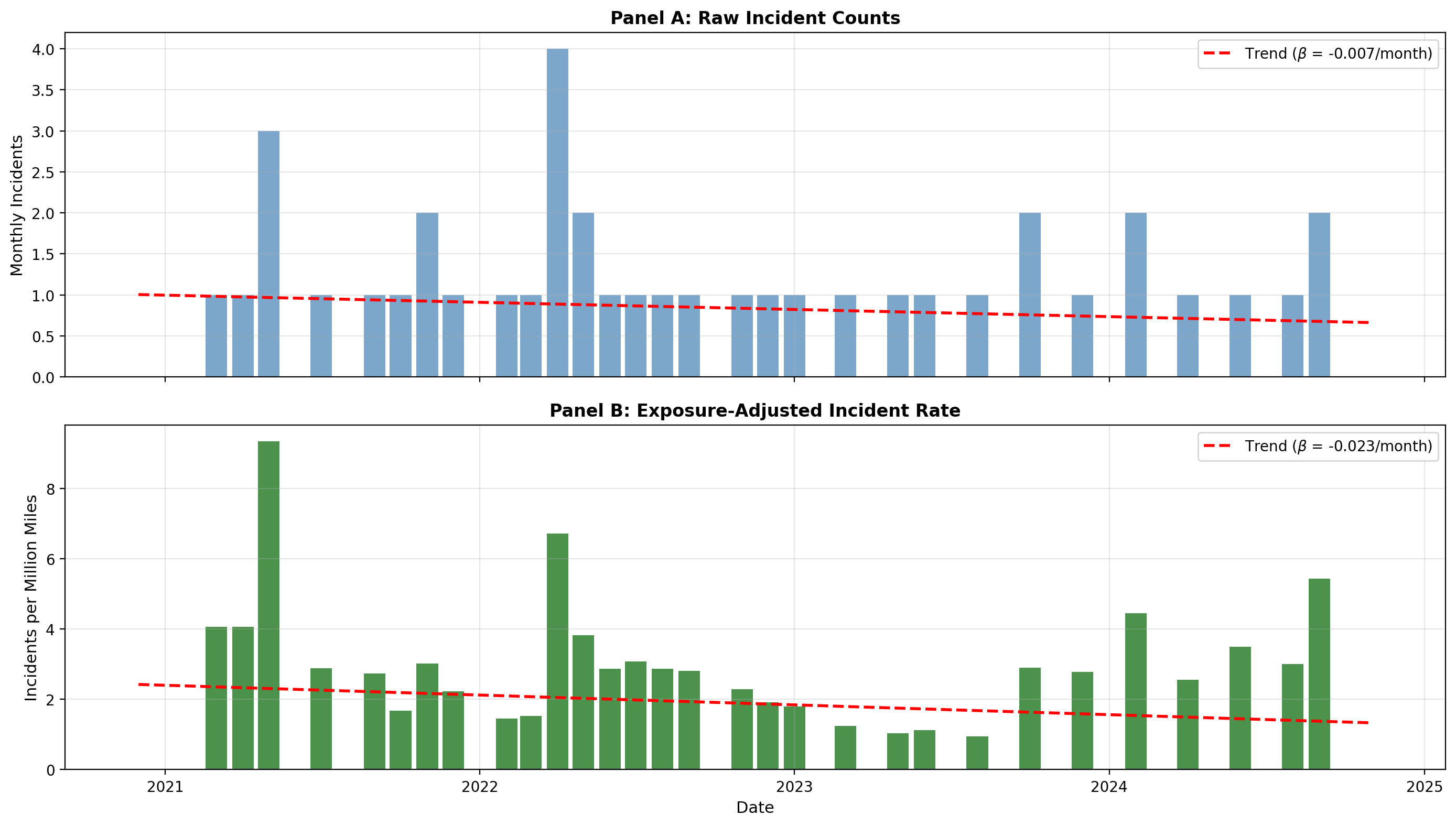}
    \caption{Raw incident counts (Panel A) vs exposure-adjusted incident rate (Panel B) during the DMV mileage coverage period (Dec 2020 -- Nov 2024). Raw counts are flat, but normalizing by autonomous miles driven reveals a declining per-exposure rate, indicating that deployment scaling --- not deteriorating safety --- accounts for observed incident volumes.}
    \label{fig:exposure_main}
\end{figure}

Finally, exposure-adjusted analysis (Figure~\ref{fig:exposure_main}) reveals that raw incident counts are essentially flat ($\beta = -0.007$/month) while the exposure-adjusted rate --- incidents per million autonomous miles driven --- shows a declining trend ($\beta = -0.023$ incidents per million miles per month, OLS on monthly rate, $p = 0.274$). \textbf{This suggests that deployment growth, not increasing risk, drives observed incident counts, a distinction that is invisible without the DMV exposure denominator.}

Taken together, these results demonstrate that the pipeline applied to AIID alone cannot recover the ground-truth phase structure: the correlation is weak even with optimal lag correction, the phase agreement is indistinguishable from chance ($\kappa = 0.062$), the detected regime boundaries diverge substantially, and the exposure context needed to distinguish deployment growth from risk growth is absent from the AIID record. This motivates the need for expert augmentation of the AIID data with domain-specific knowledge, as explored in Section~\ref{sub:3_human_centered}.

Additional model-centric analyses --- including Hidden Markov Model state assignments, PELT changepoint detection, intervention impact analysis, and a meta-reporting card prototype for AI governance --- are reported in Appendix~\ref{sec:b_autonomous_vehicles}.

\subsection{Phase Classification}
\label{sub:3_phase_classification}
Phase determination in principle requires a rate, not a count: a classifier that operates on raw counts cannot distinguish a phase transition driven by rising per-deployment risk from one driven by more vehicles on the road. An exposure denominator such as miles driven, in the AV case, is therefore a prerequisite for phase inference rather than an optional robustness check. In practice, California DMV mileage data covers only December 2020 through November 2024, so the classification that follows operates on raw counts for the earlier portion of the series. Phase outputs in domains without a mandatory exposure denominator should therefore be interpreted as characterizations of reporting dynamics rather than of per-exposure risk.

\begin{figure}[htbp]
    \centering
    \includegraphics[width=\textwidth]{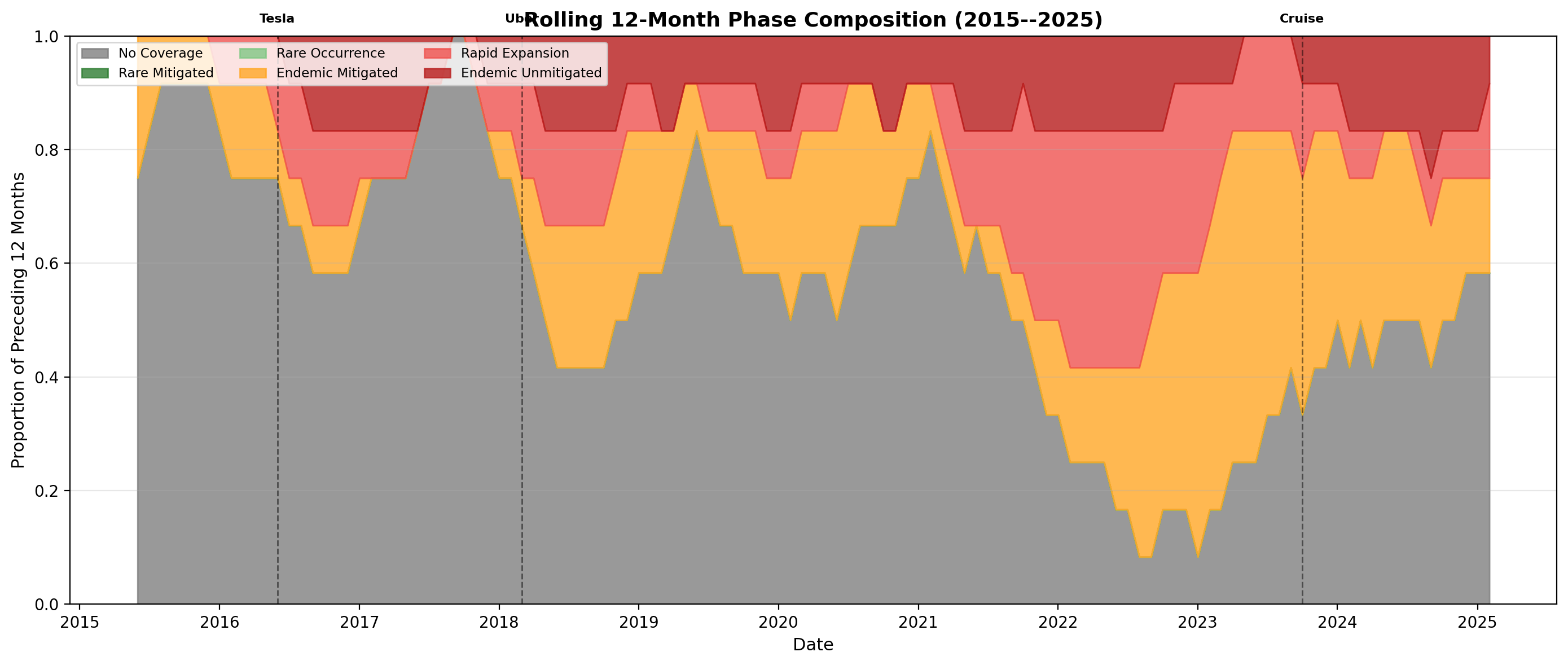}
    \caption{Rolling 12-month phase composition for AV incidents (2015--2025). For each month, the chart shows what fraction of the preceding 12 months fell into each governance phase. The oscillatory crisis-recovery pattern is visible as periodic expansions of the Rapid Expansion (pink) and Endemic Unmitigated (dark red) bands following the Tesla (2016), Uber (2018), and Cruise (2023) events, each followed by return to baseline dominance (grey).}
    \label{fig:phase_timeline_av}
\end{figure}

Applying the rule-based classification criteria from Table~\ref{tab:phase_criteria} --- which assign phases based on standardized risk ($z$-score) and local trend ($\tau$), independently of the HMM state posteriors --- yields Figure~\ref{fig:phase_timeline_av}. As established in Section~\ref{sub:3_insufficiency}, these AIID-derived classifications should not be taken as definitive phase determinations; rather, they represent one input that an expert panel (Section~\ref{sub:3_human_centered}) would consider alongside domain knowledge and ground-truth data where available.

Table~\ref{tab:phase_6_av} summarizes the phase distribution. Three patterns merit attention:

\begin{table}[htbp]
\centering
\caption{Phase Distribution for AV Incidents. Phases with 0 months reflect the framework's collapse in this sparse data regime; see Appendix~\ref{sub:b_threshold_sensitivity}.}
\label{tab:phase_6_av}
\small
\begin{tabular}{lrr}
\toprule
Phase & Months & \% \\
\midrule
No Evidenced Occurrence & 73 & 57 \\
Rare Mitigated & 0 & 0 \\
Rare Occurrence & 0 & 0 \\
Endemic Mitigated & 26 & 20 \\
Rapid Expansion & 16 & 13 \\
Endemic Unmitigated & 13 & 10 \\
\bottomrule
\end{tabular}
\end{table}

\paragraph{Crisis-Recovery Cycles.} The 2016 Tesla and 2018 Uber fatalities triggered sharp transitions to Rapid Expansion, followed by return to baseline within 6--12 months. This responsiveness reflects the AV industry's exposure to regulatory action, legal liability, and reputational consequences.

\paragraph{No Sustained Endemicity.} Despite a decade of deployment, AV incidents have not stabilized into Endemic Unmitigated status. The domain spends only 10\% of months in this highest-risk phase, compared to 57\% in the No Evidenced Occurrence phase where no incidents were recorded.

\paragraph{Current Status.} The domain is currently classified as \textbf{Rapid Expansion} based on the most recent observation (February 2025), driven by deployment scaling (Waymo, Cruise successors). However, the ARIMA(0,1,1) forecast (Appendix~\ref{sub:b_arima_forecasting}) projects reversion to Endemic Mitigated within the forecast horizon, consistent with the historical pattern of crisis-recovery oscillation rather than sustained expansion.

\subsection{Human-Centered Phase Determination}
\label{sub:3_human_centered}

Having established the insufficiency of directly mapping AIID incident data to a phase determination, we now examine the question of how effective a panel of human experts well-informed about the state of autonomous car deployments could utilize the AIID data to inform phase determinations in the absence of the ground truth DMV data.

Specifically, the AIID enables identification of deployment scope within the United States. Examination of the data reveals a small number of incidents outside California corresponding to small scale test deployments. We filter out incidents involving Tesla vehicles since all incidents were subject to human supervision and are thus not ``autonomous.'' This allows us to concentrate exclusively on California incident rates.

When examining any unchanging deployment scope we can assume few regressions in safety factor (i.e., every profit-seeking solution developer avoids making their cars more dangerous). Thus AV incident emergence is determined by total miles driven (i.e., it is exposure dominated). We can therefore utilize the raw AIID data rendered in Figure \ref{fig:aiid_v_dmv_norm} to identify new deployments based on their geographies. Reading the associated reports reveals three major events marking step changes of risk. First, Waymo's earliest incident in February of 2016 (\cite{aiid:71}) marked the introduction of AV risk to California. For years after, Waymo's deployments were highly limited in total miles, but they expanded gradually until their apparent market lead spurred Cruise Automotive to deploy within California as shown by incident 293 in June of 2022 (\cite{aiid:293}). This accelerated the deployments in California until Cruise dropped from the market as a result of incident 726 in October of 2023 (\cite{aiid:726}). Notably, this series of events identified two of the most policy-important lines found within Figure \ref{fig:pelt_comparison} -- those corresponding to the period of greatest standard risk.

Reviewing company safety reports detailing miles driven and self-disclosed incident rates could potentially identify additional changes in the incident regime. We intentionally avoided using this data to see what we could identify exclusively with AIID data. We expect any panel of expert practitioners will utilize these and other data sources available for the technology and its operating contexts.

A broader caveat applies to the AV case specifically. The domain has not clearly entered later phases of the framework in the ten years of data available. With a sparse incident record (0--4 per month, 57\% zero-months) and a deployment footprint that has grown by roughly two orders of magnitude over the observation window, the risk signal is dominated by exposure rather than by per-vehicle safety regressions. The expert reasoning above therefore identifies transitions in deployment regime rather than transitions in underlying risk; the framework's later phases (Rapid Expansion, Endemic Unmitigated, Endemic Mitigated, and Rare Mitigated) are not yet populated in any robust sense for this domain. Denser-incident subfields such as Level 2 driver-assist systems, or combined Level 2 through Level 4 deployments, may better exercise the full six-phase framework; we leave that comparison for future work.

\textbf{Case study conclusion:} A human applying deductive reasoning to incident data and their knowledge of its operating context is capable of making grounded declarations of incident risk.

\section{Case Study: Deepfake}
\label{sec:4_deepfakes}
Section~\ref{sec:3_autonomous_vehicles} showed that the pipeline applied to AIID data alone is insufficient and domain-specific knowledge is needed to make phase determination actionable. We now apply the same pipeline to a domain where no external ground truth exists. Deepfakes are digital artefacts generated by often anonymous users. No regulatory body counts incidents per unit of deployment, and no recall authority exists. The deepfake case poses a further question: when no external validation denominator is available, does the pipeline produce a signal robust enough to serve as a defensible diagnostic input? And what domain knowledge is needed to assess whether the detected phase is actionable?

We analyzed 296 unique deepfake incidents (1,114 reports) from the AIID spanning March 2017 to September 2025, following the pipeline in Appendix~\ref{sub:a_data_processing_pipeline}. Table~\ref{tab:df_data_overview} summarizes data features and Table~\ref{tab:df_model_parameters} reports domain-specific parameters; full construction details and sensitivity analyses are in Appendix~\ref{sec:c_deepfakes}.

\begin{table}[htbp]
    \centering
    \caption{Deepfakes Data Overview.}
    \label{tab:df_data_overview}
    \begin{tabular}{lc}
    \toprule
    Feature & Deepfakes \\
    \midrule
    Unique incidents & 296 \\
    Media reports & 1,114 \\
    Observation period & Mar 2017 -- Sep 2025 \\
    Months observed & 103 \\
    Median reporting delay, post-2021 (days) & 1.0 \\
    Exposure proxy & GitHub Stars (8 repositories, 24-month half-life, GLM offset) \\
    \bottomrule
    \end{tabular}
\end{table}

\begin{table}[htbp]
    \centering
    \caption{Deepfakes: Model Parameters.}
    \label{tab:df_model_parameters}
    \small
    \begin{tabular}{lll}
    \toprule
    Model & Parameter & Value \\
    \midrule
    Delay model & Median lag (post-2021) & 1.0 days \\
                & Nowcast window & 4 months \\
    NB GLM      & Dispersion $\alpha$ & 1.5 \\
    \gls*{pelt} & Penalty ($\rho$) & 3.00 \\
                & Segments detected & 3 \\
    \bottomrule
    \end{tabular}
\end{table}

Following Section~\ref{sub:3_phase_classification}, the GitHub-Stars index acts as a proxy for exposure, meaning the phase outputs capture reporting dynamics in relation to tool availability rather than exposure-adjusted risk.

\begin{figure}[htbp]
    \centering
    \includegraphics[width=\textwidth]{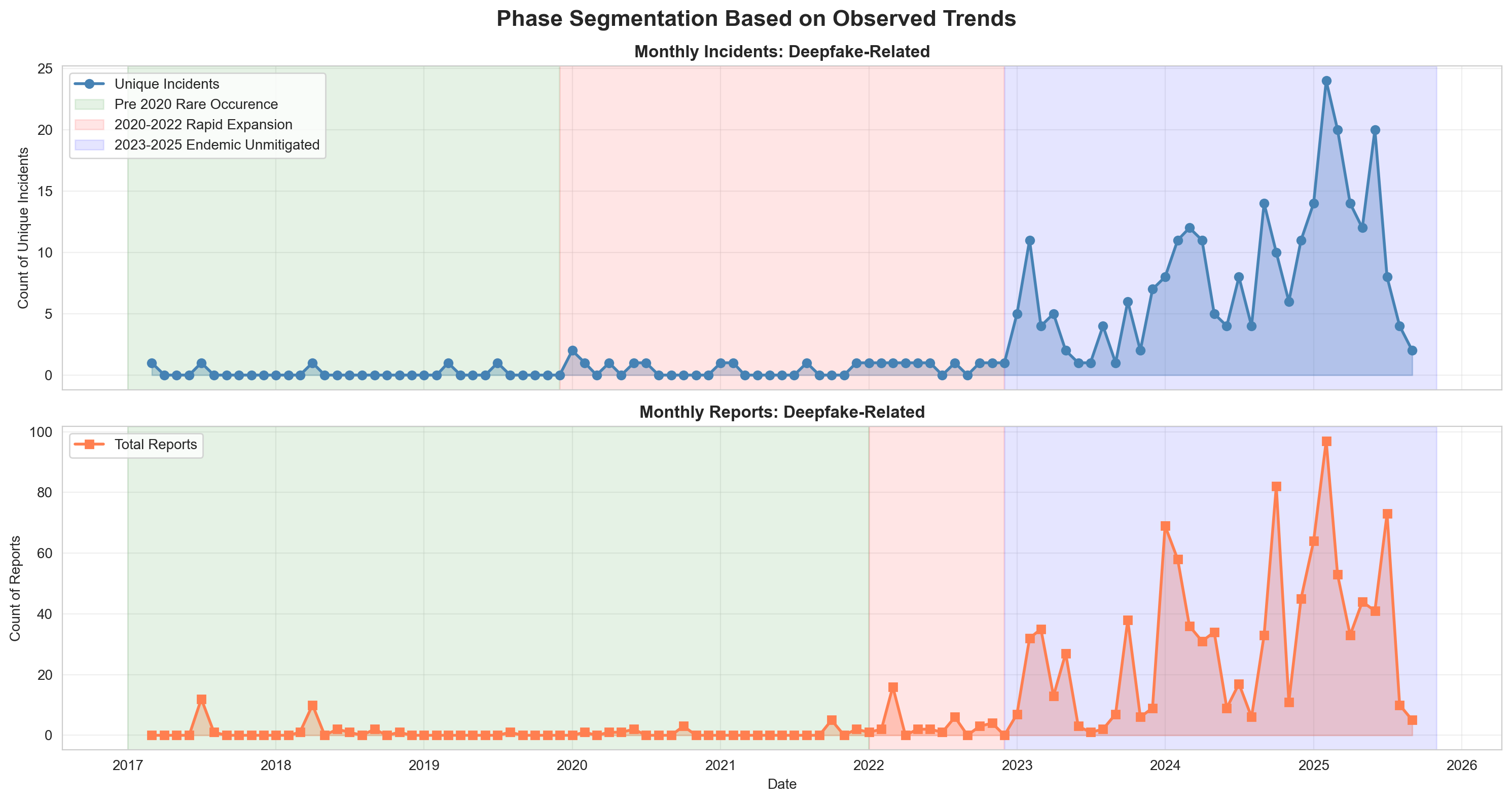}
    \caption{Raw deepfake incidents and reports by month (March 2017--September 2025). Counts remained near zero through 2022 before surging in 2023.}
    \label{fig:03_time_series_overview}
\end{figure}

Figure~\ref{fig:03_time_series_overview} shows raw incident and report counts over time. Absolute counts remained near zero through 2022 and surged in 2023 --- yet the risk-adjusted signal (Figure~\ref{fig:pelt_df}) identifies the structural break a year earlier, in March 2022, once media attention and exposure (proxied by tool availability) are controlled for. This gap shows why raw counts alone are insufficient: the underlying risk trajectory shifted before the raw series made it visible.

\subsection{Results: An Irreversible Step-Function}
\label{sub:4_results_df}

We applied changepoint detection to the media-adjusted excess risk signal (residuals from a Negative Binomial GLM with exposure as log-offset; Appendix~\ref{sub:c_media_adj_risk_nb_df}). Because the deepfake domain lacks the oscillatory dynamics seen in the AV case, we use a simplified three-phase framework (Dormant Baseline / Active Outbreak / Endemic Unmitigated); the full six-phase cross-walk is in Appendix~\ref{sub:c_phase_classification_df}. Figure~\ref{fig:pelt_df} shows the result.

\begin{figure}[htbp]
    \centering
    \includegraphics[width=\textwidth]{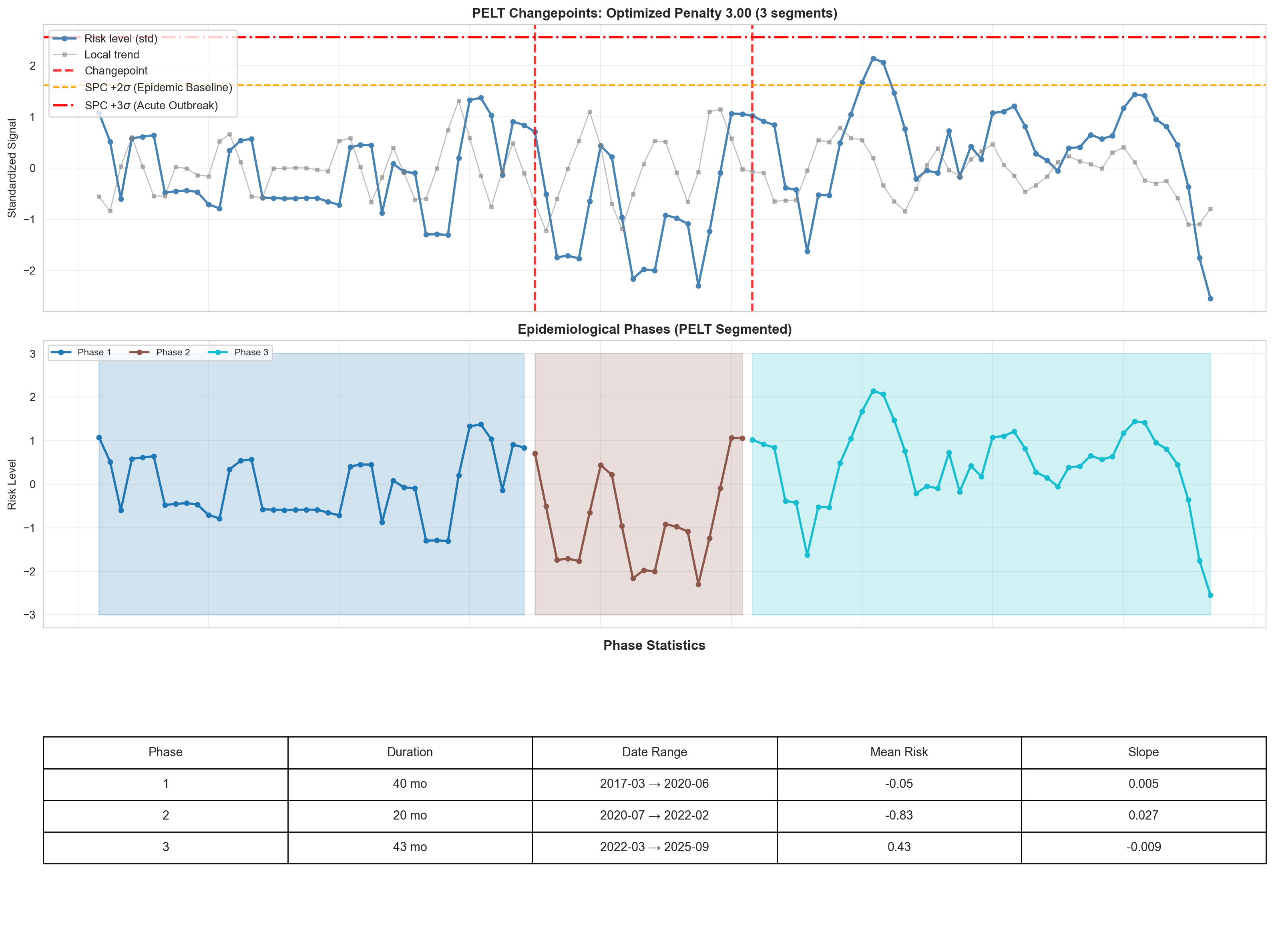}
    \caption{PELT segmentation of the media-adjusted deepfake risk signal (March 2017--September 2025). The algorithm detects two changepoints (red dashed lines), producing three structural segments. Segments 1 and 2 both fall below the Dormant threshold ($\theta = +0.14\sigma$); the phase mapping step therefore collapses them into a single Dormant Baseline. Segment 3 (March 2022 onward) sits at $+0.43\sigma$---persistently above the Dormant threshold but well below the SPC $+2\sigma$ Epidemic threshold (orange dashed line), confirming Endemic Unmitigated rather than Active Outbreak. The March 2022 break coincides with the mainstream release of high-fidelity generative tools (\textit{Roop}, \textit{FaceFusion}). Methodological details in Appendix~\ref{sec:c_deepfakes}.}
    \label{fig:pelt_df}
\end{figure}

Before March 2022, deepfake risk spent five years below the historical mean: a dormant incubation period during which incidents were rare and the ecosystem of generative tools was still forming. In March 2022, the signal breaks sharply, coinciding with the mainstream release of high-fidelity generative tools (\textit{Roop}, \textit{FaceFusion}). For 43 consecutive months following this break, risk remained elevated at $+0.43\sigma$ above baseline with a near-flat trajectory: the system did not spike, it \textit{migrated}. In addition, not a single month crosses the Active Outbreak threshold: the transition to endemicity arrived by monotonic migration.

\begin{table}[htbp]
    \centering
    \caption{Deepfake Phase Summary (2017--2025).}
    \label{tab:df_summary}
    \small
    \begin{tabular}{lrr}
    \toprule
    Phase & Months & \% \\
    \midrule
    Dormant Baseline & 60 & 58.3 \\
    Active Outbreak & 0 & 0.0 \\
    Endemic Unmitigated & 43 & 41.7 \\
    \bottomrule
    \end{tabular}
\end{table}

\begin{figure}[htbp]
    \centering
    \includegraphics[width=0.9\textwidth]{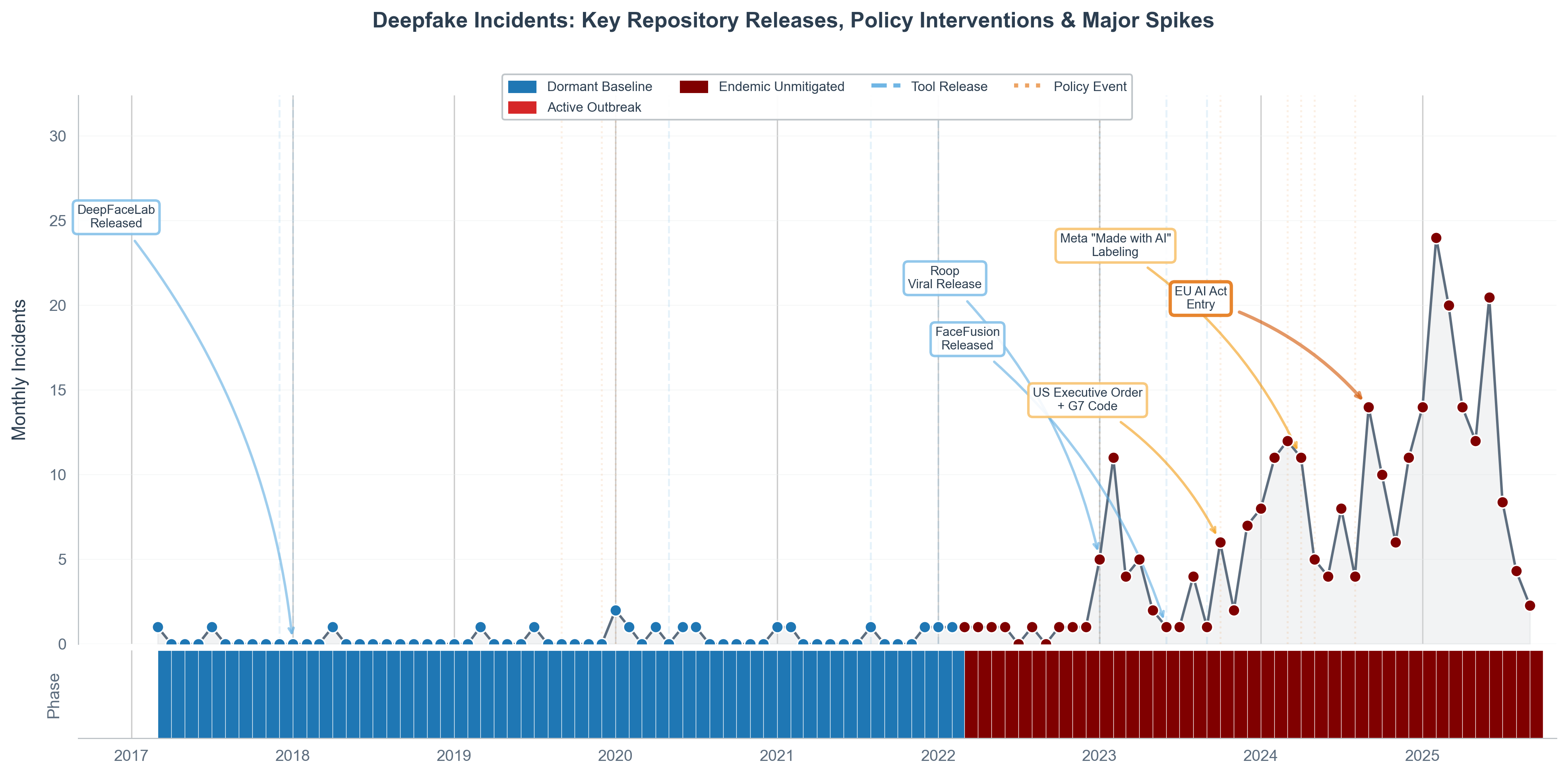}
    \caption{Deepfake incident timeline with phase classification (2017--2025). Annotated events mark incident spikes alongside landmark repository releases (\textit{Roop}, \textit{FaceFusion}). The horizontal strip shows the binary phase structure: five years of dormancy followed by 43 consecutive months of Endemic Unmitigated with no recovery.}
    \label{fig:phase_dist_df}
\end{figure}

Table~\ref{tab:df_summary} and Figure~\ref{fig:phase_dist_df} summarize the phase distribution: 58.3\% of months are Dormant Baseline and 41.7\% are Endemic Unmitigated, with zero months in Active Outbreak. Sensitivity sweeps over exposure half-life, NB dispersion, PELT penalty, and phase thresholds all confirm the binary dormant-to-endemic finding (Appendix~\ref{sub:c_pelt_sensitivity_df}, \ref{sub:c_halflife_sensitivity_df}).

The full six-phase cross-walk (Appendix~\ref{sub:c_phase_classification_df}) classifies Segment~3 as \textit{Endemic Mitigated}, because the risk level ($+0.43\sigma$) falls between $\theta_{\text{low}}$ and $\theta_{\text{high}}$ with a declining within-segment slope. We use the more conservative \textit{Endemic Unmitigated} label in the main text because no external intervention has been shown to reduce deepfake risk at a structural level. This labeling choice does not affect the binary dormant-to-endemic finding.

These model outputs are a diagnostic input, not a governance determination. Section~\ref{sub:4_endemic_trap} supplies the domain context the pipeline cannot: whether institutional structures exist to act on this phase.

\subsection{The Endemic Trap}
\label{sub:4_endemic_trap}

The contrast with the AV domain is stark. The 2018 Uber fatality triggered Uber's immediate suspension of all autonomous operations, regulatory investigations across multiple states, and a sector-wide de-escalation. This response was possible because AV deployment is centralized: a small number of corporations operate identifiable fleets under government permits. When a regulator called, someone answered.

Deepfake tools were forked thousands of times on GitHub before any governance actor existed. No single corporate actor controls the technology, and no regulatory body can issue a recall on software already distributed to anonymous users worldwide. Once the deepfake trajectory entered the Endemic phase in March 2022, it remained there for 43 consecutive months with no structural recovery, suggesting that institutional structures required for effective recall do not exist in a decentralized software ecosystem yet.

The same statistical pipeline that identified AV's oscillatory crisis-recovery cycles in Section~\ref{sec:3_autonomous_vehicles} identifies deepfake's endemic persistence here. What differs is not the pipeline's diagnostic capacity but the institutional structures available to respond (which the model cannot see). Detection and diagnosis succeed while action requires institutional structure that the data does not contain. Extended cross-domain comparison is in Appendix~\ref{sub:c_cross_domain_comparison_df}.

\subsection{What This Case Adds to the Argument}
\label{sub:4_what_this_adds}

The two case studies ask different questions.

The AV case asks: how does the pipeline compare to an external ground truth? The answer is that AIID-derived phase signals have relatively weak agreement with the DMV record ($\kappa = 0.062$). Actionable determination required domain-specific data and knowledge of identifiable institutional actors such as fleet operators and regulatory bodies who could respond once a phase was determined.

The deepfake case asks: when no external ground truth exists, is the pipeline signal robust enough to be a defensible diagnostic input? The answer is yes. The binary dormant-to-endemic transition holds across all sensitivity dimensions tested: exposure half-lives of 6, 12, and 24 months; NB dispersion values $\alpha \in \{0.5, 1.0, 1.5, 2.0\}$; PELT penalty range $\rho \in [3.0, 7.5]$; and phase threshold sweeps across the invariant zone (Appendix~\ref{sec:c_deepfakes}). But a robust signal does not mean the phase is actionable. Intervention analysis (Table~\ref{tab:wave_impact_df}) shows a clear pattern: early platform-level governance reduced risk significantly (TikTok policy, August 2020; $\Delta = -1.78\sigma$, $p_{\text{FDR}} = 0.002$); later macro-level frameworks such as the Bletchley Declaration, the EU AI Act, coordinated governance waves in 2022--2024 produced null or perverse effects. The difference is structural: interventions that reach platform-level chokepoints can suppress risk; interventions that cannot reach the distributed anonymous user base cannot, regardless of institutional weight.

Both cases support the paper's central claim: phase-based inference is defensible as a diagnostic tool. However, phases are only actionable when combined with domain knowledge about whether institutional structures exist to act on the diagnosis.

\section{Conclusion: Implications for AI Governance and Future Work}
\label{sec:5_discussion}
This study set out to address a core gap in AI governance: while incident repositories provide unprecedented visibility into AI-related harms, they do not yet supply a reliable basis for understanding how different risks evolve over time. Indeed, without immense effort to incorporate contextual information about AI deployments it is likely that incident reports will lead people to believe AI is becoming more hazardous rather than more common. These distortions mirror challenges identified in existing governance work, including the OECD’s call for standardized reporting criteria and the UK’s recognition that fragmented reporting structures leave regulators blind to emerging harms \citep{shane2024aiincident}.

By reframing classes of AI incidents through a lifecycle lens, this paper offers a conceptual and statistical structure that strengthens and complements existing governance tools. Instead of treating raw incident counts as indicators of worsening or improving safety, a phase-based approach provides a more robust, interpretable, and policy-aligned way to understand how harms emerge, escalate, stabilize, or decline and when specific interventions are warranted. 

AI incidents are not isolated failures; they progress through identifiable phases. Recognizing these phases allows governments to detect early warning signals, allocate resources efficiently, and evaluate progress toward safer AI systems. Crucially, our findings emphasize that phase classification is not a purely mechanical output of statistical modeling. It requires interpretation grounded in the operational context of each AI system class, including its deployment environment and exposure profile. Below we outline priority directions that would substantially improve the accuracy, usefulness, and fairness of phase-based monitoring.

\subsection{Implications for AI Governance}
\label{sub:5_implications_ai_governance}

The transition from a passive database of incidents to declaring incident phases carries three primary implications for governance strategy.

\subsubsection*{For Incident Reporting: Strengthening the Infrastructure and Improving Data Granularity}

Mandating standardized disclosure can be a first step towards facilitating a consistent data collection of AI incidents \citep{dixon2025ai, longpre2025inhouse}. An expansion of the EU AI Act Article 73, warranting reporting to all types of incidents, ideally classified with respect to a common taxonomy (e.g., the MIT Risk Repository's taxonomy \citep{slattery2024ai}, combined with an AI harm taxonomy \citep{hoffman2023adding}), would be a sensible path forward. It would need clear triggers (e.g., "any unwanted harm to legal rights"), as well as safe harbor provisions to encourage disclosure. In particular, it may be desirable to incentivize voluntary reporting, e.g., through offering liability protections or regulatory "good faith" credits for firms that disclose near-misses. Also, since most of the reporting effort is being performed by voluntary initiatives, policymakers could support third-party incident databases (e.g., the AIID \citep{mcgregor2021preventing}) to cross-validate industry reports, and fill gaps in understudied sectors (e.g., AI in agriculture, AI in legal aid).

In addition to improving the incident reporting frameworks as a whole, it may be useful to improve the quality of the incident reports themselves. As mentioned earlier, policymakers should endeavor to standardize incident taxonomies to enable global consistency and interoperability of reporting data \citet{perset2025framework}. Furthermore, for its phase transition identification capabilities to be optimal, our phase model requires real-time or near-real-time reporting incidents. Thus, we suggest policymakers to implement time constraints for mandatory reporting, which could be adapted to the severity of the incident (e.g., in a range from within 24 hours for the most severe incidents, and to within 15 days for the most benign). Finally, it could be useful for more fine-grained analysis to link incident data to deployment context: reporting efforts could collect metadata on system characteristics (e.g., model size, training data sources) and governance environment (e.g., whether the system was audited pre-deployment) \citep{dixon2025ai}. This would allow policymakers to better correlate phases with regulatory interventions.

\subsubsection*{For Policymakers: Resource Allocation and Addressing Structural Biases}

As policymakers operate under resource constraints, our phase model could serve as a triage mechanism: incident types classified as \emph{No Evidenced Occurrence} or \emph{Rare Occurence} may require only passive horizon scanning, while incident types classified as \emph{Rapid Expansion} or \emph{Endemic Unmitigated} warrant immediate investigative resources for risk prevention or damage control. This would allow government agencies to pivot from "policing everything" to "policing the instability."

Furthermore, to ensure global equity in AI incident reporting, it is crucial to establish regional AI safety hubs that would train local incident investigators and provide anonymous reporting channels to protect whistleblowers, taking inspiration from similar endeavors for AI ethics in Africa \citep{adams2025final}.

Finally, policymakers would do well to counteract media-driven distortions through the development of automated tools for the flagging of overrepresented incidents (e.g., viral but low-severity cases) \citep{hamborg2019automated}.

\subsection{Research Directions for Future Work}
\label{sub:5_future_work}

\subsubsection*{Methodological Development}
\begin{itemize}
    \item \textbf{Simplify Phases.} The number and focuses of phases have not been validated beyond the limited case studies of this research. A smaller number of phases may increase confidence while maintaining policy and industrial relevance.
    \item \textbf{Benchmark Alternative Generative Models.} Evaluate Hawkes processes, Bayesian structural time series, and self-exciting spatial–temporal models on both observed and simulated incident data to identify cases where each approach is most reliable.
    \item \textbf{Improve Nowcasting with Non-Stationary Delays.} Develop delay models that allow time-varying delay distributions (e.g., via covariates or hierarchical models) to better handle editorial changes or regulatory effects on reporting lag.
    \item \textbf{Probabilistic Exposure Models.} Replace deterministic synthetic exposure proxies with probabilistic exposure estimators that propagate uncertainty into phase assignments.
    \item \textbf{Decision-Theoretic Trigger Design.} Formalize trigger thresholds using decision-theory frameworks that balance the social cost of false positives and false negatives for different audiences (e.g., regulators vs. civil-society responders).
    \item \textbf{Incident Classification Framework.} Develop a systematic approach to characterize incident classes by theme to inform selection of classes for modeling.
\end{itemize}

\subsubsection*{Applied and Participatory Work}
\begin{itemize}
    \item \textbf{Co-Design Meta-Reporting Cards.} Pilot co-design workshops with regulators, CSOs, and industry to refine phase report fields, update cadence, and thresholds for action.
    \item \textbf{Cross-Validation with Administrative Data.} Collaborate with industry and regulators to validate phase assignments against privileged administrative records (e.g., internal incident tracking, verified claims settlements, or audit outcomes).
    \item \textbf{Longitudinal Intervention Studies.} Use phase assignments to identify domains for prospective intervention trials (e.g., mandated guardrails) and to evaluate their effectiveness through quasi-experimental designs.
\end{itemize}

\subsection{Closing Note}
Implementing these research priorities requires both technical work and institutional commitment to better data collection and sharing. The phase-model framework we present is operational today with publicly available data and provides governance value, but its full potential--accurate, timely, and globally representative monitoring of AI harms-depends on improved exposure reporting, richer incident timestamps, and collaborative partnerships between researchers, civil society, developers, and regulators.

A natural next step is to align the phase model within regulatory regimes that already mandate or anticipate serious-incident reporting. Under the EU AI Act, providers of high-risk and general-purpose AI systems will be required to notify regulators of serious incidents and undertake post-market monitoring; phase classification could serve as an organizing layer for these reports, indicating when a given harm type has moved from rare emergence to rapid expansion or endemic unmitigated status, and therefore warrants tighter supervision, targeted audits, or code-of-practice updates. Likewise, emerging California transparency and incident-reporting laws (e.g.\ around safety incidents in automated systems and AI services) could use phase labels as a way to prioritize enforcement and public disclosure, and to escalate reporting expectations and oversight if an incident class crosses predefined phase thresholds. 

In both contexts, future work should explore how to map legal triggers (e.g.\ ``serious incident'', ``systemic risk'',” or ``material change in risk profile'') onto phase transitions, so that regulators can use a common lifecycle vocabulary to interpret noisy incident data, coordinate responses across jurisdictions, and evaluate whether new rules actually shift specific harms from endemic unmitigated to rare and mitigated states.

Every serious incident may have antecedents in lower severity incidents. Developing a practice of responding to rapidly expanding incidents types may greatly reduce the risk of catastrophe.

\section{Acknowledgements}

We would like to thank Ben Smith for his support and oversight, and Daniel Atherton, Tommy Shaffer Shane, and Jared Leibowich for their constructive input and helpful feedback.

\section*{Contribution Statement}
\label{sec:contribution_statement}
Author Contributions (CRediT taxonomy)

\textbf{Conceptualization:} Giovanna Jaramillo-Gutierrez, Sean McGregor

\textbf{Methodology, Investigation:} Sophia Abraham, Taiye Chen, Cyril Chhun, Giovanna Jaramillo-Gutierrez

\vspace{0.2cm}
\begin{tabular}{@{}ll@{}}
$\quad$ Sophia Abraham & Sections~\ref{sec:2_conceptual_framework}, \ref{sec:3_autonomous_vehicles}, \ref{sec:5_discussion}, \ref{sec:a_methods}, \ref{sec:b_autonomous_vehicles} \\
$\quad$ Taiye Chen & Sections~\ref{sec:2_conceptual_framework}, \ref{sec:4_deepfakes}, \ref{sec:5_discussion},\ref{sec:a_methods} ,\ref{sec:c_deepfakes}\\
$\quad$ Cyril Chhun & Sections~\ref{sec:2_conceptual_framework}, \ref{sec:3_autonomous_vehicles}, \ref{sec:5_discussion}, \ref{sec:a_methods}, \ref{sec:b_autonomous_vehicles} \\
$\quad$ Giovanna Jaramillo-Gutierrez & Sections~\ref{sec:1_introduction}, \ref{sec:2_conceptual_framework}, \ref{sec:5_discussion} \\
$\quad$ Sayash Raaj & Sections~\ref{sec:1_introduction}, \ref{sec:2_conceptual_framework}, \ref{sec:5_discussion} \\
\end{tabular}
\vspace{0.2cm}

\textbf{Project Administration:} Giovanna Jaramillo-Gutierrez, Simon Mylius, Ben R. Smith

\textbf{Writing – \LaTeX:} Sophia Abraham, Taiye Chen, Cyril Chhun, Giovanna Jaramillo-Gutierrez, Simon Mylius, Sayash Raaj, Sean McGregor

\textbf{Writing – Review \& Editing:} Cyril Chhun, Giovanna Jaramillo-Gutierrez, Simon Mylius, Peter Slattery, Sean McGregor 

\textbf{Writing – Supervision:} Giovanna Jaramillo-Gutierrez, Simon Mylius,  Sean McGregor

\clearpage
\appendix
\printglossary[title=Glossary,toctitle=Glossary]
\printacronyms

\bibliographystyle{acl_natbib}
\bibliography{references}
    
\section{Statistical Methods: Technical Details}
\label{sec:a_methods}

This appendix provides technical details for the statistical methods introduced in Section~\ref{sub:2_modeling_framework}.

\subsection{Statistical Methods for Phase Inference}
\label{sub:a_statistical_methods}

Inferring latent phases from noisy incident data requires methods that can: (i) detect structural breaks in temporal dynamics, (ii) characterize distinct risk regimes, (iii) estimate trends while accounting for \gls{overdispersion}, and (iv) generate probabilistic forecasts. No single method addresses all requirements; we therefore employ a complementary suite of statistical approaches, each contributing distinct information to phase classification. Table~\ref{tab:method_rationale} summarizes each method, its purpose in the phase-inference pipeline, and why it was selected over alternatives.




\begin{table}[htbp]
    \caption{Statistical Methods: Purpose and Selection Rationale.}
    \centering
    
    
    
    
    
    
    \begin{tabular}{p{2.8cm}p{2.8cm}p{6.8cm}}
    \toprule
    Method & Purpose & Rationale \\
    \midrule
    \Gls{lognormal} delay model & Correct reporting lag & Heavy-tailed delays common in media reporting; log-normal outperformed exponential and Weibull by \glslink{aic}{AIC}. \\
    Gaussian \glslink{hmm}{HMM} & Detect latent risk states & Captures unobserved regimes with probabilistic state assignments; preferred over threshold rules because regime boundaries are uncertain. \\
    \gls{pelt} changepoint & Identify structural breaks & Optimal detection of multiple changepoints with linear complexity; interpretable and computationally efficient. \\
    \glslink{kmeans}{K-means clustering} & Validate regime detection & Independent confirmation using distance-based rather than generative assumptions; agreement patterns reveal complementary structure. \\
    \glslink{negative_binomial}{Negative Binomial} (NB) GLM & Estimate trends with \gls{exposure_offset} & Accommodates severe \gls*{overdispersion} (var/mean $>$ 12) in incident data; Poisson assumption violated. \\
    \glslink{arima}{ARIMA} & Project future volumes & Conservative baseline capturing short-term autocorrelation and mean reversion; selected via AIC. \\
    6-phase classification & Map to governance & Translates continuous risk and trend metrics into discrete, actionable phases aligned with policy levers. \\
    \bottomrule
    \end{tabular}
    \label{tab:method_rationale}
    %
\end{table}


\subsection{Why Multiple Methods?}
\label{sub:a_why_multiple_methods}

We employ methodological triangulation using multiple independent methods to characterize the same phenomenon for three reasons:

\begin{enumerate}
    \item \textbf{Robustness}: If HMM, PELT, and K-means identify similar regime structure despite different algorithmic foundations, confidence in that structure increases.
    
    \item \textbf{Complementarity}: Different methods answer different questions about incident dynamics:
    \begin{itemize}[nolistsep]
        \item HMM: \textit{What latent state is the system in?} (probabilistic state assignment)
        \item PELT: \textit{When did behavior change?} (temporal boundary detection)
        \item K-means: \textit{What risk profiles exist?} (feature-based clustering)
        \item NB GLM: \textit{What is the overall trajectory?} (trend estimation)
        \item ARIMA: \textit{Where is it heading?} (short-term forecasting)
    \end{itemize}
    
    \item \textbf{Uncertainty quantification}: Agreement across methods suggests greater confidence; disagreement signals areas requiring caution.
\end{enumerate}

Detailed diagnostics and validation results for each method are provided in Appendix~\ref{sec:b_autonomous_vehicles} and ~\ref{sec:c_deepfakes}.

\subsection{Connecting Methods to Phase Classification}
\label{sub:a_connecting_methods_to_phases}

The six-phase framework requires two inputs: \textit{risk level} (current incident intensity relative to baseline) and \textit{trajectory} (rising, stable, or declining). Our statistical methods provide these inputs as follows:

\begin{itemize}[nosep]
    \item \textbf{Risk level}: Standardized risk score ($z$-score). For AVs, this is the exposure-adjusted nowcast count, with validation from HMM state assignments (AV) and K-means cluster membership;
    \item \textbf{Trajectory}: Local trend (3-month rolling slope), validated by PELT within-segment trends and NB GLM time coefficient;
    \item \textbf{Phase assignment}: Deterministic rules applied to (risk, trajectory) pairs, with uncertainty from forecast confidence intervals and HMM state probabilities.
\end{itemize}

Table~\ref{tab:phase_criteria} specifies the classification rules.\footnote{We also evaluated zero-inflated models, Bayesian structural time series, and Hawkes processes, but did not adopt them: Negative Binomial adequately captured \gls*{overdispersion}, PELT provided comparable changepoint detection with greater interpretability, and data sparsity precluded reliable Hawkes process estimation.} Note that risk thresholds ($\theta_{\text{low}}, \theta_{\text{high}}$) should be calibrated to domain-specific volatility: for the AV domain, we set $\theta_{\text{low}} = -0.3\sigma$, $\theta_{\text{high}} = +0.3\sigma$ based on domain knowledge and quantile alignment. For the deepfake domain, we calibrate $\theta_{\text{low}} = +0.14\sigma$ as the upper 1-standard-deviation bound of the PELT-defined dormant reference distribution (see Appendix~\ref{sub:c_phase_classification_df}). The deepfake domain uses a simplified 3-phase classification because the full six-phase framework collapses to only two populated phases (Appendix~\ref{sub:c_pelt_sensitivity_df}).


\begin{table}[htbp]
\centering
\caption{Phase Classification Criteria.}
\label{tab:phase_criteria}
\small
\begin{tabular}{p{3.5cm} p{3.5cm} p{2.5cm}}
\toprule
Phase & Risk Criterion & Trend Criterion \\
\midrule
No Evidence / Coverage & $n = 0$ & --- \\
Rare Mitigated & $r < \theta_{\text{low}}$ & $\tau \leq 0.05$ \\
Rare Occurrence & $r < \theta_{\text{low}}$ & $\tau > 0.05$ \\
Endemic Mitigated & $\theta_{\text{low}} \leq r < \theta_{\text{high}}$ & $\tau \leq 0.05$ \\
Rapid Expansion & Any $r$ & $\tau > 0.05$ \\
Endemic Unmitigated & $r \geq \theta_{\text{high}}$ & $\tau \leq 0.05$ \\
\bottomrule
\multicolumn{3}{p{8.7cm}}{\footnotesize \textit{Note:} AV: $\theta_{\text{low}} = -0.3\sigma$, $\theta_{\text{high}} = +0.3\sigma$. DF: $\theta_{\text{low}} = +0.14\sigma$, $\theta_{\text{high}} = +0.54\sigma$ (see text).}
\end{tabular}
\end{table}



\subsection{Model Assumptions and Validation}
\label{sub:a_model_assumptions}

Each method relies on assumptions validated through diagnostic procedures (Table~\ref{tab:assumptions}).


\begin{table}[htbp]
    \centering
    \caption{Model Assumptions and Validation Status.}
    \label{tab:assumptions}
    \small
    \begin{tabular}{p{2.2cm}p{4.2cm}p{4.6cm}}
    \toprule
    Model & Key Assumptions & Validation Outcome \\
    \midrule
    Log-normal delay & Delays log-normally distributed; independent of severity & AIC comparison (vs.\ exponential, Weibull); 87--94\% valid lags \\
    \addlinespace
    Gaussian HMM & Gaussian emissions; first-order Markov property; time-homogeneous transitions & EM convergence (AV only); interpretable state means \\
    \addlinespace
    PELT & Piecewise constant mean; L2 cost appropriate; independent observations & Multi-penalty sweep; alignment with events (AV \& DF) \\
    \addlinespace
    K-Means & Approximately spherical clusters; Euclidean distance meaningful & Silhouette $> 0.5$ (AV), $0.39$ (DF); interpretable clusters \\
    \addlinespace
    Neg. Binomial & \Gls*{overdispersion}; log-linear mean & Likelihood ratio test ($p < 0.001$); Media RR $= 1.88$, $p = 0.121$ (DF; not significant at $\alpha=0.05$) \\
    \addlinespace
    ARIMA & Stationarity; homoscedasticity & \textbf{PARTIALLY MET}: Stationarity achieved after differencing (ADF $p < 0.001$), but residual variance remains high ($\sigma^2 = 64.85$) due to crisis spikes. \\
    \addlinespace
    Prophet & Trend decomposability; seasonality & Yearly seasonality disabled (insufficient endemic history); trend fits viral growth \\
    \addlinespace
    Gaussian Process & Smoothness (kernel-based) & Kernel optimization (length scale $\approx$ 24mo); robust uncertainty bands \\
    \bottomrule
    \end{tabular}
\end{table}

The Poisson assumption of equidispersion was \textbf{severely violated} in both case study domains (variance exceeding mean by factors of 12--14), necessitating the Negative Binomial specification throughout. The \gls{arima} model achieved stationarity after first differencing (ADF $\tau = -9.13$, $p < 0.001$), but exhibits high residual variance ($\sigma^2 = 64.85$) reflecting the ``punctuated equilibrium'' dynamic of crisis episodes. We retain ARIMA as a conservative baseline forecast for mean reversion dynamics, but note that its confidence intervals may underestimate uncertainty during volatile periods. The Negative Binomial GLM provides more reliable trend estimation, while regime-detection methods (HMM, PELT) better characterize the crisis dynamics that ARIMA treats as noise.

\subsection{Parameter Selection}
\label{sub:a_parameter_selection}

All model parameters were selected via principled procedures such as information criteria, cross-validation, or grid search rather than arbitrary specification (Table~\ref{tab:param_selection}).

\begin{table}[htbp]
\centering
\caption{Parameter Selection Methods.}
\label{tab:param_selection}
\footnotesize
\setlength{\tabcolsep}{4pt}
\begin{tabularx}{\textwidth}{l l l X}
\toprule
Parameter & Method & Criterion & Candidates / Notes \\
\midrule
Delay distribution & Model comparison & Min AIC & Exp., Weibull \\
HMM states ($n$) & Model comparison & BIC + interpretability & $n = 2,\dots,8$ \\
PELT penalty & Adaptive tuning & Event alignment (AV); stability (DF) & Sweep $[0.5,10]$ \\
K-means ($K$) & Internal validation & Max silhouette & $K = 3\text{--}6$ (main); $2\text{--}7$ (sens.) \\
NB dispersion ($\alpha$) & Grid search & Max log-likelihood & $\alpha \in [0.1,5]$ \\
ARIMA $(p,d,q)$ & Model comparison & Min AIC & 9 specifications \\
Phase thresholds & Calibration & Quantile match (AV); $\mu_d+\sigma_d$ (DF) & Invariant region verified \\
\bottomrule
\end{tabularx}
\end{table}

\subsection{Data Processing Pipeline}
\label{sub:a_data_processing_pipeline}

We follow a standardized five-stage preprocessing pipeline:

\begin{enumerate}
    \item \textbf{Ingestion}: Extract incidents from AIID by subdomain classification; parse incident and report dates; link reports to parent incidents. Filtering: subdomain match, valid date fields, non-duplicate records.
    
    \item \textbf{Temporal aggregation}: Aggregate report counts by year-month; create continuous time index spanning full observation period; zero-fill gaps for months without incidents.
    
    \item \textbf{Delay correction}: Calculate reporting lag (report\_date $-$ incident\_date); exclude invalid observations (negative lags or $>$5 years; typically 8--13\% of records); apply nowcasting adjustment to recent months based on expected unreported fraction. Window length is data-driven (95th percentile of empirical lag distribution): 6 months for the AV case; 4 months for the Deepfake case.
    
    \item \textbf{Exposure integration}: Merge domain-specific exposure proxies (California DMV autonomous miles for AVs); compute exposure-adjusted rates where data permit.
    
    \item \textbf{Feature engineering}: Compute standardized risk ($z = (x - \mu) / \sigma$), local trend (3-month rolling OLS slope), media intensity (Google Trends index), and apply 6-phase classification rules.
\end{enumerate}






\subsection{Sensitivity Analysis}
\label{sub:a_sensitivity_analysis}

We assessed robustness of phase classifications to parameter perturbations across both case studies.

\textbf{AV domain.} A threshold sensitivity sweep across $\theta_{\text{low}} \in [-0.5\sigma, 0.0\sigma]$ and $\theta_{\text{high}} \in [+0.1\sigma, +0.5\sigma]$ (Appendix~\ref{sub:b_threshold_sensitivity}) revealed a step-function transition at $\theta_{\text{high}} = +0.5\sigma$: 26 months (20\% of the series) shift between Endemic Mitigated and Endemic Unmitigated at this boundary, while $\theta_{\text{low}}$ has no effect across the tested range because all non-zero months have standardized risk scores above $0.0\sigma$. The framework's six phases therefore collapse to four operative phases for AV data, with Rare Mitigated and Rare Occurrence never assigned. Major crisis periods were detected across all PELT penalty levels tested. Governance-relevant conclusions such as the dominance of No Evidenced Occurrence months (57\%, invariant) and the current Rapid Expansion classification were stable across all parameter ranges. The oscillatory crisis-recovery pattern was consistent across all parameter ranges tested.

\textbf{Deepfake domain.} A threshold sensitivity sweep across $\theta_{\text{low}} \in [-0.8\sigma, +0.5\sigma]$ identified an \textit{invariant zone} $\theta \in (-0.05\sigma, +0.43\sigma)$ within which the binary dormant-to-endemic classification is completely stable. Both the AV-aligned threshold ($-0.3\sigma$) and the previous proposal ($-0.5\sigma$) fall outside this zone, misclassifying the pre-commercial baseline (0.2 incidents/month) as endemic. The data-driven threshold ($+0.14\sigma$) lies within the invariant zone. Additionally, the 3-segment PELT structure (plateau stability) and 5-segment decomposition both yield the same qualitative finding of an irreversible step-function transition. Full sensitivity results, including 5-segment decomposition and six-phase classification, are reported in Appendix~\ref{sub:c_pelt_sensitivity_df}.
\section{Autonomous Vehicles: Additional Analyses}
\label{sec:b_autonomous_vehicles}

\subsection{K-Means Validation}
\label{sub:b_kmeans_validation}

To validate the \gls{pelt} segmentation, we applied \glslink{kmeans}{K-means clustering} as an independent regime detection method. Features consisted of standardized risk level and local trend (3-month rolling slope, upweighted $2\times$ to emphasize trajectory). We evaluated $K = 3$ to $K = 7$ clusters using \Gls{silhouette} score (Table~\ref{tab:kmeans_selection}).

\begin{table}[htbp]
\centering
\caption{K-means Cluster Selection Metrics.}
\label{tab:kmeans_selection}
\begin{tabular}{ccc}
\toprule
$K$ & Silhouette & Calinski-Harabasz \\
\midrule
3 & 0.434 & 64.1 \\
4 & 0.481 & 82.4 \\
5 & 0.525 & 111.1 \\
6 & 0.567 & 121.9 \\
7 & 0.594 & 127.3 \\
\bottomrule
\end{tabular}
\end{table}

The $K = 3$ solution identified three distinct risk states (Table~\ref{tab:kmeans_clusters}): a dominant baseline cluster (75.8\% of months), a moderate-risk cluster with rising trend (23.4\%), and a rare extreme spike cluster (0.8\%).

\begin{table}[htbp]
\centering
\caption{K-means Cluster Features.}
\label{tab:kmeans_clusters}
\begin{tabular}{clcccp{3.5cm}}
\toprule
Cluster & Risk ($\sigma$) & Trend & $n$ & \% & Interpretation \\
\midrule
0 & $+0.06$ & $+0.84$ & 30 & 23.4\% & Elevated/rising \\
1 & $-0.06$ & $-0.25$ & 97 & 75.8\% & Baseline (stable) \\
2 & $+3.89$ & $+4.59$ & 1 & 0.8\% & Extreme spike \\
\bottomrule
\end{tabular}
\end{table}

\begin{figure}[htbp]
    \centering
    \includegraphics[width=\textwidth]{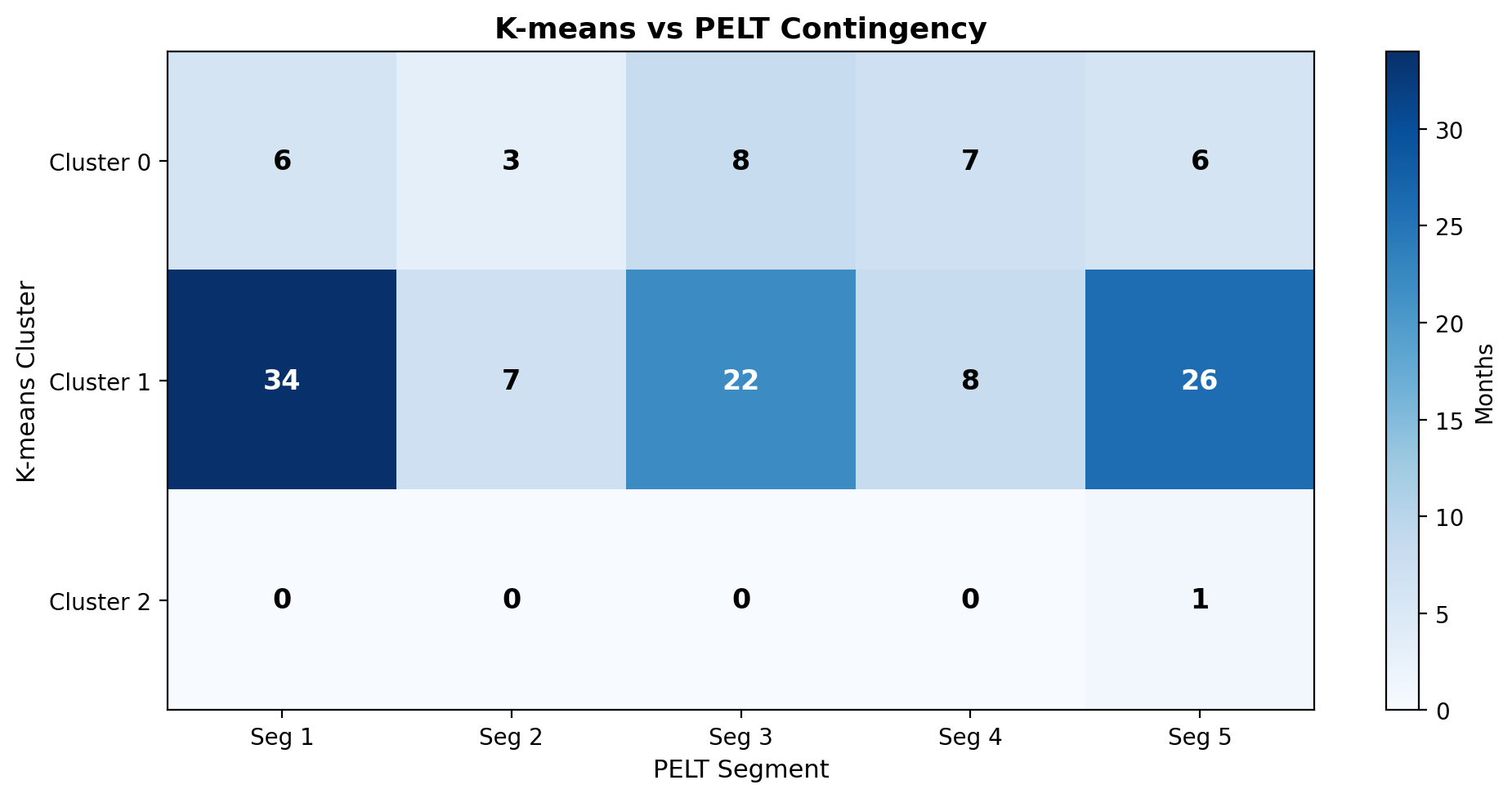}
    \caption{Cross-validation contingency matrix: K-means clusters (rows) vs \gls*{pelt} segments (columns). Cell values indicate months assigned to each combination. The baseline cluster (Cluster 1) spans all segments, while elevated months appear episodically across segments. ARI $= 0.025$, NMI $= 0.094$, reflecting complementary rather than contradictory structure.}
    \label{fig:kmeans_validation}
\end{figure}

Cross-validation against \gls*{pelt} segments yielded an Adjusted Rand Index (ARI) of 0.025 and Normalized Mutual Information (NMI) of 0.094, indicating weak agreement between the two methods (Figure~\ref{fig:kmeans_validation}). However, this divergence is substantively meaningful rather than methodologically problematic: \gls*{pelt} identifies \textit{temporal regimes} (contiguous periods), while K-means identifies \textit{risk states} (which may be scattered across time). The contingency analysis reveals that the baseline cluster (Cluster 1) spans all \gls*{pelt} segments, while elevated months (Cluster 0) appear episodically within multiple segments.

This pattern suggests that AV incidents follow a ``punctuated equilibrium'' dynamic: extended baseline periods are interrupted by brief, intense crisis episodes rather than sustained elevated regimes. The \gls*{pelt} segments capture \textit{when} the system's central tendency shifted, while K-means captures \textit{what} risk state each month occupies---together providing a more complete characterization than either method alone.

\paragraph{Methodological Triangulation.} The combination of HMM (6 states), \gls*{pelt} (5 segments over the full 2014--2025 series), and K-means (3 clusters) reveals consistent structure despite methodological differences. All three methods identify: (1) a dominant low-activity baseline comprising 57--76\% of observations, (2) rare extreme events ($<$1\% of months) generating the highest spikes, and (3) intermediate elevated states during crisis periods. The convergence across unsupervised learning, changepoint detection, and latent state modeling strengthens confidence in the identified regime structure.

\subsection{Lagged Cross-Correlation}
\label{sub:b_ccf}

\begin{figure}[htbp]
    \centering
    \includegraphics[width=\textwidth]{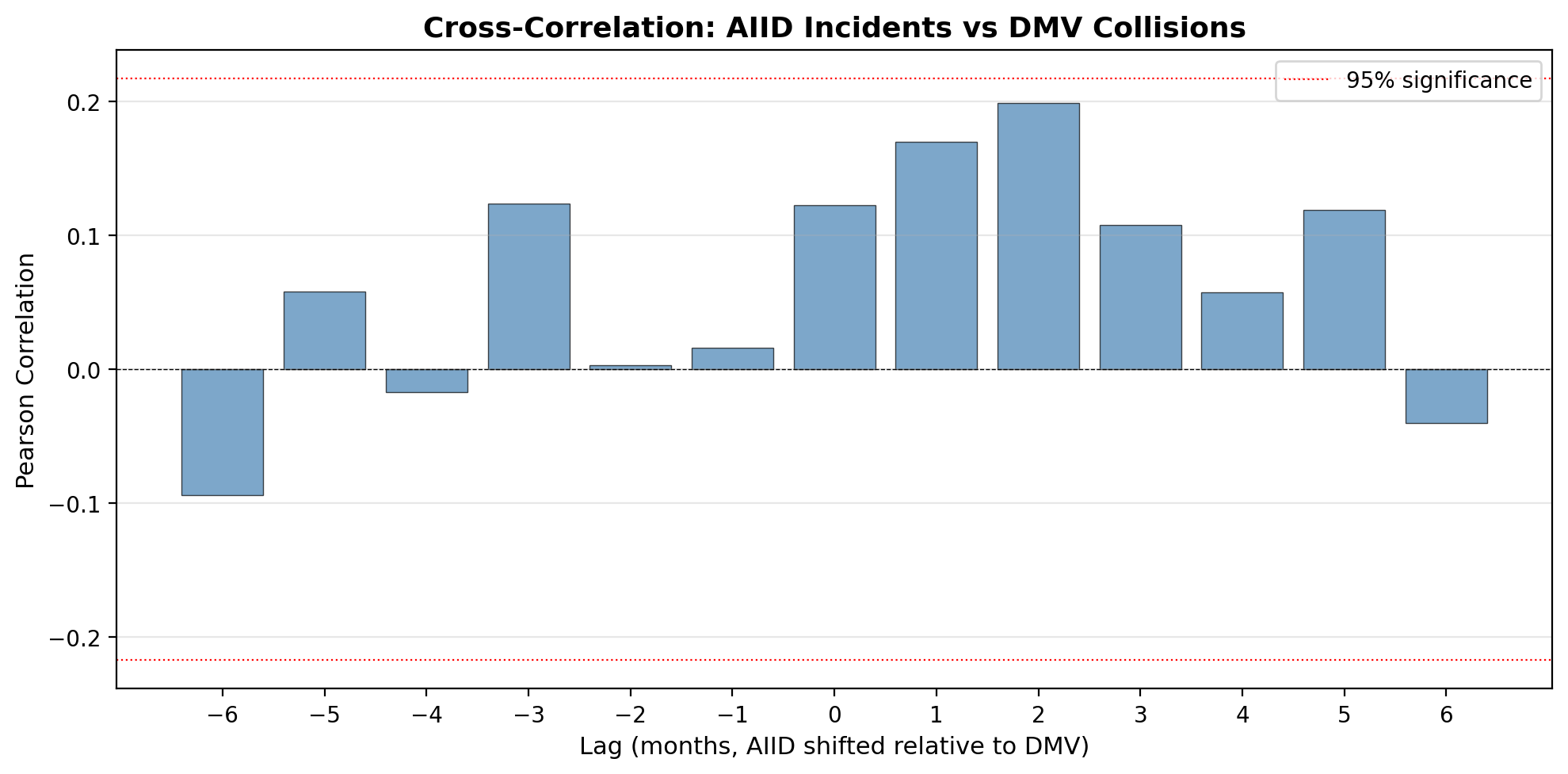}
    \caption{Cross-correlation function between AIID incident counts and DMV collision counts across lags of $-6$ to $+6$ months. Positive lags indicate AIID shifted forward (lagging DMV). The peak at lag $+2$ ($r = 0.199$) suggests a two-month media reporting delay, though no lag reaches statistical significance (red dashed lines).}
    \label{fig:ccf}
\end{figure}

To rule out the possibility that the weak contemporaneous correlation between AIID and DMV series reflects simple reporting latency, we computed Pearson correlations across lags of $-6$ to $+6$ months (Figure~\ref{fig:ccf}). The maximum correlation occurs at lag $+2$ months ($r = 0.199$), consistent with media reporting delay: collisions that enter the DMV record immediately may only attract media attention weeks later. However, no lag within the tested range achieves statistical significance at $\alpha = 0.05$, indicating that the AIID--DMV divergence cannot be explained by reporting delay alone.

\subsection{Intervention Impact Analysis}
\label{sub:b_intervention_impact}

We analyzed the impact of 17 key events on AV incident reporting, including fatal crashes (n=3), regulatory actions (n=8), company decisions (n=3), and deployment milestones (n=3). For each intervention, we compared mean standardized risk in the six months before and after the event using independent samples t-tests, with \glslink{fdr}{False Discovery Rate (FDR)} correction for multiple comparisons \citep{benjamini2001control}.

\begin{table}[htbp]
\centering
\caption{Selected AV Intervention Events and Impact Analysis}
\label{tab:av_interventions}
\small
\begin{tabular}{llcccl}
\toprule
Intervention & Date & Type & $\Delta\sigma$ & $p_{\text{FDR}}$ & Effect \\
\midrule
Tesla Autopilot Fatality & 2016-05 & Fatal & $+0.84$ & 0.705 & Shock \\
NHTSA AV Policy 1.0 & 2016-09 & Regulatory & $-1.17$ & 0.705 & Mitigation \\
NHTSA AV Policy 2.0 & 2017-09 & Regulatory & $+1.97$ & 0.705 & Shock \\
Uber ATG Fatality & 2018-03 & Fatal & $-0.68$ & 0.705 & Mitigation \\
Uber Suspends Testing & 2018-03 & Company & $-0.68$ & 0.705 & Mitigation \\
Waymo Driverless SF & 2022-03 & Deployment & $+0.37$ & 0.705 & Shock \\
Cruise Pauses Operations & 2023-10 & Company & $+0.37$ & 0.705 & Shock \\
\bottomrule
\end{tabular}
\end{table}

\begin{figure}[htbp]
    \centering
    \includegraphics[width=\textwidth]{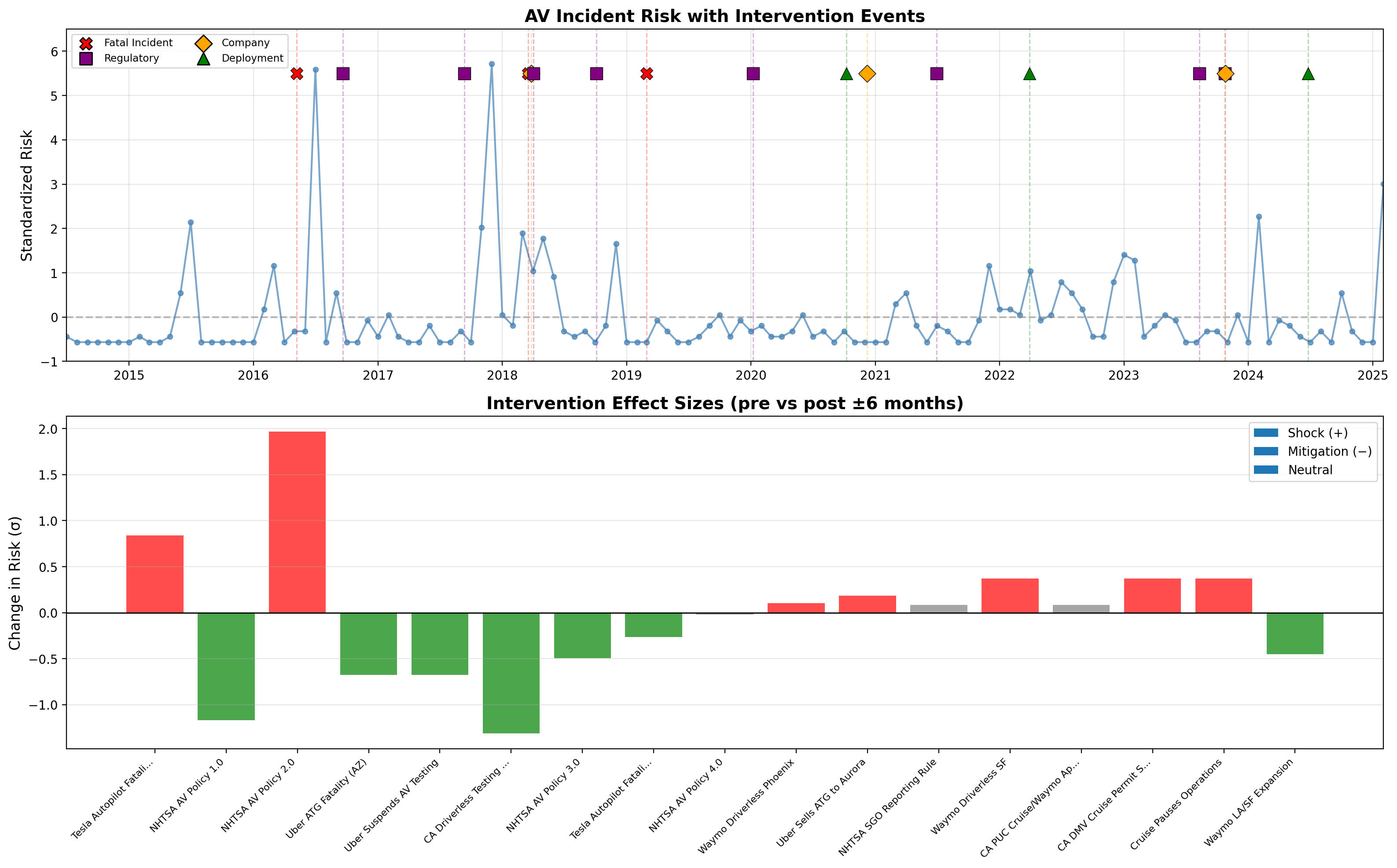}
    \caption{Intervention Impact Analysis. \textit{Top}: Risk timeline with intervention events marked by type. \textit{Bottom}: Effect sizes showing change in standardized risk; no effects achieved significance after FDR correction.}
    \label{fig:intervention_av}
\end{figure}

No interventions achieved statistical significance after FDR correction ($p_{\text{FDR}} < 0.05$). Notably, the October 2023 Cruise suspension appears in the AIID record as a positive shock ($+0.37\sigma$) because media coverage spiked, even though the DMV record shows Cruise collisions dropping to near zero after the suspension (see Section~\ref{sub:b_company_dmv}) --- a direct illustration of the AIID--DMV divergence documented in Section~\ref{sub:3_insufficiency}. More broadly, this null result is substantively informative rather than methodologically problematic (Figure~\ref{fig:intervention_av}). Analysis of expected versus observed effects revealed systematic discordance: of 9 interventions expected to produce shocks, 4 showed mitigation effects; conversely, of 8 interventions expected to mitigate risk, 4 showed shock effects (Table~\ref{tab:intervention_confusion}).

\begin{table}[htbp]
\centering
\caption{Intervention Expected vs.\ Actual Effects}
\label{tab:intervention_confusion}
\begin{tabular}{lccc|c}
\toprule
& \multicolumn{3}{c|}{Actual Effect} & \\
Expected & Mitigation & Neutral & Shock & Total \\
\midrule
Mitigation & 3 & 1 & 4 & 8 \\
Shock & 4 & 2 & 3 & 9 \\
\midrule
Total & 7 & 3 & 7 & 17 \\
\bottomrule
\end{tabular}
\end{table}

This pattern suggests a \textit{rapid response cycle} in the AV domain: fatal incidents trigger immediate operational pauses (Uber suspended testing within 9 days of the Tempe fatality), such that the post-intervention window captures both shock and response. The Uber ATG fatality, for example, shows net mitigation ($\Delta = -0.68\sigma$) because the company's suspension began almost immediately.

By contrast, the deepfake domain (see Section~\ref{sec:4_deepfakes}) shows significant intervention effects because technology shocks (generative AI releases) are not followed by comparable industry self-regulation. This divergence illustrates a key finding: AI incident trajectories depend critically on whether deployment is controlled by identifiable actors subject to regulatory and reputational pressure (as with AVs) or distributed across anonymous users (as with deepfakes).

\subsection{Negative Binomial Regression}
\label{sub:b_nb_glm}

The Poisson model assumption of equidispersion (variance equals mean) was violated in our data: the observed variance (0.75) exceeded the mean (0.62) by a factor of 1.2. While this overdispersion is modest compared to the report-date-based analysis, we fitted a Negative Binomial GLM for methodological completeness. The best-fitting model ($\alpha = 0.5$) achieved a log-likelihood of $-132.39$, representing only marginal improvement over the Poisson specification (likelihood ratio $= 0.18$, $p > 0.05$). Model comparison metrics are presented in Table~\ref{tab:nb_comparison}.

\begin{table}[htbp]
\centering
\caption{Model Comparison: Poisson vs.\ Negative Binomial GLM.}
\label{tab:nb_comparison}
\begin{tabular}{lcc}
\toprule
Metric & Poisson & Negative Binomial \\
\midrule
Log-Likelihood & $-132.49$ & $-132.39$ \\
AIC & 268.97 & 268.79 \\
\bottomrule
\end{tabular}
\end{table}

\begin{table}[htbp]
\centering
\caption{Negative Binomial GLM Coefficient Estimates.}
\label{tab:nb_glm}
\begin{tabular}{lcccc}
\toprule
Parameter & Estimate & 95\% CI & Rate Ratio & $p$-value \\
\midrule
Intercept & $-0.545$ & $[-0.810, -0.281]$ & --- & $<0.001$ \\
Time trend & $+0.362$ & $[+0.100, +0.624]$ & 1.436 & $0.007$ \\
\bottomrule
\end{tabular}
\end{table}

\begin{figure}[htbp]
    \centering
    \includegraphics[width=\textwidth]{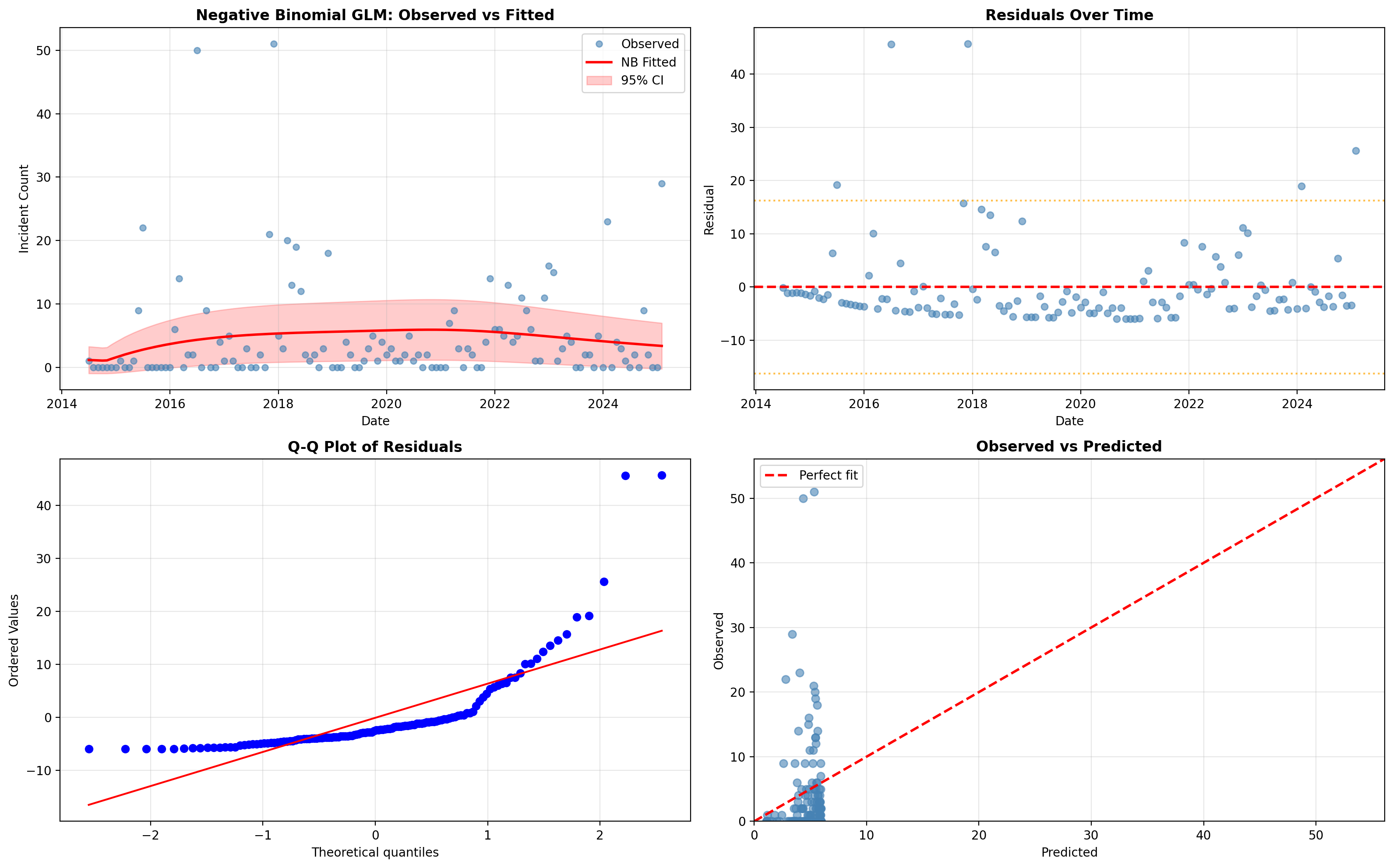}
    \caption{Negative Binomial GLM Diagnostics. \textit{Top left}: Observed vs.\ fitted values showing the model captures the central tendency but not extreme spikes. \textit{Top right}: Residuals over time with no systematic pattern. \textit{Bottom left}: Q-Q plot revealing heavy right tails from crisis episodes. \textit{Bottom right}: Observed vs.\ predicted scatter.}
    \label{fig:nb_glm}
\end{figure}

The Negative Binomial model identifies a significant positive time trend: the rate ratio of 1.436 (95\% CI: [1.105, 1.868]) indicates that for each standard deviation increase in time (approximately 37 months), the expected incident rate increases by 43.6\%. This positive trend contrasts with the declining trend observed under report-date counting, and likely reflects increasing incident visibility over time as the AV industry scales deployment and reporting infrastructure matures.

The minimal overdispersion (factor 1.2) and negligible improvement over the Poisson model indicate that the incident-date-based series is well-approximated by a Poisson process---consistent with the sparse count distribution (57\% zeros, maximum of 4 incidents per month). Note that the NB GLM's significant positive trend (rate ratio 1.436, $p = 0.007$) uses standardized time over the full 128-month series, while the OLS trends reported in Sections~\ref{b_severity_weighting} and \ref{sub:b_exposure_adjusted} use unstandardized monthly slopes ($\beta = +0.006$/month, $p = 0.775$); the two frameworks are not directly comparable.

\subsection{ARIMA Forecasting}
\label{sub:b_arima_forecasting}

To project future incident trajectories, we fitted autoregressive integrated moving average (ARIMA) models to the nowcast-adjusted time series. The augmented Dickey-Fuller test confirmed stationarity after first differencing ($\tau = -7.78$, $p < 0.001$). Model selection via AIC identified ARIMA(0,1,1) as optimal (AIC $= 329.99$; Table~\ref{tab:arima_selection}).

\begin{table}[htbp]
\centering
\caption{ARIMA Model Selection.}
\label{tab:arima_selection}
\begin{tabular}{lccc}
\toprule
Model & AIC & BIC & Log-Likelihood \\
\midrule
ARIMA(1,0,0) & 332.61 & 341.17 & $-163.31$ \\
\textbf{ARIMA(0,1,1)} & \textbf{329.99} & \textbf{335.68} & $\mathbf{-163.00}$ \\
ARIMA(1,1,1) & 331.96 & 340.49 & $-162.98$ \\
ARIMA(2,1,2) & 332.18 & 346.40 & $-161.09$ \\
\bottomrule
\end{tabular}
\end{table}

The fitted ARIMA(0,1,1) model exhibits strong moving average mean reversion ($\theta_1 = -0.954$), indicating that shocks to the incident rate dissipate almost completely within one period. The residual variance ($\sigma^2 = 0.75$) is consistent with the low-count Poisson-like dynamics of the incident-date series.

\begin{table}[htbp]
\centering
\caption{12-Month ARIMA Forecast for AV Incidents.}
\label{tab:arima_forecast}
\begin{tabular}{lccc}
\toprule
Period & Forecast & 95\% CI & Projected Phase \\
\midrule
Mar 2025 & 0.8 & $[-0.9, 2.5]$ & Endemic Mitigated \\
Apr--Dec 2025 & 0.8 & $[-0.9, 2.5]$ & Endemic Mitigated \\
Jan--Feb 2026 & 0.8 & $[-0.9, 2.5]$ & Endemic Mitigated \\
\bottomrule
\end{tabular}
\end{table}

\begin{figure}[htbp]
    \centering
    \includegraphics[width=\textwidth]{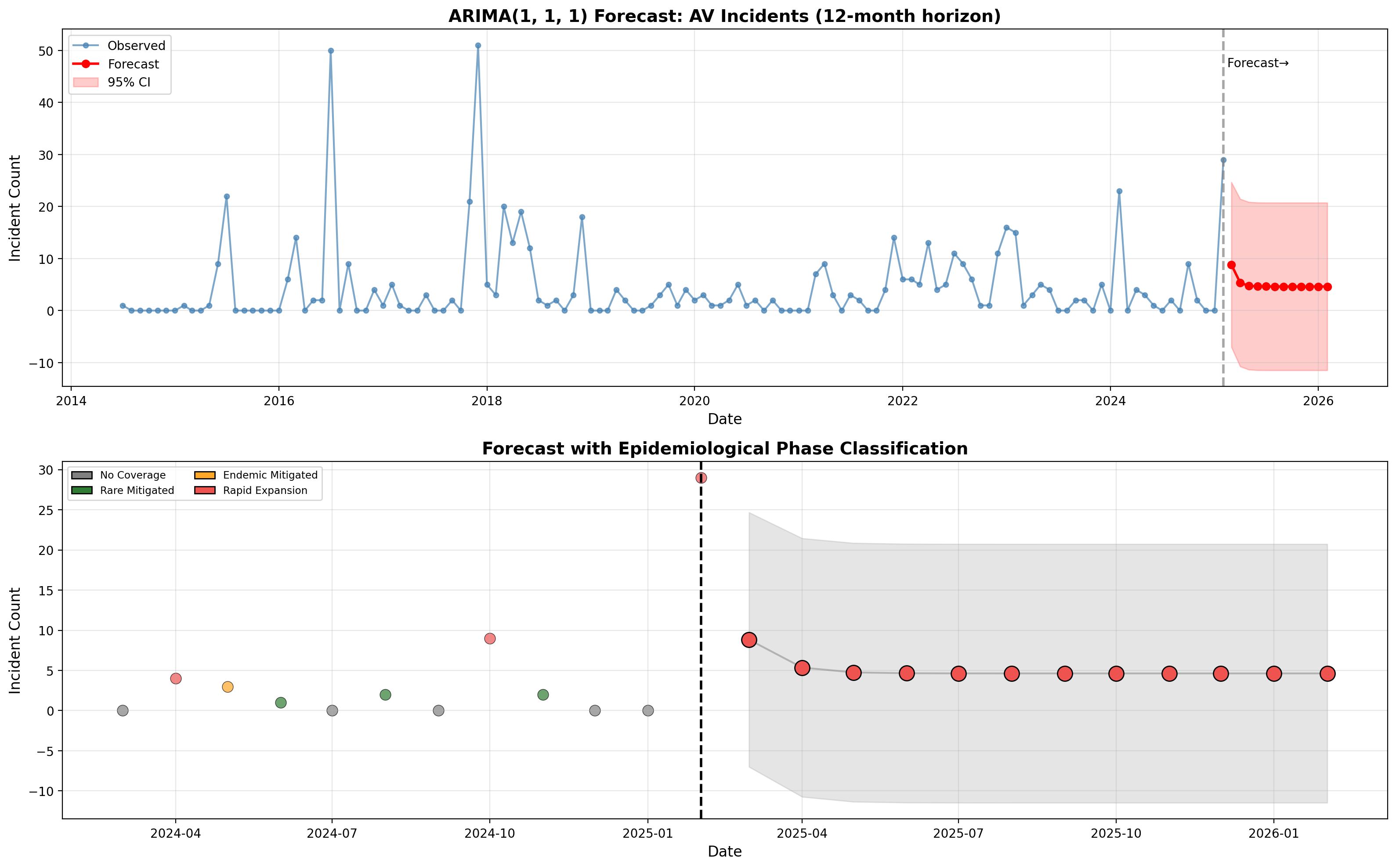}
    \caption{ARIMA(0,1,1) Forecast with Epidemiological Phase Classification. \textit{Top}: Historical observations and 12-month forecast with 95\% confidence intervals. \textit{Bottom}: Forecast points colored by projected phase, with recent history for context. The model projects rapid mean reversion from the February 2025 nowcast-adjusted count toward the long-term average.}
    \label{fig:arima_forecast}
\end{figure}

Figure~\ref{fig:arima_forecast} presents the 12-month forecast with 95\% confidence intervals. The model projects rapid mean reversion from the February 2025 nowcast-adjusted count (4 incidents) toward the long-term average of approximately 0.8 incidents per month. The constant forecast reflects the near-unit moving average coefficient ($\theta_1 = -0.954$), which effectively cancels the most recent innovation. The narrow confidence intervals (approximately $\pm$1.7 incidents) reflect the low variance of the incident-date series, though they include negative values --- a modeling artifact for count data that underscores the need for count-specific models such as the Negative Binomial GLM for probabilistic inference.

\subsection{Severity Weighting Analysis}
\label{b_severity_weighting}

We assigned numeric weights to each severity level based on a modified KABCO-inspired scale (Table~\ref{tab:severity-weights}). This weighting scheme reflects the assumption that higher-severity incidents warrant proportionally greater attention in risk assessment.

\begin{table}[h]
\centering
\caption{Severity Weight Assignments for AV Incidents.}
\label{tab:severity-weights}
\begin{tabular}{lrrr}
\toprule
\textbf{Severity Level} & \textbf{Weight} & \textbf{Count} & \textbf{\%} \\
\midrule
1 Negligible & 1 & 33 & 44.0\% \\
2 Minor & 3 & 39 & 52.0\% \\
3 Substantial & 10 & 3 & 4.0\% \\
4 Severe & 50 & 0 & 0.0\% \\
\midrule
\textbf{Total} & --- & 75 & 100.0\% \\
\bottomrule
\end{tabular}
\end{table}

For each month $t$, we computed:
\begin{itemize}
    \item \textbf{Raw count}: $C_t = \sum_{i \in t} 1$;
    \item \textbf{Severity sum}: $S_t = \sum_{i \in t} w_i$, where $w_i$ is the severity weight for incident $i$;
    \item \textbf{Standardized series}: $z_t = (x_t - \bar{x}) / \sigma_x$ for both count and severity.
\end{itemize}

\begin{figure}[h]
    \centering
    \includegraphics[width=\textwidth]{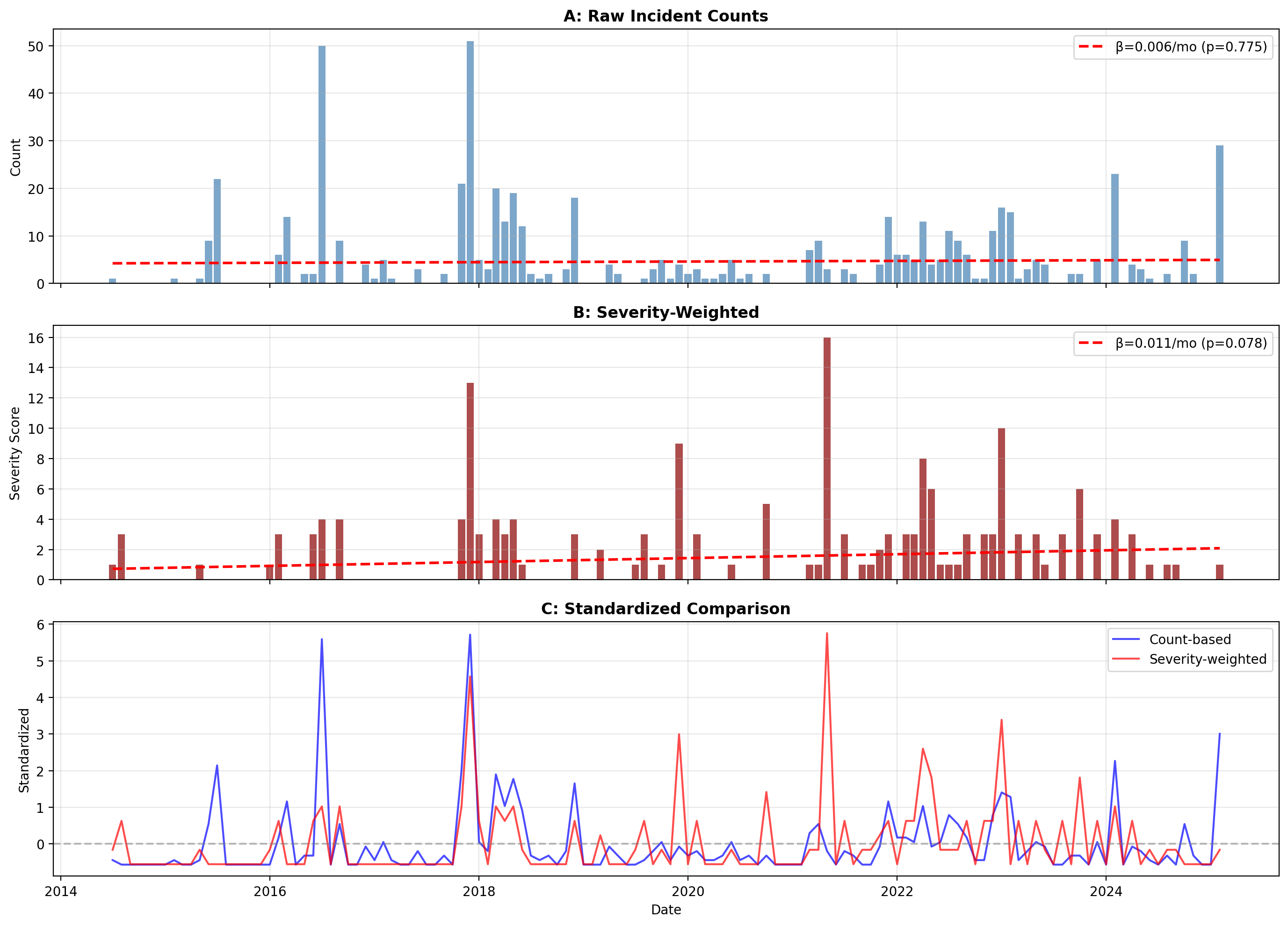}
    \caption{Comparison of Count-Based and Severity-Weighted AV Incident Series (2014--2025). Panel A shows raw monthly incident counts with trend line. Panel B shows severity-weighted monthly scores. Panel C overlays the standardized versions of both series, revealing periods of divergence where severity spikes exceed count-based signals (e.g., 2021) or vice versa (e.g., 2018).}
    \label{fig:severity-comparison}
\end{figure}

We then compared linear trends using ordinary least squares regression of each series on time. Figure~\ref{fig:severity-comparison} presents the comparison between count-based and severity-weighted incident series. The key findings are:

\begin{enumerate}
    \item \textbf{Trend divergence}: The count-based trend is flat and non-significant ($\beta = +0.006$/month, $p = 0.775$), while the severity-weighted trend shows marginal positive acceleration ($\beta = +0.011$/month, $p = 0.078$).

    \item \textbf{Moderate correlation}: The two series exhibit moderate correlation ($r = 0.47$), indicating that high-count months do not always correspond to high-severity months.

    \item \textbf{High-severity incidents}: Only three incidents were classified as ``Substantial'' (weight = 10):
    \begin{itemize}
        \item Waze wildfire navigation incident (December 2017)
        \item Tesla Autopilot/FSD misrepresentation allegations (May 2021)
        \item Cruise pedestrian dragging incident (January 2023)
    \end{itemize}
\end{enumerate}

\paragraph{Limitations of the Severity Proxy.}
The NatSec Overall Impact field presents important limitations as a severity proxy:

\begin{enumerate}
    \item \textbf{No fatalities coded as ``Severe''}: Known fatal AV incidents---including the Tesla Autopilot fatality (May 2016) and the Uber ATG fatality (March 2018)---are not classified as ``4 Severe'' in the available data. This suggests the NatSec field captures regulatory or societal impact rather than physical harm severity.

    \item \textbf{Subjective classification}: The NatSec ratings reflect analyst judgment about national security implications, which may not align with traditional crash severity metrics used in transportation safety (e.g., the KABCO scale: K=fatality, A=incapacitating injury, B=non-incapacitating injury, C=possible injury, O=property damage only).

    \item \textbf{Concentration in low-severity categories}: With 96\% of incidents rated as Negligible or Minor, the severity-weighted analysis has limited discriminatory power.
\end{enumerate}

\paragraph{Implications.} Despite these limitations, the severity analysis provides two governance-relevant insights:

\begin{enumerate}
    \item \textbf{Count-based trends may understate risk evolution}: The marginally significant positive trend in severity-weighted incidents ($p = 0.078$) suggests that while incident frequency has stabilized, the character of incidents may be shifting toward higher-impact events. This warrants continued monitoring even during periods of apparent count stability.

    \item \textbf{Need for standardized severity classification}: Future incident reporting frameworks should incorporate validated severity scales (e.g., KABCO, AIS) to enable more precise severity-adjusted trend analysis. The EU AI Act's serious incident reporting requirements (Article 73) and NHTSA's Standing General Order data could provide complementary severity information for cross-validation.
\end{enumerate}

\subsection{Exposure Data Source}

The California Department of Motor Vehicles requires companies testing autonomous vehicles to submit annual Autonomous Vehicle Disengagement Reports, which include total miles driven in autonomous mode. We aggregated monthly mileage from all reporting companies (2020--2024) to construct an exposure time series.

\begin{table}[h]
\caption{California DMV Autonomous Vehicle Exposure Data Summary.}
\centering
\small
\begin{tabular}{lr}
\toprule
\textbf{Metric} & \textbf{Value} \\
\midrule
Data source & California DMV \\
Coverage period & Dec 2020 -- Nov 2024 \\
Total autonomous miles & 23.4M \\
Peak monthly miles & 1.05M (Aug 2023) \\
AIID incidents in period & 39 \\
Mean rate (per million miles)\textsuperscript{a} & 1.67 \\
Exposure-adjusted trend & $\beta$ = --0.023, $p$ = 0.274 \\
\bottomrule
\end{tabular}
\label{tab:exposure-summary}
\end{table}
{\footnotesize \textsuperscript{a}Period-aggregate rate (total incidents $\div$ total miles). Monthly rates vary; Figure~\ref{fig:exposure-adjusted} Panel C shows month-by-month values.}

\subsection{Exposure-Adjusted Results}
\label{sub:b_exposure_adjusted}

\begin{figure}[h]
\centering
\includegraphics[width=\textwidth]{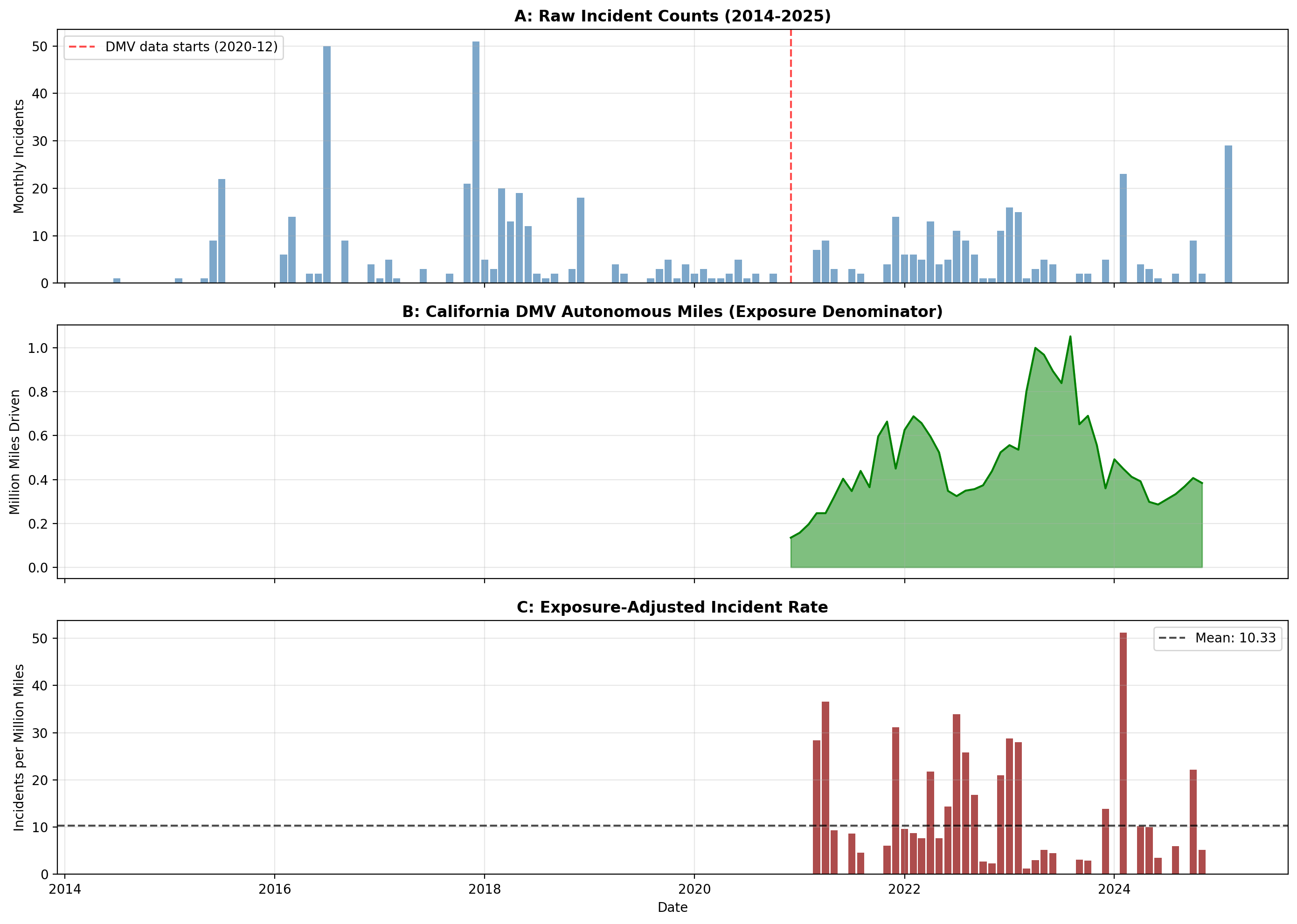}
\caption{Exposure-Adjusted Incident Analysis. Panel A shows raw AIID incident counts with vertical line indicating start of DMV exposure data. Panel B shows California autonomous miles driven monthly. Panel C shows the resulting exposure-adjusted rate.}
\label{fig:exposure-adjusted}
\end{figure}

Figure~\ref{fig:exposure-adjusted} presents three panels: (A) raw incident counts spanning our full observation period (2014--2025), (B) monthly autonomous miles driven in California, and (C) the exposure-adjusted incident rate (incidents per million autonomous miles).

Key findings from the exposure-adjusted analysis:

\begin{enumerate}
    \item \textbf{Trend divergence}: Raw incident counts show a negligible positive trend ($\beta$ = +0.006/month, $p$ = 0.775), while exposure-adjusted rates show a \textit{negative} trend ($\beta$ = --0.023 incidents per million miles/month, $p$ = 0.274). Neither trend reaches statistical significance, but the directional divergence is substantively important: it suggests that deployment growth, not increasing risk, drives observed incident counts.

    \item \textbf{Rate magnitude}: The mean incident rate of 1.67 incidents per million autonomous miles is comparable to human-driven vehicle crash rates ($\sim$2.1 police-reported crashes per million VMT nationally). However, this comparison requires caution---the AIID captures near-misses, software anomalies, and minor incidents that would not trigger police reporting for human drivers, while the denominator covers only California miles whereas incidents are drawn from multiple jurisdictions.

    \item \textbf{Temporal coverage gap}: Exposure data covers only December 2020--November 2024 (48 months), while our incident database spans 2014--2025. The pre-2020 period includes formative incidents (e.g., the 2016 Tesla fatality, 2018 Uber fatality) that cannot be exposure-normalized, limiting historical trend analysis.
\end{enumerate}

\subsection{HMM Model Selection}
\label{sub:b_hmm_selection}

The sparse AV count distribution (5 unique values, 57\% zeros, maximum of 4 incidents per month) is not well-suited to Gaussian HMM fitting: the model produces degenerate solutions in which the emission distributions collapse toward point masses at each integer value, yielding per-observation log-likelihoods greater than 2 (i.e., probability densities exceeding 8 per observation). We therefore do not report formal model selection results for the HMM. The six-state specification used in Section~\ref{sub:3_phase_classification} is retained for alignment with the six-phase classification framework rather than for statistical optimality. A Poisson or categorical HMM would be more appropriate for this data regime and is left for future work.

\subsection{Phase-Threshold Sensitivity}
\label{sub:b_threshold_sensitivity}

\begin{figure}[htbp]
    \centering
    \includegraphics[width=0.8\textwidth]{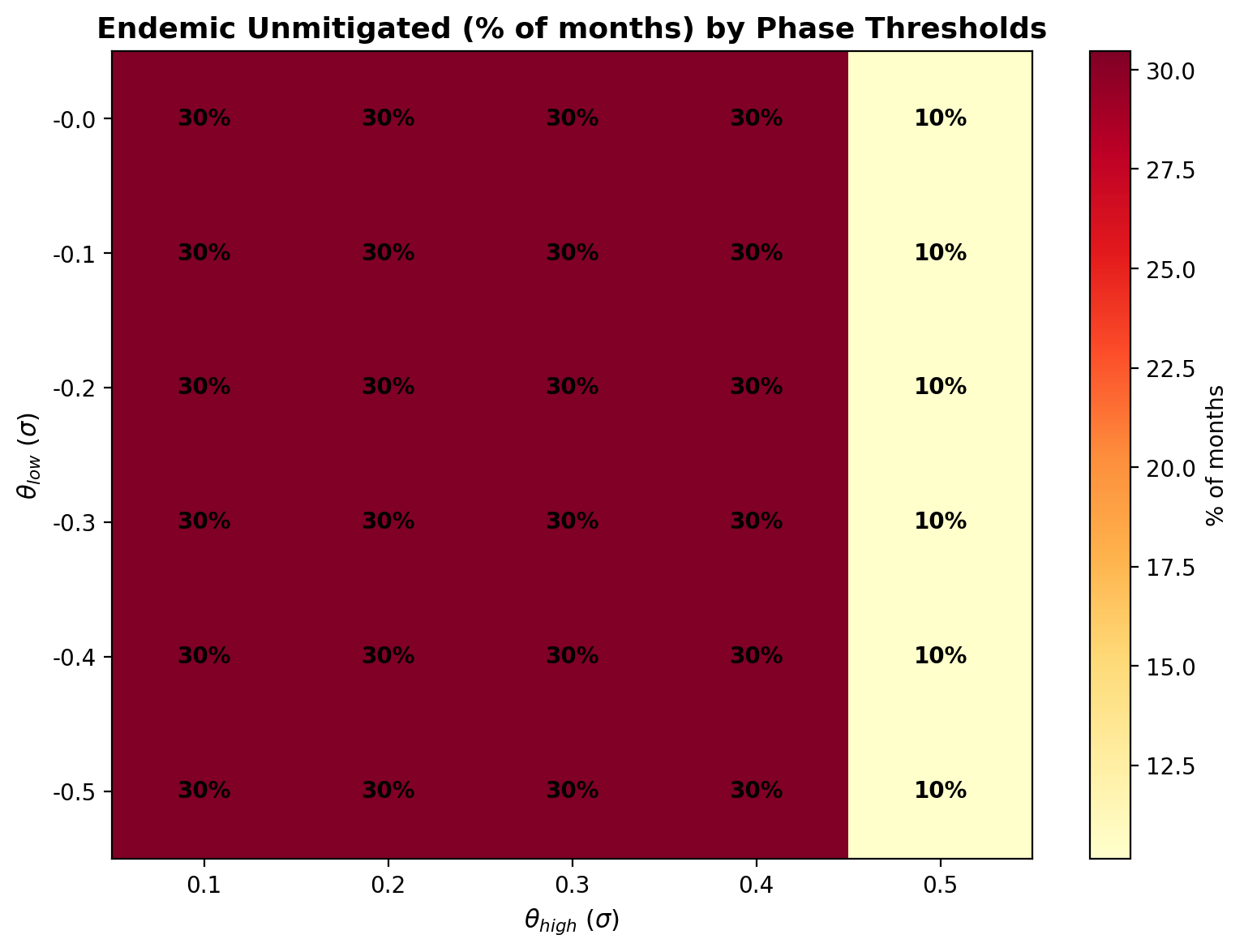}
    \caption{Percentage of months classified as ``Endemic Unmitigated'' across the $(\theta_{\text{low}}, \theta_{\text{high}})$ parameter grid. The classification exhibits a step-function sensitivity at $\theta_{\text{high}} = 0.5\sigma$: below this threshold, 30\% of months are classified as Endemic Unmitigated; at $\theta_{\text{high}} \geq 0.5\sigma$, this drops to 10\%. The low-risk threshold $\theta_{\text{low}}$ has no effect across the tested range.}
    \label{fig:threshold_sensitivity}
\end{figure}

We swept the phase classification thresholds $\theta_{\text{low}} \in [-0.5\sigma, 0.0\sigma]$ and $\theta_{\text{high}} \in [+0.1\sigma, +0.5\sigma]$ to assess the sensitivity of ``Endemic Unmitigated'' assignments (Figure~\ref{fig:threshold_sensitivity}).

Two findings emerge. First, $\theta_{\text{low}}$ has no effect on any phase assignment across the tested range --- all non-zero months have standardized risk scores above $0.0\sigma$ given the low baseline count (mean $= 0.6$), so the low-risk threshold never triggers. Second, the classification of ``Endemic Unmitigated'' depends entirely on a single step at $\theta_{\text{high}} = 0.5\sigma$: 26 months shift between Endemic Mitigated and Endemic Unmitigated at this boundary. Governance-relevant conclusions --- the current Rapid Expansion classification (invariant across all threshold combinations) and the dominance of No Evidenced Occurrence months (57\%, invariant) --- are robust. A corollary is that the framework's six phases effectively collapse to four operative phases for AV data: no month has a standardized risk score below $\theta_{\text{low}}$, so the Rare Mitigated and Rare Occurrence phases are never assigned. This reflects the framework's sensitivity to domain data characteristics rather than a design flaw; higher-volume domains (e.g., deepfakes) activate all six phases.

\subsection{Company-Level DMV Decomposition}
\label{sub:b_company_dmv}

\begin{figure}[htbp]
    \centering
    \includegraphics[width=\textwidth]{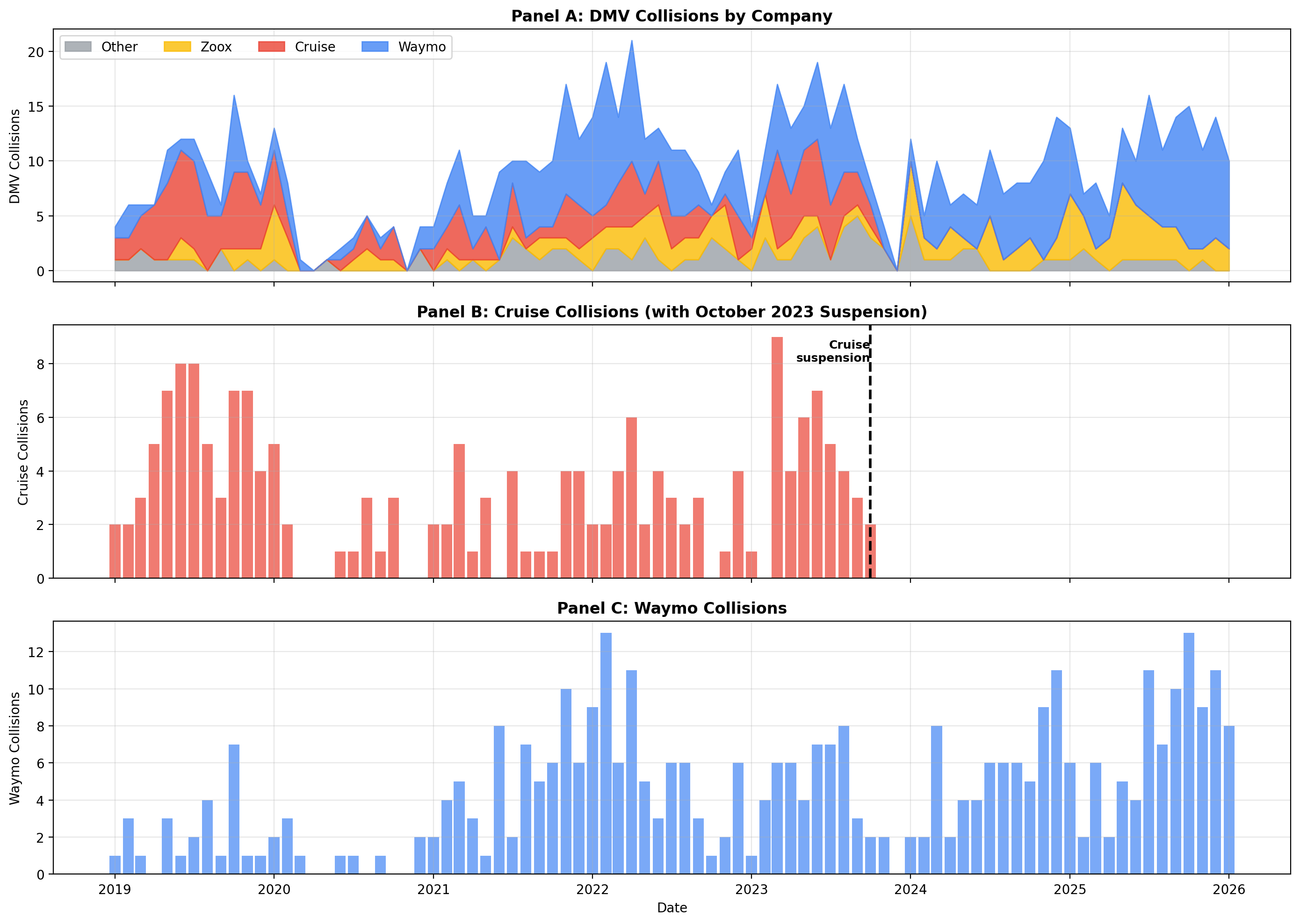}
    \caption{DMV collision reports disaggregated by company. Panel A: stacked area showing all companies. Panel B: Cruise-specific series with the October 2023 operational suspension marked. Panel C: Waymo-specific series for comparison. The Cruise suspension produces a visible structural break: collisions drop from 6--8/month to near zero.}
    \label{fig:dmv_company}
\end{figure}

We disaggregated the 793 California DMV collision reports by manufacturer to examine company-specific dynamics (Figure~\ref{fig:dmv_company}). Waymo and Cruise account for the majority of reported collisions, with distinct temporal profiles.

Applying \gls*{pelt} to the Cruise-specific series detects a changepoint within three months of the October 2023 suspension (4 segments detected, breakpoint at index 57 vs suspension at index 57), confirming that governance interventions produce detectable structural breaks in mandatory reporting data. This structural break is not visible in the AIID record, providing a concrete example of why model-centric analysis of media-reported data alone misses governance-relevant regime shifts that are apparent in mandatory reporting.


\section{Deepfake: Additional Analyses}
\label{sec:c_deepfakes}

This appendix provides full methodological details and supplementary analyses for the deepfake case study (Section~\ref{sec:4_deepfakes}). Subsections follow the logical order of the pipeline: exposure and risk construction first, then phase classification rules and model parameters, followed by sensitivity analyses, extended interpretive notes, and the optional policy artifact.

\subsection{Media-Adjusted Risk Construction via Negative Binomial Regression}
\label{sub:c_media_adj_risk_nb_df}

\subsubsection*{Raw Incident Series}

Figure~\ref{fig:03_time_series_overview} in the main text shows the raw incident and report counts before any adjustment. The raw series is used as input to the nowcasting correction and subsequently to the NB GLM described below.

\subsubsection*{Reporting Lag and Nowcasting}
A feature of this domain is reporting speed. While incidents before 2019 often had lags of several years (retrospective discovery), the post-2021 period has a median discovery lag of just 1.0 day (mean 17.0 days), suggesting the database captures near-real-time public concern. To address possible delays in recent months, we applied a nowcasting correction based on the observed distribution of reporting lags. The empirical CDF shows that 97.7\% of post-2021 incidents are captured within 4 months of occurrence. We therefore apply corrections over a 4-month nowcast window, inflating raw counts by up to $5\times$ for the most recent month. This prevents the artificial ``drop-off'' in late 2025, ensuring that observed declines reflect a more realistic trends rather than incomplete reporting.

\subsubsection*{Exposure and Media Index Construction}

A challenge in analyzing deepfake incidents is disentangling \textit{true risk escalation} from \textit{media amplification}. Unlike physical AI systems (e.g., autonomous vehicles) where incident rates can be normalized by deployment scale (vehicle-miles), deepfakes lack a clear exposure denominator. We addressed this via a two-step normalization strategy with the resources listed in Table~\ref{tab:exposure_media_deepfake}: (1) constructing a data-driven exposure index based on open-source tool availability, and (2) modeling the relationship between incidents, exposure, and media attention using count regression.

\begin{table}[htbp]
    \caption{Exposure and Media Adjustment Strategies for the Deepfake Case Study.}
    \label{tab:exposure_media_deepfake}
    \centering
    \small
    \begin{tabular}{p{2.5cm}p{4.2cm}p{4.3cm}}
    \toprule
    Case & Exposure Proxy & Media Adjustment \\
    \midrule
    \textbf{Deepfakes} &
    GitHub Stars for top 8 repositories with 24-month half-life depreciation to create ``Depreciated Installed Base.'' Represents the persistent technical capacity and persistent tool availability in the wild. Used as generalized linear model offset. Scaled to [10, 100] for interpretability. &
    Google Trends index for ``deepfake'' (worldwide, monthly) used as covariate in Negative Binomial GLM. Media amplification detected but not statistically significant after controlling for exposure and quadratic trend (Rate Ratio = 1.88, $p = 0.121$): a 1 SD increase in search volume nearly doubles expected incident count, suggesting discovery/reporting bias, though the effect does not reach conventional significance. \\
    \bottomrule
    \end{tabular}
\end{table}

We constructed the exposure index based on the ``Depreciated Installed Base'' of GitHub stars for top 8 deepfake repositories (e.g., \textit{FaceSwap}, \textit{DeepFaceLab}, \textit{Roop}, \textit{FaceFusion}). We applied a 24-month half-life depreciation to account for tool obsolescence, ensuring the index reflects the currently active arsenal of accessible tools rather than just historical interest. This proxy reflects the hypothesis that the proliferation of accessible software lowers the barrier to entry, thereby elevating future risk potential. A sensitivity analysis across half-lives of 6, 12, and 24 months confirms that the binary dormant-to-endemic structural break is robust to this choice (Appendix~\ref{sub:c_halflife_sensitivity_df}). Additionally, we integrated Google Trends data for the search term ``deepfake'' (worldwide, monthly) as a media visibility index, capturing the hype cycle's influence on reporting propensity. The media index ranges from 0 (no search interest) to 100 (peak interest), with a mean of 39.4 and standard deviation of 24.3 over the observation period.

\subsubsection*{Model Selection and Overdispersion}

We initially fitted a Poisson GLM with log link, using nowcasted incidents as the outcome and log-exposure as an offset term. However, the Poisson assumption of equidispersion was violated: the Pearson chi-squared dispersion ratio was 1.85 (Pearson $\chi^2 = 185.2$, df~$= 99$), and the Poisson log-likelihood of $-169.71$ was substantially worse than the Negative Binomial fit, confirming over-dispersion. This over-dispersion reflects the ``viral'' nature of deepfake incidents, where a single high-profile case (e.g., the Taylor Swift deepfake, January 2024) can generate an order-of-magnitude spike in global attention.

We therefore fitted a Negative Binomial GLM, which introduces an additional dispersion parameter $\alpha$ to accommodate extra-Poisson variation. We set $\alpha = 1.5$ based on preliminary model exploration balancing goodness-of-fit against residual structure. Although a grid search over $\alpha \in [0.5, 2.5]$ identifies $\alpha = 0.5$ as the log-likelihood maximizer ($\text{LL} = -151.2$ vs.\ $-163.5$ at $\alpha = 1.5$), values below 1.0 produce Pearson dispersion ratios near zero, leaving residuals too smooth for downstream changepoint detection. Since the NB residuals serve as the input to the PELT algorithm, we selected a specification that adequately corrects for overdispersion while preserving enough residual structure for regime detection. A sensitivity check confirms that the binary dormant-to-endemic structural break is robust across $\alpha \in \{0.5, 1.0, 1.5, 2.0\}$ (Section~\ref{sub:c_halflife_sensitivity_df}). The selected model achieved a log-likelihood of $-163.48$, substantially outperforming the Poisson specification (Table~\ref{tab:nb_comparison_df}).

\begin{table}[htbp]
    \centering
    \caption{Model Comparison: Poisson vs. Negative Binomial GLM for Deepfakes.}
    \label{tab:nb_comparison_df}
    \begin{tabular}{lcc}
    \toprule
    Metric & Poisson & Negative Binomial (Quad) \\
    \midrule
    Log-Likelihood & $-169.71$ & $-163.48$ \\
    Pearson Dispersion Ratio & 1.85 & 0.48 \\
    \bottomrule
    \end{tabular}
\end{table}

The NB model's Pearson dispersion ratio of 0.48 indicates residual underdispersion, implying that the estimated standard errors are modestly conservative and reported $p$-values slightly overstated.

Preliminary model diagnostics showed that deepfake risk is \textit{accelerating}. We therefore extended the Negative Binomial model to include both linear and quadratic time terms, centered at the mean observation date to reduce multicollinearity.

The final specification is:
\[
\log(\mu_t) = \alpha + \beta_1 \cdot t_{\text{centered}} + \beta_2 \cdot t_{\text{centered}}^2 + \beta_3 \cdot \text{Media}_{\text{std}} + \log(\text{Exposure}_t)
\]

where $\mu_t$ is the expected incident count at time $t$, and Media$_{\text{std}}$ is the standardized Google Trends index (mean-centered, unit variance).

\begin{table}[htbp]
    \centering
    \caption{Negative Binomial GLM Coefficient Estimates (Quadratic Model).}
    \label{tab:nb_glm_df}
    \small
    \begin{tabular}{l c c c c}
    \toprule
    Parameter & Estimate & Std. Error & Rate Ratio & $p$-value \\
    \midrule
    Intercept & $-3.801$ & 0.268 & --- & $<0.001$ \\
    Time (linear) & $0.112$ & 0.177 & 1.119 & 0.526 \\
    Time (quadratic) & $0.068$ & 0.035 & 1.070 & $0.055$ \\
    Media Index (std) & $0.633$ & 0.408 & 1.883 & $0.121$ \\
    \bottomrule
    \end{tabular}
\end{table}

Table~\ref{tab:nb_glm_df} presents the fitted coefficients. Three findings warrant emphasis:

\begin{enumerate}
    \item \textbf{Accelerating Risk}: The quadratic time term is positive and borderline significant ($\beta_2 = 0.068$, $p = 0.055$), providing suggestive evidence that the rate of incidents is \textit{accelerating}. Each unit increase in centered time-squared multiplies the expected rate by 1.07. Over the 103-month observation window, this quadratic structure captures the explosive post-2023 trajectory that a linear model would underestimate. The marginal significance is consistent with the model already absorbing much of the trend through the exposure offset.

    \item \textbf{Media Amplification}: The standardized media index shows a substantial effect size ($\beta_3 = 0.633$, rate ratio of 1.88) but does not reach conventional significance ($p = 0.121$). This implies that a 1-standard-deviation increase in Google Trends search volume (approximately 24 points on the 0--100 scale) is associated with an 88\% increase in expected incidents, holding exposure and time constant. While the wide confidence interval prevents definitive causal claims, the magnitude suggests that discovery propensity is meaningfully influenced by media attention cycles---a confound we must account for to assess truth risk.

    \item \textbf{Linear Time Insignificance}: The linear time term is non-significant ($\beta_1 = 0.112$, $p = 0.526$). This is consistent with a ``hockey stick'' growth pattern: once the quadratic component captures the post-2023 acceleration and the exposure offset absorbs baseline growth, the linear term's marginal contribution vanishes.

\end{enumerate}

\paragraph{Media-Adjusted Excess Risk.} Using the fitted model, we constructed the \textbf{media-adjusted excess risk} signal as the difference between observed and expected risk in log space:

\[
\text{Risk}_{\text{excess}}(t) = \log\left(\frac{y_t}{\text{Exposure}_t}\right) - \log\left(\frac{\hat{\mu}_t}{\text{Exposure}_t}\right)
\]

where $y_t$ is the nowcasted incident count and $\hat{\mu}_t$ is the model-predicted count. Positive values indicate that observed incidents exceed what the media hype and tool availability would predict (i.e., real risk escalation). Negative values indicate that the system is ``safer than expected'' given current tools and attention levels.

The deepfake domain's positive quadratic trend ($p = 0.055$) contrasts with the AV domain's negative linear trend (rate ratio $= 0.357$, $p < 0.001$). This divergence reflects different governance dynamics: centralized AV deployment enables incident rate reductions via technical improvements and operational pauses, whereas decentralized deepfake distribution exhibits runaway growth once generative AI tools democratized.

\subsection{Phase Classification Rules}
\label{sub:c_phase_classification_df}

Building on the PELT segmentation, we classified each segment into one of three epidemiological phases based on mean risk level and slope. We used the pre-2022 reference period (Segments~1--2) as a baseline distribution. The threshold $\theta_{\text{low}} = +0.14\sigma$ equals the upper 1-standard-deviation bound of this baseline ($\mu_d + \sigma_d$, where $\mu_d = -0.26\sigma$ and $\sigma_d = 0.40\sigma$). Values above this threshold represent a statistically unusual departure from baseline.

Under standard SPC doctrine, $+2\sigma$ defines the \textit{Epidemic Threshold} and $+3\sigma$ an \textit{Acute Outbreak}.\footnote{The SPC reference window extends to end-2022 ($<$ January 2023), which includes the first 10 months of PELT Segment~3 (March--December 2022). In a strictly SPC-pure design, the baseline should be limited to ``in-control'' data only (i.e., pre-March 2022). However, this boundary choice does not affect conclusions: the highest-risk month in Segment~3 reaches only $+0.43\sigma$ (standardized), while the SPC Epidemic Threshold ($+2\sigma$) and Acute Outbreak threshold ($+3\sigma$) remain well above any observed value under either baseline definition. Tightening the baseline to pre-March 2022 shifts $\mu$ by approximately $+0.12\sigma$, which does not alter whether any month is considered to breach the SPC thresholds.}

The three phases are defined as follows:

\begin{enumerate}
    \item \textbf{Dormant Baseline} (Risk $< +0.14\sigma$): The historical incubation period where incidents remained constrained to early technical demonstrations or localized misuse. Corresponds to PELT Segments~1--2 and the low-risk K-Means macro-band.
    \item \textbf{Active Outbreak} (Risk $\ge +2\sigma$ \textit{or} Slope $> \tau_{\text{rapid}}$): Months exhibiting explosive growth or actively breaching the SPC Epidemic Threshold. The slope threshold $\tau_{\text{rapid}} = \max(\text{P}_{75}(\text{slopes}), 0.05)$ per month is calibrated to the observed segment distribution; for this dataset $\tau_{\text{rapid}} = 0.05$, matching the fixed value used in sensitivity analyses.
    \item \textbf{Endemic Unmitigated} (Risk $\ge +0.14\sigma$, not meeting Outbreak criteria): The elevated stabilization state where risk persists structurally above the dormant reference distribution but has ceased rapid escalation. Corresponds to PELT Segment~3 and the high-risk K-Means macro-band.
\end{enumerate}

A sensitivity sweep confirms that the binary dormant-to-endemic classification is stable across $\theta \in (-0.05\sigma, +0.43\sigma)$ (Appendix~\ref{sub:c_pelt_sensitivity_df}).

\subsubsection*{Cross-Walk with the Six-Phase Framework}

To verify consistency with the AV methodology, we applied the full six-phase classification framework (Table~\ref{tab:phase_criteria}) to the primary 3-segment PELT model. Of six possible phases, only two are populated: \textit{No Evidenced Occurrence} (58.3\%) and \textit{Endemic Mitigated} (41.7\%). The remaining four phases are never occupied, confirming that the deepfake domain lacks the oscillatory dynamics that populate four of six phases in the AV domain (full details in Appendix~\ref{sub:c_pelt_sensitivity_df}). The 5-segment sensitivity model populates three 6-phase labels (adding Rare Mitigated for the Oct 2021--Dec 2022 transitional segment), but confirms the same binary conclusion: all pre-2022 months fall into low-risk phases, and the post-2023 period is the sole elevated-risk epoch.

A clarification on labeling: the six-phase system classifies Segment~3 as \textit{Endemic Mitigated} because its risk ($+0.43\sigma$) falls between $\theta_{\text{low}}$ and $\theta_{\text{high}}$ with a declining slope, whereas the simplified three-phase system used in the main text classifies it as \textit{Endemic Unmitigated} because any risk above $\theta_{\text{low}}$ qualifies. Both systems agree that Segment~3 represents a structurally elevated endemic regime; they differ only on whether the modest within-segment decline constitutes evidence of mitigation. We adopt the more conservative three-phase label (\textit{Endemic Unmitigated}) for the primary analysis because no external intervention has been shown to reduce deepfake risk (Section~\ref{sub:c_df_limitations}).

\subsubsection*{Observations on the Phase Distribution}

Two features of the simplified distribution merit emphasis. First, the absence of months classified as Active Outbreak (0\%) is a substantive finding. Unlike the AV domain, where the 2018 Uber fatality generated a brief but unmistakable spike, the deepfake trajectory never exhibits a single-month catastrophic breach. Instead, the risk signal climbed above baseline and stayed there, \textit{bypassing} the Outbreak phase entirely and settling directly into Endemic status---a 43-month plateau at $+0.43\sigma$ (Segment~3).

Second, the 58.3\% Dormant classification reflects a long pre-2022 incubation period (Segments~1--2). The deepfake domain spent nearly five years in dormancy before transitioning into endemicity in March 2022. This contrasts with the AV domain, reflecting the industry's persistent capacity for phase regression after crisis events.

\subsection{Model Parameters Quick Reference}
\label{sub:c_model_parameters_df}

Table~\ref{tab:df_model_params} consolidates the delay model and forecasting parameters that inform the pipeline but are not required for reading the main results.

\begin{table}[htbp]
    \centering
    \caption{Deepfake Model Parameters: Delay and Forecasting.}
    \label{tab:df_model_params}
    \small
    \begin{tabular}{l l l}
    \toprule
    Model & Parameter & Value \\
    \midrule
    Delay model & Median lag (post-2021) & 1.0 days \\
                & Mean lag (post-2021) & 17.0 days \\
                & Nowcast window & 4 months \\
                & Empirical CDF $P(\text{lag} \le 4\text{m})$ & 97.7\% \\
                & Max nowcast correction cap & $5\times$ \\
    \addlinespace
    Forecasting ensemble & Models used & ARIMA(1,1,1), Prophet, GPR \\
                         & Divergence (Max--Min)/Max & 46.7\% \\
                         & Ensemble mean (12-month) & 9.3 incidents/month \\
    \bottomrule
    \end{tabular}
\end{table}

\subsection{PELT Sensitivity and Six-Phase Classification}
\label{sub:c_pelt_sensitivity_df}

\subsubsection{PELT Plateau Stability Analysis}

The PELT changepoint algorithm was applied with an adaptive penalty sweep across $\rho \in [0.5, 10.0]$. The plateau stability analysis identifies the most robust segmentation as the one persisting across the widest range of penalty values:

\begin{table}[htbp]
    \centering
    \caption{PELT Plateau Stability Analysis for Deepfake Incidents.}
    \label{tab:pelt_plateau_df}
    \small
    \begin{tabular}{ccc}
    \toprule
    Segments & Appearances & Penalty Range \\
    \midrule
    8 & 1 & $\rho = 0.5$ \\
    5 & 4 & $\rho \in [1.0, 2.5]$ \\
    \textbf{3} & \textbf{10} & $\boldsymbol{\rho \in [3.0, 7.5]}$ \\
    2 & 5 & $\rho \in [8.0, 10.0]$ \\
    \bottomrule
    \end{tabular}
\end{table}

The 3-segment solution dominates across 50\% of all penalty values tested, indicating it represents the most stable macro-regime structure. The 5-segment solution, while less stable, provides useful sub-regime detail and is reported below.

\subsubsection{5-Segment PELT Decomposition}

The 5-segment decomposition ($\rho = 1.00$) reveals internal structure within the 3-segment macro-regimes:

\begin{table}[htbp]
    \centering
    \caption{5-Segment PELT Decomposition for Deepfake Incidents ($\rho = 1.00$).}
    \label{tab:pelt_5seg_df}
    \small
    \begin{tabular}{c l c c c l}
    \toprule
    Seg & Period & Months & Risk ($\sigma$) & Slope & Interpretation \\
    \midrule
    1 & Mar 2017 -- Jun 2020 & 40 & $-0.05$ & $+0.005$ & Pre-commercial baseline \\
    2 & Jul 2020 -- Sep 2021 & 15 & $-1.01$ & $-0.062$ & COVID-era suppression \\
    3 & Oct 2021 -- Dec 2022 & 15 & $-0.05$ & $+0.046$ & Rising baseline \\
    4 & Jan 2023 -- Jan 2025 & 25 & $+0.69$ & $-0.030$ & The 2023 Shock \\
    5 & Feb 2025 -- Sep 2025 & 8 & $+0.05$ & $-0.572$ & Late-period decline \\
    \bottomrule
    \end{tabular}
\end{table}

Compared to the 3-segment model (Table~\ref{tab:df_summary}), the 5-segment decomposition splits the dormant macro-regime into three sub-phases: a pre-commercial baseline, a COVID-era suppression, and a rising baseline. And it splits the endemic macro-regime into a 2023 shock and a late-period decline. Also, the qualitative finding is unchanged: the system transitioned irreversibly from a dormant state to an elevated state, with no structural recovery.

Segment~5 (February--September 2025) merits interpretive caution. Despite 13.2 incidents per month (the highest absolute count in the dataset), its risk-adjusted signal is only $+0.05\sigma$ (near the historical mean) because exposure (as proxied by GitHub stars of deepfake repositories) has increased substantially. This means that the incident rate \textit{per unit of tool availability} has returned to baseline levels, even as absolute incident counts remain high. Whether this reflects real risk normalization or simply the denominator outpacing the numerator remains an open question.

\subsubsection{Six-Phase Classification (AV-Aligned)}

To verify cross-domain consistency, we applied the full six-phase classification framework (Table~\ref{tab:phase_criteria}) to both the 3-segment and 5-segment PELT models. For the deepfake domain, we set $\theta_{\text{low}} = +0.14\sigma$ (upper 1$\sigma$ bound of the dormant reference distribution) and $\theta_{\text{high}} = +0.54\sigma$ (upper 2$\sigma$ bound).

\paragraph{3-Segment Model + 6 Phases.}

\begin{table}[htbp]
    \centering
    \caption{Six-Phase Classification Applied to 3-Segment PELT Model (Deepfakes).}
    \label{tab:6phase_3seg_df}
    \small
    \begin{tabular}{c l c c l}
    \toprule
    Seg & Period & Risk ($\sigma$) & Incidents/mo & 6-Phase Label \\
    \midrule
    1 & Mar 2017 -- Jun 2020 & $-0.05$ & 0.2 & No Evidenced Occurrence \\
    2 & Jul 2020 -- Feb 2022 & $-0.83$ & 0.3 & No Evidenced Occurrence \\
    3 & Mar 2022 -- Sep 2025 & $+0.43$ & 6.5 & Endemic Mitigated \\
    \bottomrule
    \end{tabular}
\end{table}

\begin{table}[htbp]
    \centering
    \caption{Phase Distribution: 3-Segment + 6-Phase (Deepfakes).}
    \label{tab:6phase_3seg_dist_df}
    \small
    \begin{tabular}{lrr}
    \toprule
    Phase & Months & \% \\
    \midrule
    No Evidenced Occurrence & 60 & 58.3 \\
    Rare Mitigated & 0 & 0.0 \\
    Rare Occurrence & 0 & 0.0 \\
    Endemic Mitigated & 43 & 41.7 \\
    Rapid Expansion & 0 & 0.0 \\
    Endemic Unmitigated & 0 & 0.0 \\
    \bottomrule
    \end{tabular}
\end{table}

Of six possible phases, only two are ever populated. Segment~3 is classified as ``Endemic Mitigated'' (rather than ``Endemic Unmitigated'') because its risk level ($+0.43\sigma$) falls between $\theta_{\text{low}}$ (+0.14) and $\theta_{\text{high}}$ (+0.54), with a declining slope ($-0.009$). All pre-2022 months are classified as ``No Evidenced Occurrence'' due to average incident counts below 0.5 per month.

\paragraph{5-Segment Model + 6 Phases.}

\begin{table}[htbp]
    \centering
    \caption{Six-Phase Classification Applied to 5-Segment PELT Model (Deepfakes).}
    \label{tab:6phase_5seg_df}
    \small
    \begin{tabular}{c l c c c l}
    \toprule
    Seg & Period & Risk ($\sigma$) & Slope & Inc/mo & 6-Phase Label \\
    \midrule
    1 & Mar 2017 -- Jun 2020 & $-0.05$ & $+0.005$ & 0.2 & No Evidenced Occurrence \\
    2 & Jul 2020 -- Sep 2021 & $-1.01$ & $-0.062$ & 0.3 & No Evidenced Occurrence \\
    3 & Oct 2021 -- Dec 2022 & $-0.05$ & $+0.046$ & 0.7 & Rare Mitigated \\
    4 & Jan 2023 -- Jan 2025 & $+0.69$ & $-0.030$ & 6.7 & Endemic Unmitigated \\
    5 & Feb 2025 -- Sep 2025 & $+0.05$ & $-0.572$ & 13.2 & Rare Mitigated \\
    \bottomrule
    \end{tabular}
\end{table}

The 5-segment model uses three of six phases, with Segment~4 now classified as ``Endemic Unmitigated'' (risk $+0.69\sigma > \theta_{\text{high}} = +0.54\sigma$). Segment~3 is classified as ``Rare Mitigated'' because its incident rate (0.7/month) slightly exceeds 0.5 but its risk remains below $\theta_{\text{low}}$. Segment~5 is also ``Rare Mitigated'' despite 13.2 incidents/month, because the risk-adjusted signal ($+0.05\sigma$) is below $\theta_{\text{low}}$.

\begin{table}[htbp]
    \centering
    \caption{Phase Distribution: 5-Segment + 6-Phase (Deepfakes).}
    \label{tab:6phase_5seg_dist_df}
    \small
    \begin{tabular}{lrr}
    \toprule
    Phase & Months & \% \\
    \midrule
    No Evidenced Occurrence & 55 & 53.4 \\
    Rare Mitigated & 23 & 22.3 \\
    Rare Occurrence & 0 & 0.0 \\
    Endemic Mitigated & 0 & 0.0 \\
    Rapid Expansion & 0 & 0.0 \\
    Endemic Unmitigated & 25 & 24.3 \\
    \bottomrule
    \end{tabular}
\end{table}

\subsubsection{Cross-Domain Comparison}

Table~\ref{tab:av_df_6phase_comparison} contrasts the six-phase distributions across both domains.

\begin{table}[htbp]
    \centering
    \caption{Six-Phase Distribution: AV vs.\ Deepfake (3-Segment).}
    \label{tab:av_df_6phase_comparison}
    \small
    \begin{tabular}{lrr}
    \toprule
    Phase & AV (\%) & DF (\%) \\
    \midrule
    No Evidenced Occurrence & 57 & 58.3 \\
    Rare Mitigated & 0 & 0.0 \\
    Rare Occurrence & 0 & 0.0 \\
    Endemic Mitigated & 20 & 41.7 \\
    Rapid Expansion & 12 & 0.0 \\
    Endemic Unmitigated & 10 & 0.0 \\
    \bottomrule
    \end{tabular}
\end{table}

While the AV domain populates four of six phases (No Evidenced Occurrence, Endemic Mitigated, Rapid Expansion, and Endemic Unmitigated), the deepfake domain populates only two (No Evidenced Occurrence and Endemic Mitigated), reflecting an irreversible step-function transition with no recovery mechanism. This collapse of the six-phase framework to a binary structure suggests the empirical basis for the simplified three-phase classification used in the main text.

\subsubsection{Phase Threshold Sensitivity Sweep}

We swept $\theta_{\text{low}}$ across $[-0.8\sigma, +0.5\sigma]$ and classified each PELT segment under the 3-phase framework. Table~\ref{tab:threshold_sweep_df} reports the results for the 3-segment model.

\begin{table}[htbp]
    \centering
    \caption{Phase Classification Sensitivity to $\theta_{\text{low}}$ (3-Segment Model).}
    \label{tab:threshold_sweep_df}
    \small
    \begin{tabular}{r c c c r r}
    \toprule
    $\theta_{\text{low}}$ & Seg 1 & Seg 2 & Seg 3 & Dormant \% & Endemic \% \\
    \midrule
    $-0.50$ & E & D & E & 19.4 & 80.6 \\
    $-0.30$ & E & D & E & 19.4 & 80.6 \\
    $-0.10$ & E & D & E & 19.4 & 80.6 \\
    \textbf{+0.14} & \textbf{D} & \textbf{D} & \textbf{E} & \textbf{58.3} & \textbf{41.7} \\
    $+0.30$ & D & D & E & 58.3 & 41.7 \\
    $+0.40$ & D & D & E & 58.3 & 41.7 \\
    $+0.50$ & D & D & D & 100.0 & 0.0 \\
    \bottomrule
    \end{tabular}

    \vspace{0.5em}
    {\footnotesize \textit{Note:} D = Dormant Baseline, E = Endemic Unmitigated. The phase flip occurs at $\theta = -0.05\sigma$ (Seg~1 risk level). The invariant zone spans $\theta \in (-0.05\sigma, +0.43\sigma)$, within which all classifications are identical. The selected threshold $\theta_{\text{low}} = +0.14\sigma$ lies within this zone.}
  \end{table}

The binary transition finding is robust: for any $\theta_{\text{low}}$ within the invariant zone $(-0.05\sigma, +0.43\sigma)$, the classification assigns exactly 58.3\% of months to Dormant and 41.7\% to Endemic. Thresholds below $-0.05\sigma$ (including the AV-aligned $-0.3\sigma$) misclassify Segment~1 (0.2 incidents/month) as endemic---epidemiologically implausible for a period with near-zero incidence. Thresholds above $+0.43\sigma$ classify all segments as dormant, which is equally implausible given Segment~3's sustained 6.5 incidents/month.

\subsection{Exposure Half-Life Sensitivity}
\label{sub:c_halflife_sensitivity_df}

The exposure index uses a 24-month exponential half-life to depreciate GitHub Stars and construct the Depreciated Installed Base. To verify that this choice does not drive the binary dormant-to-endemic finding, we re-estimated the full pipeline---NB GLM, residual extraction, and PELT changepoint detection (penalty $\rho = 3.0$)---under half-lives of 6, 12, and 24 months. Table~\ref{tab:halflife_sensitivity_df} reports the PELT results for each specification.

\begin{table}[htbp]
    \centering
    \caption{PELT Binary Finding Across Exposure Half-Lives.}
    \label{tab:halflife_sensitivity_df}
    \small
    \begin{tabular}{l c l c c c}
    \toprule
    Half-Life & Segs & Main Endemic Break & Dormant Seg Risk & First Endemic Risk & Finding \\
    \midrule
    6 months  & 4 & Feb 2022 & $-0.59\sigma$ & $+0.88\sigma$ & Confirmed \\
    12 months & 4 & Feb 2022 & $-0.58\sigma$ & $+0.57\sigma$ & Confirmed \\
    24 months & 4 & Sep 2021 & $-0.64\sigma$ & $+0.50\sigma$ & Confirmed \\
    \bottomrule
    \end{tabular}
\end{table}

\subsection{K-Means Supplementary Analysis}

As a supplementary consistency check on the PELT segmentation, we applied K-means clustering to a two-dimensional feature space: standardized risk level and local trend (gradient). Both features were independently standardized to prevent one from dominating. We tested $K=3$ to $K=6$, and the algorithm selected $K=5$ based on the highest \Gls{silhouette} score (0.393).

The silhouette score of 0.393 is moderate, reflecting the noisy monthly risk signal and the inherent difficulty of separating overlapping clusters in two dimensions. K-means is therefore treated as a complementary descriptive tool. Despite the modest cluster separation, the five clusters collapse into \textbf{two macro-bands} distinguished by risk level, with variation within each band driven by trend direction (Table~\ref{tab:kmeans_df}).

\begin{table}[htbp]
    \centering
    \caption{K-Means Cluster Characteristics (2D: Level + Trend, $K=5$, Silhouette = 0.393)}
    \label{tab:kmeans_df}
    \small
    \begin{tabular}{c c c c c l}
    \toprule
    Cluster & Risk ($\sigma$) & Trend & $n$ & \% Months & Interpretation \\
    \midrule
    \multicolumn{6}{l}{\textit{Low-Risk Macro-Band ($\le -0.5\sigma$)}} \\
    C3 & $-1.03$ & $-0.769$ & 14 & 13.6 & Low risk, declining \\
    C4 & $-0.86$ & $+0.758$ & 12 & 11.7 & Low risk, rising \\
    C1 & $-0.70$ & $-0.055$ & 25 & 24.3 & Low risk, declining \\
    \midrule
    \multicolumn{6}{l}{\textit{High-Risk Macro-Band ($\ge +0.5\sigma$)}} \\
    C2 & $+0.77$ & $+0.281$ & 32 & 31.1 & High risk, rising \\
    C0 & $+0.88$ & $-0.512$ & 20 & 19.4 & High risk, declining \\
    \bottomrule
    \end{tabular}
\end{table}

\begin{figure}[htbp]
    \centering
    \includegraphics[width=\textwidth]{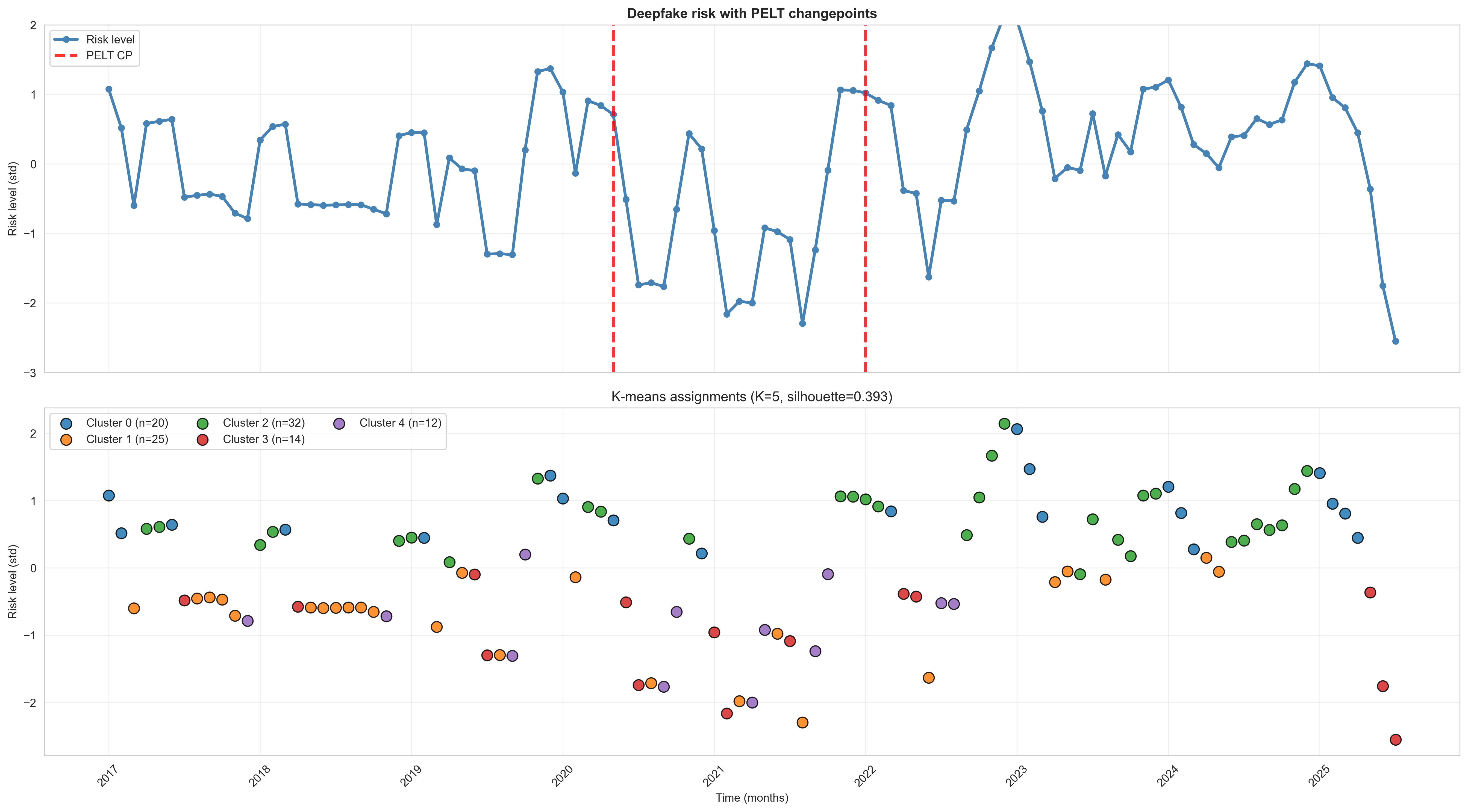}
    \caption{K-means supplementary analysis of PELT segmentation. \textit{Top Panel}: Risk trajectory with PELT changepoints. \textit{Bottom Panel}: K-means cluster assignments ($K=5$) showing the dominance of high-risk clusters after March 2022.}
    \label{fig:kmeans_validation_df}
\end{figure}

The Low-Risk Macro-Band (clusters C1, C3, C4; 49.5\% of months) captures months where risk operates below $-0.5\sigma$, collectively representing the pre-2022 incubation period. The High-Risk Macro-Band (clusters C0, C2; 50.5\% of months) captures months at or above $+0.5\sigma$. This macro-band structure aligns closely with the PELT segmentation. Segments 1--2 are dominated by low-risk clusters, whereas Segment 3 is predominantly high-risk, with 30 of 43 months classified into high-risk clusters (Figure~\ref{fig:kmeans_validation_df}).

This two-band structure differs from AV, where the HMM captured an oscillatory crisis-recovery pattern (see Section~\ref{sec:3_autonomous_vehicles} for the full cross-domain comparison). Once the deepfake system crossed into the High-Risk Macro-Band in early 2022, it remained there for 43 consecutive months (PELT Segment~3) with no structural recovery.

As a further robustness check, a standalone sensitivity analysis using the same 2D standardized clustering approach (standardized risk level and trend, $K \in [2,7]$) applied to an independently reconstructed risk signal confirmed an optimal clustering that separates a low-risk pre-2022 macro-band from a high-risk post-2022 macro-band, consistent with the main binary structural finding.\footnote{These differences test whether the binary structural finding depends on specific pipeline choices. Despite these variations, the qualitative two-band separation---low-risk pre-2022 vs.\ high-risk post-2022---is consistent across both implementations.}

\subsection{Intervention Impact Analysis}

We analyzed the impact of 22 key governance events on deepfake risk, ranging from state laws (e.g., Texas SB 751) to platform policies (e.g., TikTok bans) and international frameworks (e.g., EU AI Act) (see Appendix~\ref{subsec:c_policy_df} for the full list of policies). For each intervention, we compared the mean media-adjusted excess risk in the 3-month window before and after the event. This window size was selected based on the autocorrelation decay of the risk signal. We applied independent samples t-tests with false discovery rate (FDR) correction to account for multiple comparisons. This before--after design captures temporal associations: interventions are not randomly assigned, and the 3-month window may overlap with concurrent events or secular trends. Results should be interpreted as descriptive evidence of whether risk trajectories shifted in the vicinity of governance milestones.

\begin{table}[htbp]
    \centering
    \caption{Selected deepfake intervention impact analysis}
    \label{tab:df_interventions}
    \small
    \begin{tabular}{l l c c c c l}
    \toprule
    Intervention & Date & Pre-Mean & Post-Mean & $\Delta\sigma$ & $p_{\text{FDR}}$ & Direction \\
    \midrule
    Texas SB 751 & 2019-09 & $-0.54$ & $-1.08$ & $-0.54$ & 0.305 & None \\
    Facebook Detection Chal. & 2019-12 & $-1.42$ & $+0.35$ & $+1.77$ & \textbf{0.006} & \textbf{Deterioration} \\
    CA AB 730 & 2020-01 & $-1.08$ & $+0.01$ & $+1.09$ & 0.146 & None \\
    TikTok Policy Update & 2020-08 & $+0.05$ & $-1.73$ & $-1.78$ & \textbf{0.002} & \textbf{Mitigation} \\
    Bletchley Decl. & 2023-11 & $-0.41$ & $+0.03$ & $+0.44$ & 0.288 & None \\
    EU AI Act (Entry) & 2024-08 & $-0.43$ & $-0.14$ & $+0.29$ & 0.068 & None \\
    \bottomrule
    \end{tabular}
\end{table}

Only a minority of explicit interventions showed statistical significance after FDR correction ($p_{\text{FDR}} < 0.05$) (Table~\ref{tab:df_interventions}). The Facebook Deepfake Detection Challenge paradoxically coincided with a statistically significant escalation in risk ($\Delta = +1.77\sigma$, $p_{\text{FDR}}=0.006$), likely reflecting heightened adversarial discovery rather than suppression. While TikTok's early policy update (August 2020) demonstrated strong mitigation ($\Delta = -1.78\sigma$, $p_{\text{FDR}}=0.002$), major macro-milestones such as the Bletchley Declaration and the EU AI Act failed to structurally suppress risk.

\begin{figure}[htbp]
    \centering
    \includegraphics[width=\textwidth]{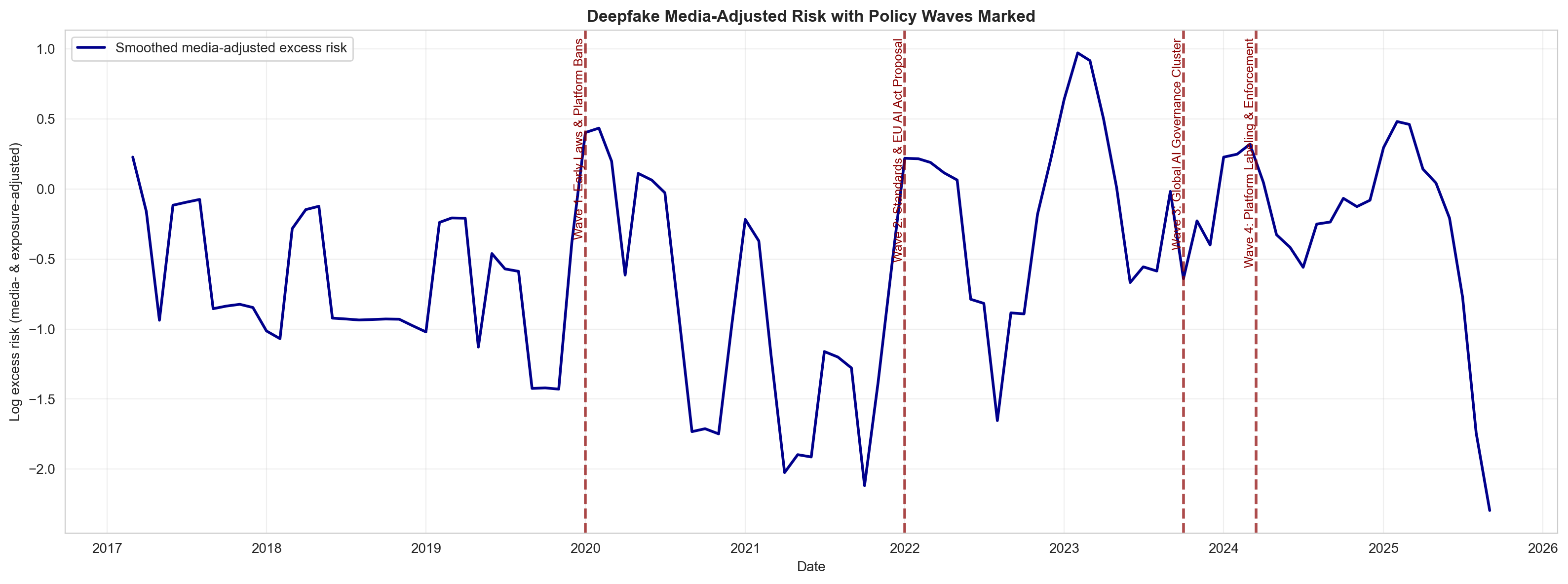}
    \caption{Deepfake Media-Adjusted Risk with Policy Waves Marked. The vertical lines indicate the timing of major governance clusters. Waves 1 and 2 coincide with periods of significant risk deterioration, while Wave 4 coincides with a tentative but non-significant decline.}
    \label{fig:risk_rate_waves}
\end{figure}

To account for potential synergistic effects, we aggregated interventions into four temporal ``waves'' corresponding to coordinated governance efforts, using a 6-month impact window based on the ACF-derived horizon as shown in Figure~\ref{fig:risk_rate_waves}.

Wave-level analysis shows a sobering pattern (Table~\ref{tab:wave_impact_df}). Waves 1 and 2 exhibited statistically significant deteriorations in risk after FDR correction ($p_{\text{FDR}} = 0.007$ for both). Wave 1 (Early Laws \& Platform Bans) saw the mean risk signal jump from $-0.97\sigma$ to $+0.03\sigma$ ($\Delta = +1.00\sigma$), while Wave 2 (Standards \& EU AI Act Proposal) saw risk rise from $-1.29\sigma$ to $-0.17\sigma$ ($\Delta = +1.12\sigma$). Both deteriorations are consistent with a pre-existing secular upward trend during the structural transition to endemicity. Wave 3 (Global AI Governance Cluster, October 2023) showed a nominal $+116.3\%$ risk increase that did not reach statistical significance ($p_{\text{FDR}} = 0.377$), suggesting that neither the governance response nor the risk signal produced a detectable regime shift at this juncture.

\begin{table}[htbp]
    \centering
    \caption{Wave-Level Intervention Impact on Deepfake Risk.}
    \label{tab:wave_impact_df}
    \small
    \begin{tabular}{l l c c c c l}
    \toprule
    Wave & Date & Pre-Mean & Post-Mean & $\Delta\sigma$ & $p_{\text{FDR}}$ & Direction \\
    \midrule
    Wave 1: Early Bans& 2020-01 & $-0.97$ & $+0.03$ & $+1.00$ & \textbf{0.007} & \textbf{Deterioration} \\
    Wave 2: Standards & 2022-01 & $-1.29$ & $-0.17$ & $+1.12$ & \textbf{0.007} & \textbf{Deterioration} \\
    Wave 3: Global Gov & 2023-10 & $-0.22$ & $+0.04$ & $+0.26$ & 0.377 & None \\
    Wave 4: Platform Label & 2024-03 & $-0.13$ & $-0.29$ & $-0.16$ & 0.378 & None \\
    \bottomrule
    \end{tabular}
\end{table}

Wave 4 (Platform Labeling, March 2024) showed a negative effect size ($\Delta = -0.16\sigma$, moving the mean risk from $-0.13\sigma$ to $-0.29\sigma$), but this did not reach statistical significance ($p_{\text{FDR}} = 0.378$). This marginal direction of effect in both Wave 3 and Wave 4 falling below the significance threshold, suggests that by 2023--2024, neither the governance wave nor the platform-labeling wave produced detectable regime shifts at the wave level. The absolute risk level has not structurally reset back to the pre-2022 Dormant Baseline, indicating that the phase transition to endemicity has not been reversed.

This pattern contrasts with the AV domain, where incidents triggered immediate operational pauses by identifiable corporations (see Section~\ref{sec:3_autonomous_vehicles}). The key distinction is not that governance is universally ineffective in the deepfake domain: early platform-level intervention did work (TikTok, August 2020; $\Delta = -1.78\sigma$, $p_{\text{FDR}} = 0.002$). What has failed is macro-level governance (e.g., international frameworks and legislative milestones) that cannot reach the distributed anonymous user base. The Endemic plateau has proven resilient to interventions that lack enforcement chokepoints.

\subsection{Forecasting}

To project future incident volumes, we fitted a multi-model ensemble to the \textbf{nowcast-adjusted incident time series}. Unlike regime detection (which used the media-adjusted risk signal), forecasting requires count-level outputs for operational planning. We used the multi-model strategy from Appendix~\ref{sub:a_statistical_methods} to bound forecast uncertainty.

  \begin{table}[htbp]
      \centering
      \caption{ARIMA Model Selection for Deepfake Incidents.}
      \label{tab:arima_selection_df}
      \small
      \begin{tabular}{lccc}
      \toprule
      Model & AIC & Log-Likelihood & Rank \\
      \midrule
      \textbf{ARIMA(1,1,1)} & \textbf{500.65} & $-247.32$ & 1 \\
      ARIMA(2,1,1) & 502.46 & $-247.23$ & 2 \\
      ARIMA(1,1,2) & 502.49 & $-247.25$ & 3 \\
      ARIMA(0,1,1) & 503.28 & $-249.64$ & 4 \\
      ARIMA(2,1,2) & 503.79 & $-246.90$ & 5 \\
      ARIMA(0,1,2) & 504.13 & $-249.07$ & 6 \\
      ARIMA(1,1,0) & 504.73 & $-250.36$ & 7 \\
      ARIMA(2,1,0) & 505.71 & $-249.86$ & 8 \\
      \bottomrule
      \end{tabular}
  \end{table}

\begin{table}[htbp]
    \centering
    \caption{12-Month Forecast Comparison for Deepfake Incidents.}
    \label{tab:forecast_comparison_df}
    \begin{tabular}{lccp{5cm}}
    \toprule
    Model & Projection & 95\% CI & Interpretation \\
    \midrule
    ARIMA(1,1,1) & 8.8/mo & [5, 12] & Stabilization at elevated baseline \\
    Prophet & 16.4/mo & [12, 21] & Continued trend extrapolation \\
    GPR & 2.6/mo & [0, 6] & Peak passed; gradual decline \\
    \midrule
    \textit{Ensemble Mean} & \textit{9.3/mo} & --- & \textit{Central estimate} \\
    \bottomrule
    \end{tabular}
\end{table}

\textbf{Model Specification:}
\begin{itemize}
    \item \textbf{ARIMA}: We evaluated eight ARIMA specifications---$(1,1,0)$, $(0,1,1)$, $(1,1,1)$, $(2,1,0)$, $(0,1,2)$, $(2,1,1)$, $(1,1,2)$, and $(2,1,2)$---on the nowcast-adjusted incident series. ARIMA(1,1,1) was selected as the optimal specification by minimum AIC ($\text{AIC} = 500.65$; next-best: $502.46$).

    \item \textbf{Prophet \citep{taylor2018forecasting}}: We disabled yearly seasonality due to the structural break. The changepoint prior was set to 0.2 to capture the rapid late-2023 acceleration without overfitting.

    \item \textbf{Gaussian Process (GPR) \citep{rasmussen2003gaussian}}: We optimized a composite kernel (Constant $\times$ RBF + WhiteKernel) via marginal likelihood maximization to detect nonlinear cyclical limits.
\end{itemize}

\begin{figure}[htbp]
    \centering
    \includegraphics[width=\textwidth]{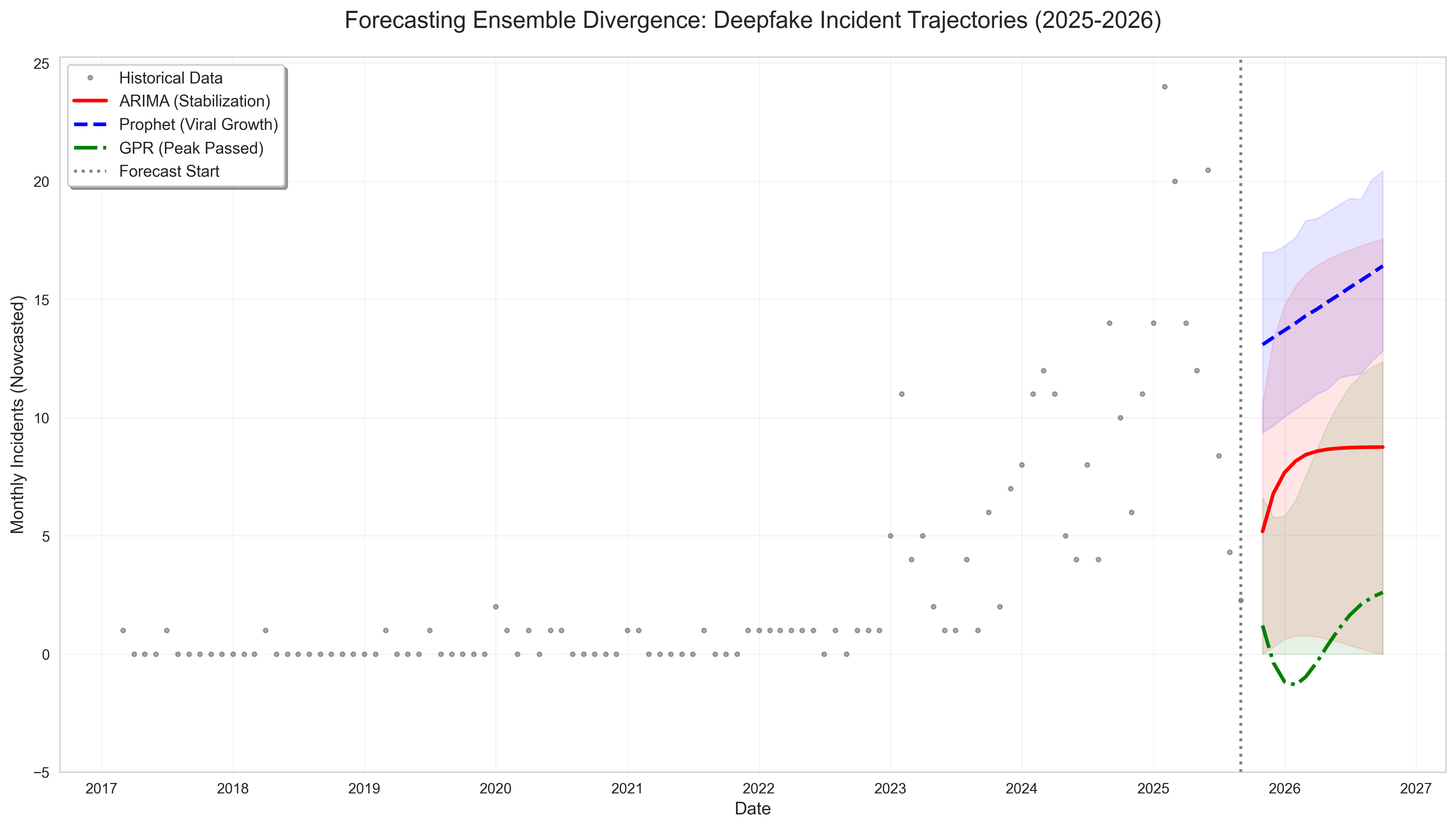}
    \caption{Combined Forecast Results. \textit{Red}: ARIMA(1,1,1) projects stabilization at ~8.8 incidents/month. \textit{Green}: Gaussian Process projects a decline to ~2.6/month. \textit{Blue}: Prophet projects continued acceleration to ~16.4/month. The divergence reflects fundamental uncertainty regarding ecological risk saturation.}
    \label{fig:forecast_combined_df}
\end{figure}

As shown in Table~\ref{tab:forecast_comparison_df} and Figure~\ref{fig:forecast_combined_df}, the three models diverge substantially. This disagreement signals large uncertainty about the system's trajectory.

\begin{itemize}[nosep]
    \item \textbf{Stabilization scenario (ARIMA)}: Projects 8.8 incidents/month [5, 12]. ARIMA treats the recent surge as a transient deviation around a new elevated mean, consistent with a stationary endemic plateau.

    \item \textbf{Decline scenario (GPR)}: Projects 2.6 incidents/month [0, 6]. The GPR's mean-reverting kernel assumes the peak has passed, producing the most optimistic projection. However, GPR's tendency toward prior-mean reversion over long horizons limits its reliability for structural forecasts.

    \item \textbf{Growth scenario (Prophet)}: Projects ~16.4 incidents/month, extrapolating the 2023--2025 trend. Prophet interprets the surge as the start of a sustained growth phase, yielding the most pessimistic projection.
\end{itemize}

The three models bracket a wide range (2.6--16.4 incidents/month), reflecting uncertainty about whether deepfake risk has plateaued or continues to accelerate. ARIMA and GPR both project below-current levels but for different structural reasons (mean reversion vs.\ stationarity), so their numerical overlap should not be interpreted as independent convergence on stabilization. The ensemble mean of 9.3 incidents/month provides a central planning estimate, but the spread underscores that scenario-based planning is the appropriate operational posture.

\subsection{Cross-Domain Comparison Notes}
\label{sub:c_cross_domain_comparison_df}

The contrast between the AV and deepfake risk trajectories presents a structural distinction between centralized and decentralized AI deployment.

In the AV case, the 2018 Uber ATG fatality in Tempe, Arizona triggered an immediate and coordinated institutional response: Uber suspended all autonomous operations, the National Transportation Safety Board launched an investigation, multiple states tightened testing requirements, and the sector collectively reduced public road testing. Risk returned to baseline within 12 months (Segment 3 in the AV PELT decomposition). This response was structurally possible because AV deployment is centralized: a small number of corporations (e.g., Uber, Waymo, Cruise) operate identifiable fleets under government-issued permits. Regulators knew whom to call; the called parties had legal and reputational incentives to comply.

The deepfake case shows the opposite structural profile. Deepfake tools proliferated via open-source repositories (GitHub and other platforms) and were forked thousands of times before any governance framework existed. The eight major repositories analyzed here collectively accumulated millions of stars; each fork represents an independently deployable copy. No single corporate actor controls the technology stack, and no regulatory body possesses jurisdiction over anonymous repository users globally. The Facebook Deepfake Detection Challenge (December 2019) and the EU AI Act (August 2024) both produced statistically null or perverse effects on the risk signal (Appendix~\ref{sub:c_media_adj_risk_nb_df}), consistent with the hypothesis that macro-level governance cannot reach the distributed user base that generates the risk.

This structural asymmetry explains why the same phase-detection pipeline produces qualitatively different outputs from the two domains:
\begin{itemize}
    \item AV: crisis-recovery cycles (Rapid Expansion $\to$ baseline within 12 months)
    \item Deepfake: monotonic endemic persistence (43 consecutive months above threshold, no recovery)
\end{itemize}

The implication for governance design is that effective intervention in decentralized AI ecosystems requires reaching infrastructure chokepoints (cloud compute, model API providers, platform hosting).

\subsection{Methodological Limitations and Threats to Validity}
\label{sub:c_df_limitations}

While PELT and SPC provide a rigorous framework for analyzing deepfake risk, the reliance on crowdsourced databases introduces structural limitations:

\begin{enumerate}
    \item \textbf{Selection Bias and the ``Hype Cycle'' Confound}: The AI Incident Database relies on voluntary reporting and journalistic discovery. Consequently, the observed incident surge in 2023 is inextricably linked to the intense global media attention surrounding generative AI. Although our Negative Binomial regression framework explicitly controls for public attention using a normalized Google Trends media index, no empirical adjustment can perfectly disentangle the true proliferation of risk from the heightened vigilance of database contributors searching for deepfake incidents during an AI hype cycle.

    \item \textbf{Left-Censorship of the Baseline}: The period from 2018 to 2022 serves as the historical baseline for calculating the SPC epidemic control limits ($\pm 2\sigma$). However, prior to 2023, the majority of deepfake misuses involved localized deployment scattered across underground forums. Because early incidents received less mainstream news coverage than recent political and financial disruptions, they are underrepresented in the dataset. This ``left-censorship'' artificially depresses the historical variance ($\sigma$). Thus, the magnitude of the control limit breach during the ``2023 Shock'' may be partially inflated by the artificial suppression of the historical baseline.

    \item \textbf{Qualitative Compression and Severity Agnosticism}: To enable objective mathematical time-series modeling, our incident-centric pipeline implicitly assigns equal statistical weight to all registered incidents. An early academic demonstration video is mathematically indistinguishable from a coordinated multi-million dollar international fraud campaign. Therefore, our definition of an ``Epidemic'' phase captures a \textit{quantitative} shift in incident frequency, and does not differentiate based on the qualitative severity of societal harm.

    \item \textbf{Right-Censoring of the Reporting Tail}: Incident databases such as AIID exhibit systematic right-censoring: incidents from the most recent months are underrepresented because the full cycle of occurrence, discovery, reporting, and editorial curation has not yet completed. For the deepfake corpus, the post-2021 empirical lag distribution has a median of 0 months and a 95th percentile of 4 months, so the pipeline applies an empirical CDF nowcasting correction to the final 4-month window (inflating raw counts by up to 5$\times$ for the most recent month). This hard cap prevents implausible corrections for months with very few observations but introduces a conservative bias: if the true reporting delay distribution has a longer right tail, the capped correction will undercount recent incidents.

    This correction addresses the level of incident counts but does not fully eliminate the risk of \textit{false phase regression}. In principle, a brief tail of artificially low nowcasted counts could trigger a spurious PELT changepoint near the analysis endpoint and cause a short recent segment to be misclassified into a lower-risk phase. The primary structural safeguard is that PELT computes phase labels from \textit{segment means}, not individual monthly values; Segment~3 spans 43 months, so the 4-month nowcast tail carries less than 10\% weight in the mean, making the current \textit{Endemic Unmitigated} classification robust to right-censoring. For the present snapshot analysis (endpoint September 2025), Segment~3 mean risk of $+0.43\sigma$ comfortably exceeds the $+0.14\sigma$ Endemic threshold even after removing the final 4 months. This caveat becomes more material in a live streaming deployment, where a minimum-duration confirmation window (requiring a declining trend to persist for a specified number of months before accepting a phase downgrade) would be a prudent additional safeguard.
\end{enumerate}

Despite these limitations, the persistence of the high-risk Endemic plateau across sensitivity sweeps shows that the post-2022 phase transition represents a structural shift in the socio-technical ecosystem, rather than observational noise.

\subsection{Top 8 Deepfake Repositories on Github as of December 2025.}
\label{subsec:top_8_github}

\begin{itemize}
    \item facefusion/facefusion (The current dominant ``one-click'' face swapper)
    \item s0md3v/roop (The tool that triggered the 2023 viral explosion)
    \item hacksider/Deep-Live-Cam (Real-time face swapping, popular in 2024-2025 fraud)
    \item iperov/DeepFaceLab (The historical standard for high-fidelity ``pro'' deepfakes)
    \item deepfakes/faceswap (The original repository that started the community)
    \item iperov/DeepFaceLive (Live streaming implementation of DeepFaceLab)
    \item RVC-Project/Retrieval-based-Voice-Conversion-WebUI (The leading voice cloning tool)
    \item CorentinJ/Real-Time-Voice-Cloning (An early influential voice cloning repository)
\end{itemize}

\subsection{22 Major Interventions on Deepfakes.}
\label{subsec:c_policy_df}

\begin{itemize}
    \item 2019-09-01: Texas SB 751 (Election Deepfakes), State Law, mitigation.
    \item 2019-12-01: Facebook Deepfake Detection Challenge, Tech Initiative, mitigation.

    \item 2020-01-01: California AB 730/602 Effective, State Law, mitigation.
    \item 2020-01-06: Facebook Deepfake Ban Policy, Platform Policy, mitigation.
    \item 2020-08-01: TikTok Policy Updates, Platform Policy, mitigation.

    \item 2021-04-21: EU AI Act Proposal, Regulatory, mitigation.
    \item 2022-01-26: C2PA Technical Spec v1.0, Standards, mitigation.

    \item 2023-01-10: China Deep Synthesis Provisions Effective, Regulatory, mitigation.
    \item 2023-03-07: Twitch Deepfake Porn Ban, Platform Policy, mitigation.
    \item 2023-04-19: TikTok Synthetic Media Policy Update, Platform Policy, mitigation.
    \item 2023-07-21: White House Voluntary AI Commitments, Standards, mitigation.
    \item 2023-08-01: Minnesota Deepfake Law Effective, State Law, mitigation.
    \item 2023-10-30: US Executive Order on AI, Regulatory, mitigation.
    \item 2023-10-30: G7 Hiroshima Code of Conduct, Standards, mitigation.
    \item 2023-11-01: Bletchley Declaration (AI Safety Summit), Standards, mitigation.

    \item 2024-01-31: UK Online Safety Act (Deepfake Offence), Regulatory, mitigation.
    \item 2024-03-18: YouTube AI Labeling Requirement, Platform Policy, mitigation.
    \item 2024-03-21: UN General Assembly AI Resolution, Standards, mitigation.
    \item 2024-04-05: Meta ``Made with AI'' Labeling Update, Platform Policy, mitigation.
    \item 2024-05-09: TikTok Auto-Labeling for AI Content, Platform Policy, mitigation.
    \item 2024-08-01: EU AI Act (Entry into Force), Regulatory, mitigation.
    \item 2025-02-02: EU AI Act (Prohibitions Effective), Regulatory, mitigation.

\end{itemize}

\end{document}